\documentclass[12pt,a4paper]{article}
\pdfoutput=1
\usepackage{amsfonts}
\usepackage{amssymb}
\usepackage[centertags]{amsmath}
\usepackage{graphicx}
\usepackage{hyperref}
\usepackage{booktabs}
\usepackage{theorem}
\usepackage{array}
\usepackage{epsfig}
\usepackage{colordvi}
\usepackage{color}
\usepackage{ulem}
\usepackage[small,bf]{caption}
\usepackage{cite}
\usepackage{color}
\usepackage{subfigure}
\usepackage{pdflscape}

\def\1{\mathbf{1}}
\def\3{\mathbf{3}}
\def\2{\mathbf{2}}

\def\th{\theta}

\newcommand{\hbeta}{$\mbox{}^3 {\rm H}$ $\beta$-decay \ }

\def\ltap{\ \raisebox{-.4ex}{\rlap{$\sim$}} \raisebox{.4ex}{$<$}\ }
\def\gtap{\ \raisebox{-.4ex}{\rlap{$\sim$}} \raisebox{.4ex}{$>$}\ }

\allowdisplaybreaks[1]
\def\th{\theta}

\def\3{\mathbf{3}}

\usepackage{a4wide}

\newcommand{\bec}{\begin{cases}}
\newcommand{\eec}{\end{cases}}
\newcommand{\beq}{\begin{equation*}}
\newcommand{\eeq}{\end{equation*}}
\newcommand{\be}{\begin{equation}}
\newcommand{\ee}{\end{equation}}
\newcommand{\ba}{\begin{eqnarray}}
\newcommand{\ea}{\end{eqnarray}}

\DeclareMathOperator{\diag}{diag}

\makeatletter

\newcommand{\Rmnum}[1]{\expandafter\@slowromancap\romannumeral #1@}
\makeatother

\begin{document}
%
%
\vspace*{-15mm}
\begin{flushright}
SISSA 59/2017/FISI\\
IPMU17-0167\\
\end{flushright}
\vspace*{0.7cm}

\begin{center}
{\bf{\Large Discrete Flavour Symmetries, Neutrino Mixing and \\
\vspace*{0.3cm}
Leptonic CP Violation}}\\
%
[4mm]
\vspace{0.4cm} 
S. T. Petcov$\mbox{}^{a,b)}$ 
\\[1mm]
\end{center}
\vspace*{0.50cm}
\centerline{$^{a}$ \it SISSA/INFN, Via Bonomea 265, 34136 Trieste, Italy}
\vspace*{0.2cm}
\centerline{$^{b}$ \it Kavli IPMU (WPI), University of Tokyo,
5-1-5 Kashiwanoha, 277-8583 Kashiwa, Japan}
\vspace*{0.8cm}

\begin{abstract}
\noindent
The current status of our knowledge of the 
3-neutrino mixing parameters and of 
the CP violation in the lepton sector is summarised. 
The non-Abelian discrete symmetry approach to understanding the 
observed pattern of neutrino mixing and the related predictions 
for neutrino mixing angles and leptonic Dirac CP violation are reviewed. 
Possible tests of these predictions using the existing data on neutrino 
mixing angles as well as prospective data from current and future 
neutrino oscillation experiments (T2K, NO$\nu$A, Daya Bay, T2HK, 
T2HKK, DUNE) are also discussed.

\end{abstract}

%
\setcounter{footnote}{0}

\vspace{-0.4cm}

%
\section{Introduction}
\label{aba:sec1}
%
%
  Understanding the origins of the patterns of neutrino mixing 
and of neutrino mass squared differences, 
revealed by the data obtained in the neutrino oscillation experiments
(see, e.g., \cite{Olive:2016xmw}), 
is one of the most challenging problems in neutrino physics.
It is part of the more general fundamental problem
in particle physics of understanding the origins of
flavour, i.e., of the patterns of quark, charged lepton and 
neutrino masses, and of the quark and lepton mixing.

 We believe, and we are not alone in holding this view,
that with the observed pattern of neutrino mixing 
Nature is ``sending'' us a Message. 
The Message is encoded in the 
values of the neutrino mixing angles, leptonic 
CP violation (CPV) phases in the Pontecorvo, Maki, Nakagawa 
and Sakata (PMNS) neutrino mixing matrix  \cite{BPont57,MNS62,BPont67} 
and neutrino masses.
We do not know at present what is the content 
of Nature's Message. However, on the basis of the current 
ideas about the  possible origins of the observed 
pattern of neutrino mixing, the Nature's Message 
can have two completely different contents, each of which can be 
characterised by one word: ANARCHY or SYMMETRY.
In the ANARCHY approach \cite{Anarchy}
to understanding the pattern of neutrino mixing 
it is assumed that Nature ``threw dice'' when 
 ``choosing'' the values of the neutrino masses, 
mixing angles and leptonic CPV phases. 
The main prediction of the ANARCHY explanation of  
the pattern of  neutrino mixing is the absence of 
whatever correlations between the values 
of the neutrino masses, between the values 
of the neutrino mixing angles, and 
between the values of the neutrino mixing angles 
and the CPV phases, all of them 
being random quantities. 
As a consequence, no specific values of, e.g., neutrino 
mixing angles are predicted: the predictions of these 
(and other leptonic) measurable quantities 
are in the form of distributions.
In contrast, one of the main characteristic 
features of the SYMMETRY approach to neutrino mixing 
is the prediction of the values of some of the mixing angles 
and/or of the existence of correlations 
between the values of at least some of 
the observables (angles, CPV phases) of the 
the neutrino mixing matrix.

  Within the SYMMETRY approach, 
the observed pattern of neutrino mixing, which 
differs drastically from the quark mixing pattern,
can be naturally understood 
on the basis of specific class of symmetries - 
the class of non-Abelian discrete flavour symmetries 
(see, e.g., \cite{Altarelli:2010gt,Ishimori:2010au,Tanimoto:2015nfa}).
Thus, the specific form of the neutrino mixing can have its origin in 
the existence of new fundamental symmetry in the lepton sector.
The most distinctive feature of the approach to neutrino mixing  
based on non-Abelian discrete flavour symmetries 
is the predictions of the values of
some of the neutrino mixing angles and leptonic CPV phases, 
and/or of existence of correlations between 
the values of at least some the neutrino mixing angles 
and/or between the values of the neutrino 
mixing angles and the Dirac CPV phase in the PMNS matrix, etc.    
(see, e.g., 
\cite{Tanimoto:2015nfa,Petcov:2014laa,Girardi:2015vha,Girardi:2015rwa,
Ge:2011qn,TanM}). 
Combining the discrete symmetry approach with 
the idea of generalised CP invariance  
\cite{Branco:1986gr,Feruglio:2012cw,Holthausen:2012dk}
~--~ a generalisation 
of the standard CP invariance requirement~--~
allows to obtain predictions also for the Majorana 
CPV phases in the PMNS matrix in the case 
of massive Majorana neutrinos 
(see, e.g., 
\cite{Ding:2013hpa,Girardi:2013sza,Ding:2013bpa,Hagedorn:2014wha,Li:2015jxa,DiIura:2015kfa,Ballett:2015wia,Yao:2016zev,Turner:2015uta,Girardi:2016zwz,Lu:2016jit,Penedo:2017vtf,Li:2017abz} and references quoted therein).
Most importantly, these predictions 
and predicted correlations, 
and thus the discrete symmetry approach itself, 
can be tested experimentally (see, e.g., \cite{Petcov:2014laa} and
\cite{Girardi:2015vha,Girardi:2013sza,Hanlon:2013ska,Ballett:2013wya,Girardi:2014faa,Agarwalla:2017wct,Petcov:2018snn}).

  In the present article we  review 
aspects of the symmetry approach to neutrino mixing  
based on the class of  non-Abelian discrete 
flavour symmetries, which is widely explored at present (see, e.g., 
\cite{Tanimoto:2015nfa,Penedo:2017vtf,Li:2017abz,King:2013eh,Meloni:2017cig} 
and references quoted therein
\footnote{For early attempts see, e.g., \cite{Early}.} 
).
We will discuss also the typical phenomenological predictions
of the approach and their possible tests in currently running 
and future planned neutrino oscillation experiments.

 Before discussing the discrete flavour symmetry approach to neutrino 
mixing we would like to review briefly the current status of our 
knowledge of neutrino masses, neutrino mixing and leptonic 
CPV phases, the remaining fundamental problems 
in neutrino physics and the future prospects in this field.

%
\section{The Three-Neutrino Mixing}
\label{aba:sec2}
%

 The experiments with solar, atmospheric, 
reactor and accelerator neutrinos
have provided compelling evidences for 
the existence of neutrino oscillations \cite{BPont57,MNS62}~--~ 
transitions in flight between the different flavour neutrinos
$\nu_e$, $\nu_\mu$, $\nu_\tau$
(antineutrinos $\bar{\nu}_e$, $\bar{\nu}_\mu$, $\bar{\nu}_\tau$)~--~ 
caused by nonzero neutrino masses and neutrino mixing 
(see, e.g., ref. \cite{Olive:2016xmw} for review of the relevant data).
The existence of flavour neutrino oscillations implies 
the presence of mixing in the weak charged lepton current:
\begin{equation}
\label{CC}
{\cal L}_{\rm CC} = - ~\frac{g}{\sqrt{2}}\,
\sum_{l=e,\mu,\tau}
\overline{l_L}(x)\, \gamma_{\alpha} \nu_{l\mathrm{L}}(x)\,
W^{\alpha \dagger}(x) + h.c.\,,~
\nu_{l \mathrm{L}}(x)
= \sum^n_{j=1} U_{l j} \nu_{j \mathrm{L}}(x)\,,
\end{equation}
%
\noindent 
where $\nu_{lL}(x)$ are the flavour neutrino fields, 
$\nu_{j \mathrm{L}}(x)$ is the left-handed (LH)
component of the field of the neutrino $\nu_j$ having a 
mass $m_j$, and $U$ is a unitary matrix - the
PMNS neutrino mixing matrix \cite{BPont57,MNS62,BPont67}. 
All compelling  neutrino oscillation data
can be described assuming 3-neutrino mixing in vacuum, $n=3$.
The number of massive neutrinos $n$ 
can, in general, be bigger than 3 if, e.g., 
there exist RH sterile neutrinos \cite{BPont67}
and they mix with the LH flavour neutrinos. 
It follows from the current data that at least 3 of 
the neutrinos $\nu_j$, say 
$\nu_1$, $\nu_2$, $\nu_3$, must be light,
i.e., must have masses smaller than roughly 1 eV, 
$m_{1,2,3} \ltap 1$ eV, and must have different 
masses, $m_1\neq m_2 \neq m_3$  
\footnote{At present there are several 
experimental inconclusive hints 
for existence of one or two light 
sterile neutrinos at the eV scale, 
which mix with the flavour neutrinos, 
implying the presence in the neutrino mixing  
of additional one or two neutrinos, $\nu_4$ or $\nu_{4,5}$, 
with masses $m_4~(m_{4,5})\sim 1~{\rm eV}$ 
(see, e.g., ref.~\cite{Giunti:2015wnd}).
For a recent discussion of these hints and of the 
related implications see, e.g., 
ref.~\cite{Gariazzo:2017fdh}.
}.

 In the case of 3 light neutrinos, the 
$3\times 3$ unitary neutrino mixing 
matrix $U$ can be parametrised, as is well known, 
by 3 angles and, depending on whether 
the  massive neutrinos $\nu_j$ are Dirac 
or Majorana particles, 
by one Dirac, or one Dirac and two Majorana,
CP violation (CPV) phases \cite{Bilenky:1980cx}: 
\begin{equation}
U= VP\,,~~~
P = {\rm diag}(1, e^{i \frac{\alpha_{21}}{2}}, e^{i \frac{\alpha_{31}}{2}})\,, 
\label{VP}
\end{equation}
%
where $\alpha_{21,31}$ 
are the two Majorana CPV 
phases and in the ``standard'' parametrisation the matrix $V$ 
is given by: 
\begin{equation} 
\begin{array}{c}
\label{eq:Vpara}
V = \left(\begin{array}{ccc} 
 c_{12} c_{13} & s_{12} c_{13} & s_{13} e^{-i \delta}  \\[0.2cm] 
 -s_{12} c_{23} - c_{12} s_{23} s_{13} e^{i \delta} 
 & c_{12} c_{23} - s_{12} s_{23} s_{13} e^{i \delta} 
 & s_{23} c_{13} 
\\[0.2cm] 
 s_{12} s_{23} - c_{12} c_{23} s_{13} e^{i \delta} & 
 - c_{12} s_{23} - s_{12} c_{23} s_{13} e^{i \delta} 
 & c_{23} c_{13} 
\\ 
  \end{array}    
\right)\,. 
\end{array} 
\end{equation}
%
\noindent 
In eq.~(\ref{eq:Vpara}),
$c_{ij} = \cos\theta_{ij}$, 
$s_{ij} = \sin\theta_{ij}$,
the angles $\theta_{ij} \in [0,\pi/2]$, 
and $\delta \in [0,2\pi)$ is the 
Dirac CPV phase. 
Thus, in the case of massive Dirac neutrinos, 
the neutrino mixing matrix $U$ is similar,
in what concerns the number of 
mixing angles and CPV phases, to the Cabibbo, Kobayashi, Maskawa 
(CKM) quark mixing matrix. 
The PMNS matrix $U$ contains two additional 
physical CPV phases if  $\nu_j$ are Majorana particles due 
to the special properties of Majorana fermions
(see, e.g., refs.~\cite{Bilenky:1980cx,BiPet87,STPNuNature2013}).
On the basis of the existing neutrino data 
it is impossible to determine whether the massive 
neutrinos are Dirac or Majorana fermions.

 The  probabilities of neutrino oscillations 
are functions of the neutrino energy, $E$, 
the source-detector distance $L$, 
of the elements of $U$ and, 
for relativistic neutrinos used in all neutrino 
experiments performed so far, 
of the neutrino mass squared 
differences $\Delta m^2_{ij} \equiv (m^2_{i} - m^2_j)$, $i\neq j$ 
(see, e.g., ref.~\cite{BiPet87}).
In the case of 3-neutrino mixing
there are only two independent $\Delta m^2_{ij}$,
say $\Delta m^2_{21}\neq 0$ and $\Delta m^2_{31} \neq 0$.
The numbering of neutrinos $\nu_j$ is arbitrary.
We will employ  the widely used convention 
which allows to associate $\theta_{13}$ with the smallest 
mixing angle in the PMNS matrix, 
and  $\theta_{12}$, $\Delta m^2_{21}> 0$, and
$\theta_{23}$, $\Delta m^2_{31}$,
with the parameters which drive the 
solar ($\nu_e$) and the dominant atmospheric 
$\nu_{\mu}$ and $\bar{\nu}_{\mu}$ 
oscillations, respectively.
In this convention $m_1 < m_2$,
$ 0 < \Delta m^2_{21} < |\Delta m^2_{31}|$, 
and, depending on ${\rm sgn}(\Delta m^2_{31})$,
we have either $m_3 < m_1$ or $m_3 > m_2$. 

\begin{table}
\centering
\renewcommand{\arraystretch}{1.2}
\begin{tabular}{lccc} 
\toprule
Parameter & Best fit value & $2\sigma$ range & $3\sigma$ range \\ 
\midrule
$\sin^2\theta_{12}/10^{-1}$ & $2.97$ & $2.65 - 3.34$ & $2.50 - 3.54$ \\
$\sin^2\theta_{13}/10^{-2}$~(NO) & $2.15$ & $1.99 - 2.31$ & $1.90 - 2.40$\\
$\sin^2\theta_{13}/10^{-2}$~(IO) & $2.16$ & $1.98 - 2.33$ & $1.90 - 2.42$\\
$\sin^2\theta_{23}/10^{-1}$~(NO) & $4.25$ & $3.95 - 4.70$ & $3.81 - 6.15$\\
$\sin^2\theta_{23}/10^{-1}$~(IO) & $5.89$ & $3.99 - 4.83 \oplus 5.33 - 6.21$ & $3.84 - 6.36$\\
$\delta/\pi$~(NO) & $1.38$ & $1.00 - 1.90$ & $0 - 0.17 \oplus 0.76 - 2$\\
$\delta/\pi$~(IO) & $1.31$ & $0.92 - 1.88$ & $0 - 0.15 \oplus 0.69 - 2$\\
\midrule
$\Delta m_{21}^{2}/10^{-5}$~eV$^2$ & $7.37$ & $7.07 - 7.73$ & $6.93 - 7.96$\\
$\Delta m_{31}^{2}/10^{-3}$~eV$^2$~(NO) & $2.56$ & $2.49 - 2.64$ & $2.45 - 2.69$\\
$\Delta m_{23}^{2}/10^{-3}$~eV$^2$~(IO) & $2.54$ & $2.47 - 2.62$ & $2.42 - 2.66$\\
\bottomrule
\end{tabular}
\caption{The best fit values, $2\sigma$ and 3$\sigma$ ranges of the 
neutrino oscillation parameters obtained in the global 
analysis of the neutrino oscillation data 
performed in~\cite{Capozzi:2017ipn}. (The Table is taken 
from ref. \cite{Penedo:2017vtf}.)
}
\label{tab:parameters}
\end{table}
%

The existing  data, accumulated over many years 
of studies of neutrino oscillations, allow us to determine 
$\Delta m^2_{21}$, $\theta_{12}$, and
$|\Delta m^2_{31(32)}|$, $\theta_{23}$ and $\theta_{13}$, 
with a relatively high precision 
\cite{Capozzi:2017ipn,Esteban:2016qun}. 
Since 2013 there are also persistent hints that the Dirac CPV phase 
$\delta$ has a value close to $3\pi/2$ (see \cite{Capozzi:2013csa}).
The best fit values (b.f.v.) and the 2$\sigma$ and 3$\sigma$ allowed ranges of 
$\Delta m^2_{21}$, $s^2_{12}$, $|\Delta m^2_{31(32)}|$, 
$s^2_{23}$, $s^2_{13}$ and $\delta$, found in the 
analysis of global 
neutrino oscillation data performed in \cite{Capozzi:2017ipn} 
are given in Table \ref{tab:parameters}.
Similar results were obtained in ref.~\cite{Esteban:2016qun}.

  In both analyses
\cite{Capozzi:2017ipn,Esteban:2016qun}
the authors find, in particular, 
that the best fit value of the 
Dirac CPV phases $\delta$ is close to $3\pi/2$:
in \cite{Capozzi:2017ipn}, for example, the authors find    
$\delta = 1.38\pi~(1.31\pi)$ for  $\Delta m^{2}_{31(32)}>0$  
($\Delta m^{2}_{31(32)} < 0$).
The absolute $\chi^2$ minimum
takes place for  $\Delta m^{2}_{31(32)}>0$.
According to ref. \cite{Capozzi:2017ipn},
the CP conserving values $\delta = 0$ or $2\pi$  
are disfavored at $2.4\sigma$ ($3.2\sigma$) 
for  $\Delta m^2_{31(32)}>0$
($\Delta m^2_{31(32)}<0$); 
the CP conserving value $\delta = \pi$ 
in the case of $\Delta m^2_{31(32)}>0$ 
($\Delta m^2_{31(32)}<0$) is statistically 
approximately $2.0\sigma$ ($2.5\sigma$) away from the best 
fit value $\delta \cong 1.38\pi~(1.31\pi)$.
In what concerns the CP violating value 
$\delta = \pi/2$, it is strongly disfavored at 
$3.4\sigma$ ($3.9\sigma$) for  $\Delta m^2_{31(32)}>0$ 
($\Delta m^2_{31(32)}<0$)~
\footnote{The quoted
confidence levels for  $\delta = 0,\pi$ and $\pi/2$
are all with respect to the absolute
$\chi^2$ minimum.}.
At $3\sigma$, $\delta/\pi$ is found to lie
in the case of $\Delta m^{2}_{31(32)}>0$  
($\Delta m^{2}_{31(32)} < 0$)
in the following intervals \cite{Capozzi:2017ipn}:
$(0.00-0.17(0.15))\oplus (0.76(0.69)-2.00))$.
The results on $\delta$ obtained in  \cite{Esteban:2016qun} 
differ somewhat from, but are compatible at $1\sigma$ confidence level 
(C.L.) with, those found in \cite{Capozzi:2017ipn}.
 
  It follows also from the results quoted 
in Table \ref{tab:parameters}
that  $\Delta m^2_{21}/|\Delta m^2_{31(32)}| \cong 0.03$.
We have $|\Delta m^2_{31}| = |\Delta m^2_{32} - \Delta m^2_{21}| 
\cong |\Delta m^2_{32}|$. 
The angle  $\theta_{12}$ is definitely smaller than $\pi/4$:
the value of $\theta_{12} = \pi/4$, i.e.,
maximal solar neutrino mixing,  
is ruled out at 
high confidence level (C.L.) by the data - 
one has  $\cos2\theta_{12} \geq 0.29$ at $99.73\%$ C.L.
The quoted results imply also that the value of 
$\theta_{23}$ can deviate by approximately $\pm 0.1$
from $\pi/4$, $\theta_{12} \cong \pi/5.4$ and 
that $\theta_{13} \cong \pi/20$. Thus, the pattern of 
neutrino mixing differs drastically from 
the pattern of quark mixing. 

It should be noted that in the more recent global analyses 
\cite{NuFITv32Jan2018,Capozzi:2018ubv}, which used, in particular, 
updated results on $\sin^2\theta_{23}$ from the NO$\nu$A experiment,  
the best fit value of $\sin^2\theta_{23}$ for NO spectrum 
was found to be larger than 0.5~
\footnote{In what concerns the other two neutrino mixing angles 
$\theta_{12}$ and $\theta_{13}$, the results 
reported in  \cite{Capozzi:2017ipn}
and in \cite{NuFITv32Jan2018,Capozzi:2018ubv} 
differ insignificantly.
}:
\be
\sin^2\theta_{23} = 0.538~(0.554)~[46]\,,~~~
\sin^2\theta_{23} = 0.551~(0.557)~[47]\,,~~~{\rm NO~(IO)}\,. 
\label{eq:s2th232018}
\ee
%
 
 Apart from the hint that 
the Dirac phase $\delta \sim 3\pi/2$,  
no other experimental information on
the Dirac and Majorana 
CPV phases in the neutrino mixing matrix is available 
at present. Thus, the status of CP symmetry 
in the lepton sector is essentially 
unknown. With $\theta_{13} \cong 0.15 \neq 0$,
the Dirac phase $\delta$ can generate
CP violating effects in neutrino 
oscillations \cite{Bilenky:1980cx,Cabibbo78,Barger:1980jm},
i.e, a difference between the probabilities of the 
$\nu_l \rightarrow \nu_{l'}$ and
$\bar{\nu}_l \rightarrow \bar{\nu}_{l'}$
oscillations, $l\neq l'=e,\mu,\tau$.
The magnitude of CP violation in
$\nu_l \rightarrow \nu_{l'}$ and
$\bar{\nu}_l \rightarrow \bar{\nu}_{l'}$
oscillations in vacuum, $l\neq l'=e,\mu,\tau$,
is determined by \cite{PKSP3nu88} the 
rephasing invariant $J_{CP}$, associated with the Dirac 
CPV phase in $U$:
\begin{equation}
J_{\rm CP} =
{\rm Im}\, \left (U_{\mu 3}\,U^*_{e3}\,U_{e2}\,U^*_{\mu 2}\right )\,.
\label{JCP02}
\end{equation}
%
It is analogous to the rephasing invariant
associated with the Dirac 
CPV phase in the CKM
quark mixing matrix \cite{CJ85}.
In the standard parametrisation of the
neutrino mixing matrix (\ref{eq:Vpara}), 
$J_{\rm CP}$ has the form:
\begin{equation}
J_{\rm CP} \equiv
{\rm Im}\,(U_{\mu 3}\,U^*_{e3}\,U_{e2}\,U^*_{\mu 2}) =
\frac{1}{8}\,\cos\theta_{13}
\sin 2\theta_{12}\,\sin 2\theta_{23}\,\sin 2\theta_{13}\,\sin \delta\,.
\label{JCPstandparam}
\end{equation}
%
Thus, given the fact that $\sin 2\theta_{12}$,
$\sin 2\theta_{23}$ and $\sin 2\theta_{13}$ have been 
determined experimentally with a relatively 
good precision, the size of CP violation
effects in neutrino oscillations depends
essentially only on the magnitude 
of the currently not well determined 
value of  the Dirac phase $\delta$.
The current data imply 
$0.026(0.027)|\sin\delta|\lesssim |J_{\rm CP}| \lesssim  0.035 |\sin\delta|$,
where we have used the $3\sigma$ ranges of 
$\sin^2\theta_{12}$, $\sin^2\theta_{23}$ and $\sin^2\theta_{13}$
given in  Table \ref{tab:parameters}.
For the current best fit values of 
$\sin^2\theta_{12}$, $\sin^2\theta_{23}$, $\sin^2\theta_{13}$
and $\delta$ we find in the case of $\Delta m^2_{31(2)} > 0$ 
($\Delta m^2_{31(2)} < 0$):  
$J_{\rm CP} \cong 0.032\sin\delta \cong -\; 0.030$ 
($J_{\rm CP} \cong  0.032\sin\delta \cong -\; 0.027$).
Thus, if the indication that $\delta$ 
has a value close to $3\pi/2$ 
is confirmed by future more precise data, 
i) the $J_{\rm CP}$ factor
in the lepton sector would 
be approximately by 3 orders of magnitude larger 
in absolute value than corresponding 
 $J_{\rm CP}$ factor in the quark sector, and
ii) the CP violation effects in neutrino 
oscillations would be relatively large and observable.

 If the neutrinos with definite masses
$\nu_i$, $i=1,2,3$, are Majorana particles,
the 3-neutrino mixing matrix contains
two additional Majorana CPV phases \cite{Bilenky:1980cx}. 
However, the flavour neutrino
oscillation probabilities
$P(\nu_l \rightarrow \nu_{l'})$ and
$P(\bar{\nu}_l \rightarrow \bar{\nu}_{l'})$,
$l,l' =e,\mu,\tau$, do not depend on
the Majorana phases\cite{Bilenky:1980cx,Lang87}.
The Majorana phases can play
important role, e.g, in $|\Delta L| = 2$
processes like neutrinoless double beta 
($(\beta\beta)_{0\nu}$-)
decay $(A,Z) \rightarrow (A,Z+2) + e^- + e^-$,
$L$ being the total lepton charge,
in which the Majorana nature of
massive neutrinos $\nu_i$ manifests itself
(see, e.g, refs. \cite{BiPet87,STPNuNature2013,BPP1}). 

 Our interest in the CPV
phases present in the neutrino mixing matrix
is stimulated also by the intriguing possibility
that the Dirac phase and/or the Majorana phases in
$U_{\rm PMNS}$ can provide the CP violation
necessary for the generation of the observed
baryon asymmetry of the Universe (BAU)
\cite{Pascoli:2006ci} 
(for specific models in which this possibility 
is realised see, e.g., 
\cite{Hagedorn:2009jy,Hagedorn:2016lva,Chen:2016ptr,Shimizu:2017vwi};
for a recent review see \cite{Hagedorn:2017wjy}).
More specifically, if, e.g., all CP violation necessary for the 
generation of BAU is due to the Dirac phase $\delta$, 
which is possible within the ``flavoured'' leptogenesis 
scenario \cite{FlavLG,nardietal} of generation of baryon asymmetry,
a necessary condition for reproducing 
the observed BAU in this scenario 
(with hierarchical heavy Majorana neutrinos)
is \cite{Pascoli:2006ci}  
$|\sin\theta_{13}\,\sin\delta|\gtap 0.09$. 
This condition is comfortably 
compatible with the measured value of $\sin\theta_{13}$ and with 
the best fit value of $\delta \sim 3\pi/2$. 

  The sign of $\Delta m^2_{31(32)}$ cannot be 
directly and unambiguously determined 
from the existing data 
\footnote{In the recent analysis of the global neutrino oscillation data 
performed in \cite{Capozzi:2018ubv} it was found that 
the case of $\Delta m^2_{31(32)} < 0$ is disfavored at 
$3.1\sigma$ with respect to the case of $\Delta m^2_{31(32)} > 0$.
}.
In the case of 3-neutrino mixing,  
the two possible signs of
$\Delta m^2_{31(32)}$ correspond to two 
types of neutrino mass spectrum.
In the convention of numbering 
of neutrinos $\nu_j$ employed by us
the two spectra read:\\
{\it i) spectrum with normal ordering (NO)}:
$m_1 < m_2 < m_3$, $\Delta m^2_{31(32)} >0$,
$\Delta m^2_{21} > 0$,
$m_{2(3)} = (m_1^2 + \Delta m^2_{21(31)})^{1\over{2}}$; \\~~
{\it ii) spectrum with inverted ordering (IO)}:
$m_3 < m_1 < m_2$, $\Delta m^2_{32(31)}< 0$, 
$\Delta m^2_{21} > 0$,
$m_{2} = (m_3^2 + \Delta m^2_{23})^{1\over{2}}$, 
$m_{1} = (m_3^2 + \Delta m^2_{23} - \Delta m^2_{21})^{1\over{2}}$.\\ 
Depending on the values of the lightest neutrino mass, 
${\rm min}(m_j)$, the neutrino mass spectrum can also be:~\\
{\it a) Normal Hierarchical (NH)}: $m_1 \ll m_2 < m_3$, 
$m_2 \cong (\Delta m^2_{21})^
{1\over{2}}\cong 8.6\times 10^{-3}$ eV,
$m_3 \cong (\Delta m^2_{31})^{1\over{2}} \cong 0.0506$ eV; or\\ 
{\it b) Inverted Hierarchical (IH)}: $m_3 \ll m_1 < m_2$, 
$m_{1} \cong (|\Delta m^2_{32}| - \Delta m^2_{21})^{1\over{2}}\cong 0.0497$ eV, 
$m_{2} \cong (|\Delta m^2_{32}|)^{1\over{2}}\cong 0.0504$ eV; or\\ 
{\it c) Quasi-Degenerate (QD)}: $m_1 \cong m_2 \cong m_3 \cong m_0$,
$m_j^2 \gg |\Delta m^2_{31(32)}|$, $m_0 \gtap 0.10$ eV.\\ 
All three types of spectrum are compatible
with the constraints on the absolute scale 
of neutrino masses. Determining the type of neutrino 
mass spectrum is one of the main goals of the future 
experiments in the field of neutrino physics
\footnote{For a brief discussion of experiments 
which can provide data on the type of 
neutrino mass spectrum see, e.g.,
ref. \cite{XQPVogel2015};
for some specific proposals see, e.g., 
ref.~\cite{NOIO1,NOIO11,NOIO2,NOIO3}.}
(see, e.g., refs.~
\cite{Olive:2016xmw,XQPVogel2015,JUNO,KM3Net2016,PINGU,DUNE2016,INO}).

 Data on  the absolute neutrino mass scale 
(or on ${\rm min}(m_j)$)
can be obtained, e.g., 
from measurements of the spectrum 
of electrons near the end point in 
\hbeta experiments \cite{Fermi34,Mainz,MoscowH3}
and from cosmological and astrophysical 
observations. The most stringent upper 
bound on the $\bar{\nu}_e$ mass 
was reported by the Troitzk~\cite{MoscowH3b} 
experiment:
\beq
m_{\bar{\nu}_e} < 2.05~\mathrm{eV}~~~\mbox{at}~95\%~\mathrm{C.L.} 
\label{H3beta}
\eeq
%
\noindent Similar result 
was obtained in the 
Mainz experiment \cite{Mainz}~: 
$m_{\bar{\nu}_e} < 2.3~\rm{eV}$ at 95\% CL.
We have $m_{\bar{\nu}_e} \cong m_{1,2,3}$
in the case of QD  spectrum. 
The KATRIN experiment~\cite{MainzKATRIN}, 
which was commissioned on June 11, 2018, 
is designed to reach sensitivity  
of  $m_{\bar{\nu}_e} \sim 0.20$~eV,
i.e., to probe the region of the QD spectrum. 

Constraints on the sum of the neutrino masses 
can be obtained from cosmological and astrophysical data 
(see, e.g., ref.~\cite{summj}).
 Depending on the model complexity and the input data used 
one typically obtains \cite{summj}:
$\sum_j m_j\ltap (0.3 - 1.3)$ eV, 95\% C.L.
Assuming the existence of
three light massive neutrinos and the validity of 
the $\Lambda$ CDM (Cold Dark Matter) model,
and using their data on the CMB temperature power spectrum 
anisotropies, polarisation,
on gravitational lensing effects and the low $l$ 
CMB polarization spectrum data (the
“low P” data), etc. the Planck Collaboration reported 
an updated upper limit on the sum of the neutrino 
masses \cite{Ade:2015xua}, which, depending on the data-set 
used, varies in the interval: 
$\sum_j m_j\, < \,(0.340 - 0.715)$ eV, 95\%~C.L.
Adding data on Baryon Acoustic Oscillations (BAO) 
lowers the limit to \cite{Ade:2015xua}: 
\begin{equation}
\sum_j m_j\, < \, 0.170~{\rm eV},~~~ 95\%~{\rm C.L.}
\label{Planck1}
\end{equation}
%

 In spite of the remarkable progress made in the last 19 years 
in establishing the existence of neutrino oscillations 
caused by non-zero neutrino masses and neutrino mixing 
and in measuring the 3-neutrino oscillation parameters, 
one has to admit that 
we are still completely ignorant about the fundamental aspects 
of neutrino mixing. We do not know whether the massive neutrinos 
are Dirac or Majorana particles, what is the neutrino mass ordering,
what is the status of the CP symmetry in the lepton sector and 
what is the absolute neutrino mass scale 
(i.e., the lightest neutrino mass).
Determining the nature~---~Dirac or Majorana~---~of massive neutrinos,
the type of spectrum the neutrino masses obey,
establishing the status of the CP symmetry in the lepton sector 
and determining the absolute neutrino mass scale  
are among the highest priority goals of the programme of 
future experimental research in neutrino physics 
(see, e.g., 
\cite{Olive:2016xmw,JUNO,KM3Net2016,PINGU,DUNE2016,INO,T2K20152016,T2HK2015,Abe:2016ero}), 
which extends beyond 2030.
The principal goal of the theoretical studies in this field 
is the understanding at a fundamental level
the mechanism giving rise to neutrino masses and mixing and to
$L_l-$non-conservation. Are the observed patterns of
$\nu$-mixing and of $\Delta m^2_{21,31}$
related to the existence of a new
fundamental symmetry of particle interactions?
Is there any relation between quark
mixing and neutrino mixing?
What is the physical origin of CPV 
phases in the neutrino mixing matrix $U$?
Is there any relation (correlation)
between the (values of) CPV 
phases and mixing angles in  $U$?
Progress in the theory of
neutrino mixing might also lead, in particular, 
to a better understanding of the
mechanism of generation of baryon
asymmetry of the Universe.
 
\vspace{-0.3cm}
%
\section{Origins of the Pattern of Neutrino Mixing: 
the Discrete Symmetry Approach
}
\label{sec3}
%
%
%
\subsection{The General Framework}

\label{subsec31}
%
%

 The observed pattern of neutrino mixing 
in the reference 3-neutrino mixing scheme 
we are going to consider in what follows 
is characterised, as we have seen,  
by two large mixing angles 
$\theta_{12}$ and $\theta_{23}$, and one small 
mixing angle $\theta_{13}$: 
$\theta_{12}\cong 33^\circ$, 
$\theta_{23} \cong 45^\circ \pm 6^\circ$ and 
$\theta_{13} \cong 8.4^\circ$.  
These values can naturally be explained 
by  extending the Standard Model (SM) with 
a flavour symmetry corresponding to 
a non-Abelian discrete (finite) group  $G_f$. This symmetry 
is supposed to exist at some high-energy scale 
and to be broken at lower energies to residual symmetries of   
the charged lepton and neutrino sectors, described respectively 
by subgroups $G_e$ and $G_{\nu}$ of  $G_f$. 
Flavour symmetry groups $G_f$ that have been used 
in this approach to neutrino mixing and lepton flavour 
include $A_4$ \cite{A4}, $S_4$ \cite{S4}, $T'$ \cite{Tprime}, 
$A_5$ \cite{A5}, $D_{n}$ (with $n=10,12$) \cite{D10,HGM}, 
$\Delta(27)$ \cite{Delta27}, the series $\Delta(6n^2)$ \cite{Delta6n2}, 
to name several
\footnote{Some of the groups $T'$, $A_5$, etc.  
can be and have been used also for a unified description 
of the quark and lepton flavours, see, e.g., 
refs. \cite{Li:2017abz,Tprime,FlavourG,Chen:2013wba,Gehrlein:2014wda} 
and references quoted therein.
} 
(see, e.g., 
ref. \cite{Ishimori:2010au} for definitions of these groups and 
discussion of their properties 
\footnote{
$A_4$ is the group of even permutations of 4 objects
and the symmetry group of the regular tetrahedron.
$S_4$ is the group of permutations of 4 objects
and the symmetry group of the cube. 
$T'$ is the double covering group of $A_4$.
$A_5$ is the icosahedron symmetry group 
of even permutations of five objects, etc.
All these groups are subgroups of 
the group $SU(3)$. 
}).
The numbers of elements, of generators and 
of irreducible representations of the groups 
$S_4$, $A_4$, $T'$, $A_5$, $D_{10}$ and $D_{12}$ are given in 
Table \ref{tab:discrG}. 
In what concerns the group $S_4$, it is well known 
that $S_4$ can be generated by two transformations, $S$ and $T$ 
(see, e.g.,\cite{Ishimori:2010au}). 
However, in the context of non-Abelian discrete symmetry approach 
to neutrino mixing it often proves convenient to use the three generators 
$S$, $T$ and $U$ of $S_4$, indicated in Table 2, and these 
generators are widely used in the literature on the subject 
(see, e.g., the review article \cite{Tanimoto:2015nfa}).
We will use the two generator formalism for the group $S_4$ 
in the analysis performed in sub-section \ref{use:subsec42}.

The choice of the non-Abelian discrete 
groups  $A_4$, $S_4$, $T'$, $A_5$, etc.
is related, in particular,  to the fact that 
they describe symmetries with respect to rotations 
on fixed large mixing angles and,  
correspondingly, lead to values of 
the neutrino mixing angles 
$\theta_{12}$ and  $\theta_{23}$,   
which can differ from the measured values at most 
by sub-leading perturbative corrections, 
with  $\theta_{13}$ typically (but not universally) predicted to be zero. 
\begin{table}[t!]
\centering
\renewcommand*{\arraystretch}{1.5}
\begin{tabular}{|c|c|c|c|}
\hline
Group & Number of elements & Generators & Irreducible representations \\
\hline
$S_4$ & $24$ & $S$, $T$ ($U$) & $\mathbf{1}$, $\mathbf{1^{\prime}}$, $\mathbf{2}$, $\mathbf{3}$, $\mathbf{3^{\prime}}$ \\
$A_4$ & $12$ & $S$, $T$ & $\mathbf{1}$, $\mathbf{1^{\prime}}$, $\mathbf{1^{\prime \prime}}$, $\mathbf{3}$ \\
$T^{\prime}$ & $24$ & $S$, $T$ ($R$) & $\mathbf{1}$, $\mathbf{1^{\prime}}$, $\mathbf{1^{\prime \prime}}$, $\mathbf{2}$, $\mathbf{2^{\prime}}$, $\mathbf{2^{\prime \prime}}$, $\mathbf{3}$ \\
$A_5$ & $60$ & $\tilde{S}$, $\tilde{T}$ & $\mathbf{1}$, $\mathbf{3}$, $\mathbf{3^{\prime}}$, $\mathbf{4}$, $\mathbf{5}$ \\
$D_{10}$ & $20$ & $A$, $B$ & $\mathbf{1}_1$, $\mathbf{1}_2$, $\mathbf{1}_3$, $\mathbf{1}_4$, $\mathbf{2}_1$, $\mathbf{2}_2$, $\mathbf{2}_3$, $\mathbf{2}_4$ \\
$D_{12}$ & $24$ & $\tilde{A}$, $\tilde{B}$ & $\mathbf{1}_1$, $\mathbf{1}_2$, $\mathbf{1}_3$, $\mathbf{1}_4$, $\mathbf{2}_1$, $\mathbf{2}_2$, $\mathbf{2}_3$, $\mathbf{2}_4$, $\mathbf{2}_5$ \\
\hline
\end{tabular}
\caption{Number of elements, generators and irreducible representations of 
some discrete groups.}
\label{tab:discrG}
\end{table}
%
The requisite corrections can most naturally be provided 
by the unitary matrix $U_e$ 
which originates from the diagonalisation of the 
charged lepton mass term and enters into the expression 
of the PMNS neutrino mixing matrix 
(see, e.g., \cite{Frampton:2004ud,Marzocca:2011dh,Marzocca:2013cr} 
and references quoted therein):
\be
U_{\rm PMNS} = U_e^\dagger\,U_\nu\,.
\label{UPMNSUeUnu}
\ee
%
where $U_\nu$ is a unitary matrix coming 
from the diagonalisation of the neutrino mass term. 
More specifically, $U_e$ diagonalises the product 
$M_e M_e^\dagger$, where $M_e$ is the charged lepton 
mass matrix in the charged lepton mass term  
$\mathcal{L}_{\ell}(x)$ (written in the 
left-right convention):
\begin{eqnarray}
\label{Le}
& \mathcal{L}_{\ell}(x) = 
-\, \overline{\tilde{l}_L(x)}\, (M_e)_{\tilde{l}\tilde{l}'}\, \tilde{l}'_R(x) 
+~{\rm h.c.}\,,\\[0.25cm] 
&U^\dagger_e\, M_e M_e^\dagger\, U_e = {\rm diag}(m^2_e,m^2_\mu,m^2_\tau)\,,
\label{UeMe}
\end{eqnarray}
%
$\tilde{l}_L(x)$ and $\tilde{l}'_R(x)$, 
$\tilde{l},\tilde{l}' =\tilde{e},\tilde{\mu},\tilde{\tau}$,  
being respectively the $SU(2)$ doublet and singlet 
left-handed (LH) and right-handed (RH) components 
of the charged lepton fields 
in the basis in which the charged lepton mass term  
$\mathcal{L}_{\ell}(x)$ is not diagonal, while
$m_e$, $m_\mu$ and $m_\tau$ are the masses of the charged leptons 
\footnote{The LH components of the fields of the electron, muon, and tauon, 
$l_L(x)$, $l=e,\mu,\tau$, are related to the fields  
 $\tilde{l}_L(x)$ via the matrix $U_e$: 
$l_L(x) = (U^\dagger_e)_{l\tilde{l}}\tilde{l}_L(x)$.
}. 
In certain classes of models, however, 
$U_e$ coincides with the unit $3\times3$ matrix and 
the requisite corrections are incorporated 
in a factor contained in the matrix $U_\nu$ (see, e.g., 
\cite{Tanimoto:2015nfa,Shimizu:2014ria}).

  We shall assume in what follows that the 
weak-eigenstate neutrino fields 
(in the basis in which charged lepton mass term is not diagonal), 
$\nu_{\tilde{e}}(x)$, $\nu_{\tilde{\mu}}(x)$ and $\nu_{\tilde{\tau}}(x)$, 
possess a Majorana mass term, $\mathcal{L}_{M}^{\nu}(x)$,
and thus the neutrinos with definite mass $\nu_1$, $\nu_2$ and 
$\nu_3$, are Majorana particles. In this case $U_\nu$ diagonalises the
neutrino Majorana mass matrix $M_{\nu}$:
\begin{eqnarray}
\label{nuMajMass}
& \mathcal{L}_{M}^{\nu}(x) =  
 ~\frac{1}{2}~\nu^{\rm T}_{\tilde{l}' L}(x)~C^{-1}~M_{\nu \tilde{l}'\tilde{l}}~\,
\nu_{\tilde{l} L}(x)~+~h.c.\,,~~~~
C^{-1}\,\gamma_{\alpha}\,C = -\,\gamma_{\alpha}^{\rm T}\,,\\[0.25cm]
& U^{{\rm T}}_\nu\, M_{\nu}\,U_\nu = {\rm diag}(m_1,m_2,m_3)\,,
\label{UnuMnu}
\end{eqnarray}
%
where $C$ is the charge conjugation matrix (see, e.g., 
\cite{BiPet87}).
 It should be noted, however, that the approach to neutrino mixing 
we are discussing can be employed also if $\nu_{1,2,3}$ are Dirac 
fermions (see, e.g., \cite{Girardi:2015rwa}), 
e.g., when the theory contains right-handed 
neutrino fields $\nu_{\tilde{l} R}(x)$ which 
form a Dirac mass term with the LH 
neutrino fields $\nu_{\tilde{l}'L}(x)$ 
$\tilde{l},\tilde{l}'=\tilde{e},\tilde{\mu},\tilde{\tau}$, and 
\footnote{The neutrino Dirac mass term in question 
originates \cite{Petcov:1976ff}  
from an $SU(2)_L\times U(1)_{Y_w}$ invariant Yukawa-type term 
in the Lagrangian after the spontaneous breaking of the 
Standard Theory $SU(2)_L\times U(1)_{Y_w}$ symmetry.} 
   the total lepton charge 
$L=L_e + L_\mu + L_\tau$ is conserved.

 In the approach under discussion it is standardly assumed that 
the LH neutrino fields, $\nu_{\tilde{l}L}(x)$, and the LH components of the 
charged lepton fields (in the basis in which charged lepton mass 
term is not diagonal) $\tilde{l}_L(x)$, which form an $SU(2)_L$ 
doublet in the Standard Theory, 
are assigned to the same r-dimensional irreducible unitary 
representation $\rho_r(g_f)$ of the Group 
$G_f$, $g_f$ being an element of $G_f$. Thus, under the action  
of $G_f$,  $\nu_{\tilde{l}L}(x)$ and $\tilde{l}_L(x)$ transform as follows:  
\begin{eqnarray}
\label{gfnu}
& \nu_{\tilde{l}L}(x)\,\xrightarrow{G_f}\, (\rho_r(g_f))_{\tilde{l}\tilde{l}'}\, 
\nu_{\tilde{l}'L}(x)\,,~~~ \quad\, g_f \in G_f\,,\\[0.25cm] 
& \tilde{l}_L(x)\,  \xrightarrow{G_f}\, (\rho_r(g_f))_{\tilde{l}\tilde{l}'} \tilde{l}'_L(x)\,,~~~
\tilde{l} = \tilde{e},\tilde{\mu},\tilde{\tau}\,.
\label{gflL}
\end{eqnarray}
%
In the cases of $G_f = A_4$, $S_4$, $T'$ and $A_5$, 
which possess 3-dimensional irreducible representations,    
$\rho(g_f)$ is standardly taken to be a 3-dimensional 
irreducible unitary representation {\bf 3}, 
$\rho_r(g_f) = \rho_3(g_f)$. This is equivalent 
to the assumption of unification 
of the three lepton families at some high energy scale. 
We are going to consider this 
choice in what follows 
\footnote{In specific models the choice $\rho_r(g_f) = \rho_3(g_f)$ 
is usually accompanied by the assumption that 
$\tilde{e}_R(x)$, $\tilde{\mu}_R(x)$ and $\tilde{\tau}_R(x)$ 
transform as singlet irreducible representations 
of  $G_f$ (see, e.g., \cite{Tanimoto:2015nfa}).
}.

   At low energies the flavour symmetry $G_f$ 
has necessarily to be broken so that the 
electron, muon and tauon as well as the three neutrinos with 
definite mass $\nu_1$, $\nu_2$ and $\nu_3$,  
can get different masses.
The breaking of $G_f$ is realised in specific models 
by scalar ``flavon'' fields, which are singlets 
with respect to the Standard Theory gauge group 
but transform under certain irreducible representations 
of $G_f$ and acquire non-zero vacuum expectation values (VEVs),
thus breaking  $G_f$ spontaneously.
The breaking of the flavour symmetry $G_f$ 
can leave certain subgroups of $G_f$, $G_e$ and $G_\nu$, 
unbroken in the charged lepton and neutrino 
sectors. The unbroken symmetries  $G_e\in G_f$ and $G_\nu \in G_f$ 
are {\it residual symmetries} of the charged lepton 
and neutrino mass matrices. 

    The existence of a residual symmetry $G_e \in G_f$
in the charged lepton sector implies that 
$M_e M_e^\dagger$ is invariant with respect 
to the action of $G_e$ on the LH components 
of the charged lepton fields $\tilde{l}_L(x)$, 
$\tilde{l} = \tilde{e},\tilde{\mu},\tilde{\tau}$:
\be
\rho_r(g_e)^{\dagger} M_e M_e^{\dagger} \rho_r(g_e) = M_e M_e^{\dagger} \,,
\label{GeMe}
\ee
%
where $g_e$ is an element of $G_e$ 
and $\rho_r(g_e)$ gives the action of $G_e$ on 
$\tilde{l}_L(x)$.

 Similarly,  if $G_\nu$ is the residual symmetry 
of the neutrino Majorana mass matrix $M_{\nu}$ one has:
\be
\rho_r(g_{\nu})^T M_{\nu} \rho_r(g_{\nu}) = M_{\nu} \,,
\label{GnuMnu}
\ee
%
where $g_\nu$ is an element of $G_\nu$
and  $\rho_r(g_{\nu})$ determines the action 
of $G_\nu$ on  $\nu_{\tilde{l}L}(x)$, 
$\tilde{l} = \tilde{e},\tilde{\mu},\tilde{\tau}$.
From eq. (\ref{GnuMnu}) we get: 
\be
\rho_r(g_{\nu})^\dagger M^\dagger_{\nu}\,M_{\nu}\,\rho_r(g_{\nu}) = 
M^\dagger_{\nu}\, M_{\nu}\,.
\label{GnuMnuMnu}
\ee
%

 It  follows from 
eqs. (\ref{GeMe}) and (\ref{GnuMnuMnu})
that  $M_e M_e^{\dagger}$ commutes with $\rho_r(g_e)$, while 
$M^\dagger_{\nu}M_{\nu}$ commutes with $\rho_r(g_{\nu})$.
This implies that 
$M_e M_e^{\dagger}$ and $\rho_r(g_e)$ are diagonalised with 
one and the same matrix $U_e$, and that 
similarly,  $M^\dagger_{\nu}M_{\nu}$ and $\rho_r(g_{\nu})$ 
are diagonalised by the same matrix $U^\circ_\nu$:
\begin{eqnarray}
\label{Uerhoge}
U^\dagger_e\,\rho_r(g_e)\,U_e = \rho_r^{\diag}(g_e)\,,\\[0.30cm]
\label{Uonurhognu}
(U^\circ_\nu)^\dagger\,\rho_r(g_{\nu})\,U^\circ_\nu = \rho_r^{\diag}(g_\nu)\,.
\end{eqnarray}
%
Given $G_f$, $\rho_r(g_f)$, and non-trivial $G_e$, 
 $\rho_r(g_e)$ is uniquely determined. As a consequence, 
the matrix $U_e$ diagonalising $\rho_r(g_e)$ (and $M_e M_e^{\dagger}$), 
which enters into the expression for the PMNS matrix $U$, is 
either completely determined or significantly constrained 
\footnote{Obviously, if $G_e$ is trivial consisting just of the unit 
element of $G_f$, i.e., if  $G_f$ is completely broken 
in the charged lepton sector, $U_e$ would not be constrained.}. 
Similarly, for given $G_f$, $\rho_r(g_f)$, and non-trivial $G_\nu$,
the matrix  $U^\circ_\nu$ disgonalising   $\rho_r(g_\nu)$ 
(and  $M^\dagger_{\nu}\,M_{\nu}$)  will either be completely 
determined or strongly constrained. One can show that the 
matrix $U_\nu$ diagonalising the neutrino Majorana mass 
matrix $M_{\nu}$ and the matrix  $U^\circ_\nu$
diagonalising $M^\dagger_{\nu}\,M_{\nu}$ are related, in general, 
in the following way:
\be
U_{\nu} = U^\circ_{\nu}\, P^\circ\,,~~~~
P^\circ={\rm diag}(1,e^{i\frac{\xi_{21}}{2}},e^{i\frac{\xi_{31}}{2}})\,. 
\label{UnuU0nu}
\ee
%
The phases $\xi_{21}$ and $\xi_{31}$ contribute respectively 
to the Majorana phases $\alpha_{21}$ and $\alpha_{31}$ of the PMNS matrix 
(see eq. (\ref{VP})).

  Thus, within the discussed approach the PMNS 
neutrino mixing matrix $U = U^\dagger_e U_{\nu}$ is 
either completely determined or else has a 
constrained form. The form of $U$ one obtains 
depends on the choices of $G_f$, $\rho_r(g_f)$, $G_e$ and $G_{\nu}$.
 
 It should be clear from the preceding discussion that 
the residual symmetries $G_e$ and $G_\nu$, in particular, 
play a crucial role in obtaining a specific form of the 
PMNS matrix. If, in particular,  $G_e \equiv G_\nu$, 
we would have  $U_e = U^\circ_{\nu}$ and the PMNS matrix 
will be trivial, which is ruled out by the data.

 The largest possible exact symmetry of the charged 
lepton Dirac mass term $\mathcal{L}_{\ell}(x)$ (mass matrix $M_e$) is 
$U(1)\times U(1) \times U(1)$. The largest possible 
exact symmetry of the neutrino Majorana mass term 
$\mathcal{L}_{M}^{\nu}(x)$,
with mass matrix $M_\nu$ having three non-zero 
non-degenerate eigenvalues, is $Z_2\times Z_2\times Z_2$.
Making the standardly used simplifying assumption
that $G_f$ is a subgroup of $SU(3)$, 
the largest possible symmetries of 
$\mathcal{L}_{\ell}(x)$ and $\mathcal{L}_{M}^{\nu}(x)$ 
reduce to $U(1)\times U(1)$ and $Z_2\times Z_2$ 
owing to the $SU(3)$ determinant condition.
The residual symmetry group $G_e$ 
should be a subgroup of $U(1)\times U(1)$, 
while $G_\nu$ should be contained in $Z_2\times Z_2$ 
($U(1)\times U(1)$) in the case of massive Majorana 
(Dirac) neutrinos. Thus, $G_e$ and  $G_\nu$ 
should be Abelian groups.

 It follows from the preceding discussion 
that the possible discrete symmetries $G_e$ 
of the charged lepton mass
term leaving  $M_e M_e^{\dagger}$ invariant are:
i)  $G_e = Z_n$, with integer $n \geq 2$, or 
ii) $Z_m \times Z_k$, with integers $m,\,k \geq 2$.
The maximal symmetry $G_{\nu}$ of the  
Majorana mass term of the LH flavour 
neutrino fields $\nu_{\tilde{l}L}(x)$
is the $Z_2 \times Z_2$ (sometimes referred to as 
the Klein four group) symmetry.
$G_{\nu}$ can obviously be just  $Z_2$.
These two possible types of 
$G_{\nu}$ are associated with two approaches 
in constructing realistic 
models of lepton flavour: 
the {\it direct} approach with 
$G_{\nu} = Z_2 \times Z_2$ and the 
{\it semi-direct} approach with 
$G_{\nu} = Z_2$. Since the neutrino Majorana 
mass term (mass matrix $M_\nu$) possesses always a 
$Z_2 \times Z_2$ symmetry,
the second $Z_2$ factor appears accidentally 
in models employing the semi-direct approach. 
The symmetry $G_f$ might be completely 
broken by the neutrino Majorana mass term 
$\mathcal{L}_{M}^{\nu}(x)$, 
i.e., the $Z_2 \times Z_2$ group of symmetry of 
$\mathcal{L}_{M}^{\nu}(x)$ might not necessarily 
be a subgroup of $G_f$. This corresponds to 
the so-called {\it indirect} approach 
in lepton flavour model building.

  The group $A_4$, for example, has  
three subgroups $Z_2$,  four subgroups  $Z_3$ and 
one subgroup  $Z_2\times Z_2$,
while $S_4$ has nine $Z_2$, four $Z_3$,
three $Z_4$ and four $Z_2 \times Z_2$ subgroups. 
The bigger groups 
$T'$, $A_5$, etc. all have a certain number of 
$Z_2$,  $Z_3$,   $Z_2 \times Z_2$, etc. subgroups 
\footnote{For complete list of the subgroups 
of the groups $T'$, $A_5$, $\Delta(6n^2)$ and of the larger 
groups employed in the discrete flavour symmetry approach 
to neutrino mixing see, e.g., ref. \cite{Ishimori:2010au}.}.
 
 As we have indicated in the Introduction, one of the 
main characteristics of the discussed approach to neutrino 
mixing based on discrete flavour symmetries is 
that it leads to  certain specific 
predictions for the values of, and/or 
correlations between, the low-energy neutrino 
mixing parameters, which can be tested experimentally.
These predictions depend on 
the chosen $G_f$, $\rho(g_f)$, $G_e$ and $G_{\nu}$. 
We give a few examples
\cite{Tanimoto:2015nfa,Petcov:2014laa,Girardi:2015vha,Girardi:2015rwa,Feruglio:2012cw,Girardi:2013sza,Ballett:2015wia,Girardi:2016zwz,Hanlon:2013ska,Altarelli:2009gn}.\\

\noindent {\bf I.} In a large class of models one gets 
$\sin^2\theta_{23} = 0.5$.\\

\noindent  {\bf II.} In different class of models one finds that 
the values of $\sin^2\theta_{23}$ and 
$\sin^2\theta_{13}$ are correlated: 
$\sin^2\theta_{23} = 0.5(1 \mp \sin^2\theta_{13} + O(\sin^4\theta_{13}))$.\\

\noindent  {\bf III.} In certain models 
$\sin^2\theta_{23}$ is predicted to have specific values 
which differ significantly from those in cases {\bf I} and {\bf II} 
\cite{Girardi:2015vha}:  
$\sin^2\theta_{23} = 0.455$; or 0.463; or 0.537; or 0.545, 
the uncertainties in these predictions being insignificant.\\

\noindent {\bf IV.} Certain class of models predict a correlation 
between the values of $\sin^2\theta_{12}$ and 
$\sin^2\theta_{13}$: 
$\sin^2\theta_{12} = 1/(3\cos^2\theta_{13}) = 
(1 + \sin^2\theta_{13} + O(\sin^4\theta_{13}))/3 
\cong 0.340$, where we have used the 
b.f.v. of $\sin^2\theta_{13}$.\\

\noindent  {\bf V.} In another class of models 
one still finds a correlation 
between the values of $\sin^2\theta_{12}$ and 
$\sin^2\theta_{13}$, which, however, differs from that 
in Case {\bf IV}: 
$\sin^2\theta_{12} = (1-3\sin^2\theta_{13})/(3\cos^2\theta_{13}) =
 (1 - 2\sin^2\theta_{13} + O(\sin^4\theta_{13}))/3 
\cong 0.319$, where we have used again the
b.f.v. of   $\sin^2\theta_{13}$.\\

\noindent  {\bf VI.} In large classes of models in which the 
elements of the PMNS matrix are predicted to be 
functions of just one real continuous 
free parameter (``one-parameter models''), 
the Dirac and the Majorana CPV phases 
have ``trivial'' CP conserving values 0 or $\pi$. 
In certain one-parameter schemes, however, 
the Dirac phase $\delta = \pi/2$ or $3\pi/2$.\\  

\noindent {\bf VII.} In theories/models in which the elements of 
the PMNS matrix 
are functions of two (angle) or three (two angle and one phase) 
parameters, the Dirac phase $\delta$ satisfies a sum rule 
by which $\cos\delta$ is expressed in terms of the three 
neutrino mixing angles $\theta_{12}$, $\theta_{23}$, $\theta_{13}$ 
and one (or more) fixed (known) parameters $\theta^{\nu}$ 
which depend on the discrete symmetry $G_f$ employed and 
on the residual symmetries $G_e$ and $G_{\nu}$ 
\cite{Petcov:2014laa,Girardi:2015vha,Girardi:2015rwa}:
\be
\cos\delta = 
\cos\delta(\theta_{12},\theta_{23},\theta_{13};\theta^{\nu}).
\label{cosdSumR}
\ee
%
In these cases the $J_{CP}$ factor which determines the magnitude of 
CP violation effects in neutrino oscillations, is also completely 
determined by the values of the three neutrino mixing angles 
and the symmetry parameter(s)  $\theta_{\nu}$:  
\be
J_{CP} = J_{CP}(\theta_{12},\theta_{23},\theta_{13},\delta)
= J_{CP}(\theta_{12},\theta_{23},\theta_{13};\theta^{\nu})\,.
\label{JCPSumR}
\ee
%
If in the model considered 
a correlation of the type corresponding to Case {\bf II} 
(case {\bf IV} or case {\bf V}) takes place, 
$\theta_{23}$ ($\theta_{12}$)  in the sum rule 
for $\cos\delta$ and the expression for the $J_{CP}$ factor 
has to be expressed in terms of $\theta_{13}$ 
using the correlation. 

The predictions listed above, and therefore the respective models,  
can be and will be tested in the currently running 
(T2K \cite{T2K20152016} and NO$\nu$A \cite{NOvA2018}) and planned future 
(JUNO \cite{JUNO}, T2HK \cite{T2HK2015}, T2HKK \cite{Abe:2016ero} and
DUNE \cite{DUNE2016}) experiments.

  As an illustration of the preceding discussion we will consider 
first the example  
of the tri-bimaximal mixing as an underlying 
symmetry form of the matrix $U_{\nu}$ ($U^\circ_{\nu}$). 
 
%
\subsection{Symmetry Forms of $U_{\nu}$: 
Tri-bimaximal Mixing}
\label{subsec32}
%

 Consider  the case of $G_f =S_4$, i.e., the group 
of permutations of four objects. 
$S_4$ is isomorphic to the group of 
rotational symmetries of the cube. It has 24 elements, 
two singlet, one doublet and two triplet irreducible 
representations. As was indicated earlier, we will assume that 
$\rho_r(g_f) = \rho_3(g_f)$, i.e., that 
$\nu_{\tilde{l}L}(x)$ and $\tilde{l}_L(x)$ transform 
under one of the two  3-dimensional irreducible unitary 
representations of $S_4$.
In what follows, with the exception of sub-section \ref{use:subsec42},  
we will work with the three generators of the group $S_4$, $S$, $T$ and $U$.
These generators satisfy 
the following presentation rules
(see, e.g., \cite{Ishimori:2010au}):
\be 
S^2 = T^3 = (ST)^3 = U^2 = (TU)^2 = (SU)^2 = (STU)^4 = {\bf 1}\,,
\label{S4STU}
\ee
%
${\bf 1}$ being the unit (identity) element of $S_4$, i.e., 
it is the $3\times3$ unit matrix  
in the case of the triplet representations of $S$, $T$ and $U$. 
In what follows we will use the basis \cite{King:2009mk} in which 
$S$, $T$ and $U$ have the following form  in the two triplet 
representations 
\footnote{As can be shown, the results one obtains 
for the form of the PMNS matrix are independent of the 
chosen basis for the generators of the discrete symmetry   
group $G_f$.}: 
\be
S = \frac{1}{3}
\begin{pmatrix}
-1 & 2 & 2 \\
2 & -1 & 2 \\
2 & 2 & -1
\end{pmatrix}\,,
\quad
T = 
\begin{pmatrix}
1 & 0 & 0 \\
0 & \omega^2 & 0 \\
0 & 0 & \omega
\end{pmatrix}
 \quad
 {\rm and}
 \quad
 U = \mp\,
 \begin{pmatrix}
 1 & 0 & 0 \\
 0 & 0 & 1 \\
 0 & 1 & 0
\end{pmatrix}\,,
\label{S4STU3drep}
\ee
%
where $\omega = e^{2\pi i/3}$.
For simplicity we use the same notation ($S$, $T$ and $U$) for 
the generators and their 3-dimensional representation matrices.

 Assume next that \cite{Lam:2008rs} (see also, e.g.,  
\cite{Tanimoto:2015nfa,Girardi:2015rwa})
\be
G_e = Z^T_3 = \{1,T,T^2\}\,,~~
G_\nu = Z^S_2\times Z^U_2 = \{1,S,U,SU\}\,,
\label{GeZ3GnuZ2Z2}
\ee
%
where $Z^T_3$ and $Z^S_2\times Z^U_2$ are two specific 
$Z_3$ and $Z_2\times Z_2$ subgroups of $S_4$~
\footnote{$S$ and $U$ are order two elements of $S_4$ 
(since  $S^2 = U^2 = {\bf 1}$) 
and they commute. Correspondingly, 
$Z^S_2\times Z^U_2 = \{1,S,U,SU\}$ is a subgroup of $S_4$.
Similarly, $T$ is order 3 element of $S_4$ 
and  $Z^T_3 = \{1,T,T^2\}$ is a subgroup of $S_4$.}. 
In this case we have, in particular: 
$\rho_3(g_e) = 1,T,T^2$, $T$ being the diagonal matrix 
given in eq. (\ref{S4STU3drep}). 
As a consequence, $U_e$, which diagonalises 
$\rho_3(g_e)= T$, is just a diagonal phase 
matrix, whose phases are unphysical 
(they can be absorbed by 
the charged lepton fields in the weak charged 
lepton current of the weak interaction Lagrangian),
while $M_e$ is a diagonal matrix with the masses 
of the electron, muon and tauon as diagonal elements.

 It follows from eq. (\ref{S4STU}) that  
$\rho_3(g_\nu) = S$ and $\rho_3(g'_\nu) = U$ 
commute. In the triplet representation of the generators of 
$S_4$ employed by us, eq. (\ref{S4STU3drep}), 
$S$ and $U$ are real symmetric matrices. 
Thus, they are diagonalised by a real orthogonal matrix.  The 
matrix which diagonalises both 
$\rho_3(g_\nu) = S$ and $\rho_3(g'_\nu) = U$,
with $S$ and $U$ given in  eq. (\ref{S4STU3drep}),
is the orthogonal tri-bimaximal (TBM) mixing matrix \cite{TBM}: 
\begin{equation}
U^\circ_{\nu}= V_{\rm TBM} =  
 \begin{pmatrix}
\sqrt{\dfrac{2}{3}} & \dfrac {1}{\sqrt{3}} & 0 \vspace{0.2cm} \\
- \dfrac{1}{\sqrt{6}} & \dfrac{1}{\sqrt{3}} &
- \dfrac{1}{\sqrt{2}} \vspace{0.2cm} \\
- \dfrac{1}{\sqrt{6}}  &
\dfrac{1}{\sqrt{3}} & \dfrac{1}{\sqrt{2}}
\end{pmatrix} \;.
\label{VTBM}
\end{equation}
%
Indeed, it is not difficult to check that
\begin{eqnarray}
\label{VTBMS}
V^\dagger_{\rm TBM}\,S\,V_{\rm TBM}  = {\rm diag}(-1,1,-1)\,,\\[0.30cm]
\label{VTBMU}
V^\dagger_{\rm TBM}\,U\,V_{\rm TBM} = \pm\, {\rm diag}(1,1,-1)\,.%
\end{eqnarray}
%
Thus, in the discussed case of $S_4$ symmetry and 
residual symmetries $G_e = Z^T_3$ and $G_\nu = Z^S_2\times Z^U_2$, 
the PMNS matrix has the TBM form 
\cite{Lam:2008rs}, 
$U = U^\circ_{\nu}\, P^\circ = V_{\rm TBM}\, P^\circ$.
We can cast $V_{\rm TBM}$ in the form:
\be
V_{\rm TBM} =  
R_{23} \left (\theta^\nu_{23} \right) R_{13}\left (\theta^\nu_{13} \right)
R_{12}\left(\theta^\nu_{12}\right)\,,~~
\theta^\nu_{23} = -\,\pi/4,~\theta^\nu_{13} = 0\,,
~\theta^\nu_{12}=\sin^{-1}\frac{1}{\sqrt{3}}\,,
\ee
%
where $R_{23}(\theta^\nu_{23})$, $R_{13}(\theta^\nu_{13})$ 
and $R_{12}(\theta^\nu_{12})$ are $3\times3$ orthogonal matrices 
describing rotations in the 2-3, 1-3 and 1-2 planes, respectively.
We see that in the case of the TBM symmetry form we have
 $\sin^2\theta^\nu_{12} = 1/3$,
$\sin^2\theta^\nu_{23} = 1/2$ and 
$\sin^2\theta^\nu_{13} = 0$. Without additional 
corrections leading to $\theta_{13} \cong 0.15 \neq 0$, 
the TBM symmetry from of the PMNS matrix is ruled out by the data. 

 We will consider next two cases of 
realistic models based on the flavour symmetries
$G_f = A_4$ and $G_f = S_4$,  
in which the corrections to the 
underlying symmetry form of the PMNS matrix 
are obtained by ``decreasing'' 
the residual symmetry 
$G_\nu$ from  $Z_2\times Z_2$ symmetry to  $Z_2$,  

%
\subsection{Neutrino Mixing from $A_4$ Symmetry}
\label{subsec33}
%
%

The group $A_4$ has two generators  $S$ and $T$, which satisfy 
the presentation rules given in eq. (\ref{S4STU}). 
In the triplet representation of interest 
and in the Altarelli-Feruglio basis \cite{Altarelli:2005yx}, 
$S$ and $T$ have the form  given in eq. (\ref{S4STU3drep}).

 Assume next that (see, e.g., \cite{Tanimoto:2015nfa,Girardi:2015rwa}) 
\be
G_e = Z^T_3 = \{1,T,T^2\}\,,~~~G_\nu = Z^S_2 = \{1,S\}\,,
\label{GeZ3GnuZ2}
\ee
%
where  
$Z^T_3$ and $Z^S_2$ are two specific 
$Z_3$ and $Z_2$ subgroups of $A_4$.
In this case we have, in particular: 
$\rho(g_e) = 1,T,T^2$, $T$ being the diagonal matrix 
given in eq. (\ref{S4STU3drep}).
As a consequence, $U_e$, which diagonalises 
$\rho(g_e)= T$, 
as in the case discussed in the preceding subsection,
is just a diagonal phase 
matrix, whose phases are unphysical,  
while $M_e$ is a diagonal matrix with the masses 
of the electron, muon and tauon as diagonal elements.

 The most general matrix which diagonalises 
$\rho(g_\nu) = S$, with $S$ given in  eq. (\ref{S4STU3drep})
has the form: 
\be
U^\circ_{\nu} = V_{\rm TBM}\,U_{13}(\theta^\nu_{13},\alpha)\,,
\label{UonuA4}
\ee
%
where $V_{\rm TBM}$ is the tri-bimaximal (TBM) mixing matrix 
given in eq. (\ref{VTBM}),
and 
\begin{equation}
U_{13}(\theta^\nu_{13},\alpha) 
= 
 \begin{pmatrix}
\cos \theta^{\nu}_{13} & 0  & \sin \theta^{\nu}_{13}\,e^{i\alpha}   \vspace{0.2cm} \\
0  &  1 & 0 \\
-\,\sin \theta^{\nu}_{13}\,e^{-i\alpha}  & 0  &
\cos \theta^{\nu}_{13}
\end{pmatrix} \;.
\label{U13}
\end{equation}
%
The angle $\theta^{\nu}_{13}$ and the phase $\alpha$ 
in $U_{13}(\theta^\nu_{13},\alpha)$
are arbitrary free parameters.
Indeed, it is not difficult to convince oneself that
\be 
S =  U^\circ_{\nu}\,{\rm diag}(-1,1,-1)\,(U^\circ_{\nu})^\dagger 
= V_{\rm TBM}\,{\rm diag}(-1,1,-1)\,V^T_{\rm TBM}\,.
\label{Sdiag}
\ee
%
Thus, the matrix $U_{13}(\theta^\nu_{13},\alpha)$ appears 
in the matrix $U^\circ_{\nu}$ diagonalising $S$ 
as a consequence of the degeneracy of the first and third 
eigenvalues of $S$.

 We see that in the $A_4$ model considered, the underlying symmetry form 
of the PMNS matrix is the tri-bimaximal mixing, $V_{\rm TBM}$. 
The matrix $U_{13}(\theta^\nu_{13},\alpha)$ provides the 
necessary corrections to  $V_{\rm TBM}$ that lead, in particular, to 
$\theta_{13} \neq 0$. Thus, the model considered contains two free 
parameters - the angle $\theta^\nu_{13}$ and the phase $\alpha$.
 
 Taking into account the results for the forms of $U_e$ and 
$U^\circ_{\nu}$ we have obtained and eq. (\ref{UnuU0nu}), 
we get the following expression for the PMNS matrix:
\be
U_{\rm PMNS} 
= U^\circ_{\nu}\,P^\circ = V_{\rm TBM}\,U_{13}(\theta^\nu_{13},\alpha)\,P^\circ  
=
 \begin{pmatrix}
\sqrt{\dfrac{2}{3}}c & \dfrac {1}{\sqrt{3}} & \sqrt{\dfrac{2}{3}}s\,e^{i\alpha} \vspace{0.2cm} \\
- \dfrac{c}{\sqrt{6}} +\dfrac{s}{\sqrt{2}}e^{-i\alpha}  & \dfrac{1}{\sqrt{3}} &
- \dfrac{c}{\sqrt{2}} - \dfrac{s}{\sqrt{6}}\,e^{i\alpha}\vspace{0.2cm} \\
  - \dfrac{c}{\sqrt{6}} - \dfrac{s}{\sqrt{2}}e^{-i\alpha} &
\dfrac{1}{\sqrt{3}} &  \dfrac{c}{\sqrt{2}} - \dfrac{s}{\sqrt{6}}\,e^{i\alpha}
\end{pmatrix}\,P^\circ\;,
\label{A4PMNS}
\end{equation}
%
where $c \equiv \cos\theta^\nu_{13}$ and $s \equiv \sin\theta^\nu_{13}$.

  We will consider next the phenomenological predictions  
of the discussed  $A_4$  model of neutrino mixing.
Comparing, for example,  
the absolute values of the elements of the first rows of the 
PMNS matrix in eq. (\ref{A4PMNS}) and in the standard parametrisation, 
eqs. (\ref{VP}) and (\ref{eq:Vpara}), we get:
\be 
\sin^2\theta_{13} = \dfrac{2}{3}\,s^2\,,~~ 
\sin^2\theta_{12} \cos^2\theta_{13} = \dfrac{1}{3}\,.
\label{A4Resth12th12}    
\ee
%
Comparing the $U_{\mu 3}$ elements and using the first relation
in the preceding equation we find:
\be 
\sin^2\theta_{23} = \dfrac{1}{c^2_{13}}\,
\large |\dfrac{c}{\sqrt{2}} +
\dfrac{s}{\sqrt{6}}\,e^{i\alpha}\, \large |^2 = 
 \dfrac{1}{2} +  \dfrac{s_{13}}{2}\, 
\dfrac{(2 - 3\,s^2_{13})^{\frac{1}{2}}}{(1 - s^2_{13})}\,\cos\alpha\,.
\label{A4s2231}
\ee
%
To leading order in $s_{13}$ we have:
\be 
 \dfrac{1}{2} -  \dfrac{s_{13}}{\sqrt{2}}
\ltap  \sin^2\theta_{23} \ltap 
\dfrac{1}{2} +  \dfrac{s_{13}}{\sqrt{2}}  
\,,~~{\rm or}~~
 0.391\ltap  \sin^2\theta_{23} \ltap 0.609\,,
\label{A4s2232}
\ee
%
where the numerical values correspond to 
the maximal allowed value of $\sin^2\theta_{13}$
at $3\sigma$ C.L. The interval of possible values of  
$\sin^2\theta_{23}$ in eq. (\ref{A4s2232}) 
lies within the $3\sigma$ ranges of experimentally allowed 
values of $\sin^2\theta_{23}$ for NO and IO spectra, quoted in 
Table  \ref{tab:parameters}.

 Further, using the constraint $|U_{\mu 2}|^2 = 1/3$ 
(or  $|U_{\tau 2}|^2 = 1/3$) following from 
the form of $U_{\rm PMNS}$ in eq. (\ref{A4PMNS}), 
we obtain the following sum rule for the Dirac phase 
$\delta$:
\be
\cos\delta = 
\dfrac{\cos2\theta_{23}\,\cos2\theta_{13}}
{\sin2\theta_{23}\,\sin\theta_{13}\,(2-3\sin^2\theta_{13})^{\frac{1}{2}}}\,, 
\label{cosdA4}
\ee
%
where we have expressed $\cos\theta_{12}\sin\theta_{12}$
in terms of $\sin\theta_{13}$ using eq. (\ref{A4Resth12th12}).  

 It follows from the preceding brief discussion that 
$\theta_{13}$ and $\theta_{23}$ of the standard 
parametrisation of the PMNS matrix are equivalent 
to the two independent parameters $\theta^\nu_{13}$ 
and $\alpha$ of the considered $A_4$ model, 
while the angle $\theta_{12}$ and the Dirac phase $\delta$ 
can be considered as functions of 
$\theta_{13}$ and $\theta_{23}$~
\footnote{Actually, any pair of the four parameters 
 $\theta_{12}$, $\theta_{23}$, $\theta_{13}$ and    
$\delta$ can play the role of the two independent 
parameters of the model.}.

 The phase $\alpha$ and the Dirac phase $\delta$ 
are related via 
\be
\sin2\theta_{23}\, \sin\delta = \sin\alpha\,.
\label{sindsinalphaA4}
\ee
%
This relation follows from the 
equality between the expressions of 
the rephasing invariant $J_{\rm CP}$, eq. (\ref{JCPstandparam}), 
in the standard parametrisation of the PMNS matrix and in  
the parametrisation defined in eq. (\ref{A4PMNS}).

 As it is not difficult to show, the phase $\alpha$ contributes also to the 
Majorana phase $\alpha_{31}$ of the 
PMNS matrix, eqs. (\ref{VP}) and (\ref{eq:Vpara}):
\be
\dfrac{\alpha_{31}}{2} = \dfrac{\xi_{31}}{2} + \alpha_{2} + \alpha_3\,,
\label{alpha31A4}
\ee
%
where 
\be
\label{alpha2alpha3}
\alpha_2 = {\rm arg}
\big(-\,\dfrac{c}{\sqrt{2}} - \dfrac{s}{\sqrt{6}}\,e^{i\alpha}\big)\,,~~~
\alpha_3 = {\rm arg}
\big(\dfrac{c}{\sqrt{2}} - \dfrac{s}{\sqrt{6}}\,e^{i\alpha}\big)\,,
\ee
%
\begin{eqnarray}
\label{alpha2}
\sin\alpha_2 = -\,\dfrac{s}{\sqrt{6}}\,\dfrac{\sin\alpha}{s_{23}\,c_{13}} = 
 -\,\tan\theta_{13}\,\cos\theta_{23}\,\sin\delta\,,\\[0.30cm]
\label{alpha3}
\sin\alpha_3 = -\,\dfrac{s}{\sqrt{6}}\,\dfrac{\sin\alpha}{c_{23}\,c_{13}} = 
 -\,\tan\theta_{13}\,\sin\theta_{23}\,\sin\delta\,,
\end{eqnarray}
%
where we have used eq. (\ref{sindsinalphaA4}). 
In  eqs. (\ref{alpha2}) and (\ref{alpha3}), 
$\sin\delta$ can be considered as a function of 
$\theta_{23}$ and $\theta_{13}$ (see  eq. (\ref{cosdA4})). 
We also have:
\be
\sin(\alpha - \alpha_2 - \alpha_3) = -\,\sin\delta\,.
\label{sindsinalphab2b3A4}
\ee
%

That the phases $\alpha_{2}$ and $\alpha_3$
contribute to the Majorana phase $\alpha_{31}$  
can be seen by casting the parametrisation of 
$U_{\rm PMNS}$ in eq. (\ref{A4PMNS}) in the standard 
parametrisation form, eqs. (\ref{VP}) and (\ref{eq:Vpara}). 
This can be done by multiplying the matrix 
in eq. (\ref{A4PMNS}) on the right by 
$P^*_{33}P_{33}$ with $P_{33}  = {\rm diag}(1,1,e^{i(\alpha_2 + \alpha_3)})$,
and absorbing $P_{33}$ in $P^\circ$. The phases $e^{-i\alpha_3}$ and 
$e^{-i\alpha_2}$, which after that appear respectively in the 
$U_{\mu 3}$ and $U_{\tau 3}$ elements of $U_{\rm PMNS}$ 
in  eq. (\ref{A4PMNS}), are removed from these elements 
by phase redefinition of the $\mu^\mp$ and $\tau^\mp$  fields  
in the weak charged lepton current (\ref{CC}). 
As a consequence of these simple manipulations 
the phase factor $e^{i\alpha_3}$ ($e^{i\alpha_2}$) appears in the 
$U_{\mu 1}$ and $U_{\mu 2}$ ($U_{\tau 1}$ and $U_{\tau 2}$)
elements of $U_{\rm PMNS}$, while the phase factor $e^{i\alpha}$
of the $U_{e 3}$ element (see eq. (\ref{A4PMNS}))
changes to $e^{i(\alpha - \alpha_2 - \alpha_3)}$.
The phases in  $P_{33}P^\circ$ contribute to the Majorana phases 
$\alpha_{21}/2$ and $\alpha_{31}/2$.

The phenomenology of neutrino mixing described by the PMNS matrix 
given in  (\ref{A4PMNS}), apart from 
the relation  (\ref{sindsinalphaA4}) and the contribution  
of the phase $\alpha$ to the Majorana phase $\alpha_{31}$, 
eqs. (\ref{alpha31A4}) -  (\ref{alpha3}), as well as of the 
relation (\ref{sindsinalphab2b3A4}), was discussed 
in \cite{Grimus:2008tt}.
The prediction for $\sin^2\theta_{12}$ in  
eq. (\ref{A4Resth12th12}) and the sum rule for the Dirac phases $\delta$, 
eq. (\ref{cosdA4}), can also be obtained from the 
general results on neutrino mixing in the case of $A_4$ lepton 
flavour symmetry derived in \cite{Girardi:2015rwa}.

Thus, the $A_4$ model considered predicts 
\footnote{The result for $\sin^2\theta_{12}$ 
and the sum rule for $\cos\delta$  
can be obtained respectively from 
eq. (58)  in subsection 4.1 and 
Table 3 (Case B1)  in \cite{Girardi:2015rwa} by setting 
$\sin^2\theta^\circ_{12} = 1/3$ and $\sin^2\theta^\circ_{23} = 1/2$.
}
i) a correlation between the values of $\sin^2\theta_{12}$ and 
$\sin^2\theta_{13}$:  $\sin^2\theta_{12} = 1/(3(1-\sin^2\theta_{13}))$,
ii) an interval of possible values of $\sin^2\theta_{23}$, 
which depends on $\sin\theta_{13}$, and
ii) a sum rule for the Dirac CPV phase $\delta$
by which $\cos\delta$ is expressed in terms of the two 
measured neutrino mixing angles $\theta_{13}$ and $\theta_{23}$. 
In this model the Majorana phases $\alpha_{21}$ and $\alpha_{31}$ 
remain undetermined due to the contribution respectively of the 
phases $\xi_{21}$ and $\xi_{31}$, which are not fixed.

 The correlation between $\sin^2\theta_{12}$ and 
$\sin^2\theta_{13}$ leads to the prediction
$\sin^2\theta_{12} = 0.340$,
where we have employed the best fit 
value of $\sin^2\theta_{13}$ in 
Table \ref{tab:parameters}.
This value lies outside the $2\sigma$, but is inside the $3\sigma$, 
currently allowed intervals of values of $\sin^2\theta_{12}$. 
Using the best fit values of $\sin^2\theta_{23}$ 
and $\sin^2\theta_{13}$ for the NO and IO neutrino mass 
spectrum, given in Table \ref{tab:parameters}, 
and the sum rule for $\cos\delta$, eq. (\ref{cosdA4}), we find: 
\begin{eqnarray}
\label{cosdA4NO}
 & \cos\delta = 0.728\,(-\,0.865)\,,~
\delta = \pm 43.32^\circ\,( 180^\circ \pm 30.07^\circ)\,,~~{\rm NO~(IO)}\,.
\label{cosdA4IO}
\end{eqnarray}
%
If instead we use the best fit values for NO (IO) spectrum of 
$\sin^2\theta_{23} = 0.538~(0.554)$ 
and $\sin^2\theta_{13} = 0.02206~(0.02227)$ 
reported in \cite{NuFITv32Jan2018} we get:
\begin{eqnarray}
\label{cosdA4NO2018}
& \cos\delta = -\,0.353\,(-500)\,,
~\delta = 180^\circ \pm 69.4^\circ\,
(180^\circ \pm 60.0^\circ)\,,~~{\rm NO~(IO)}\,.
\label{cosdA4IO2018}
\end{eqnarray}
%

Thus, as a consequence primarily 
of the fact that $\cos\delta\propto \cos2\theta_{23}$,
the predictions for $\cos\delta$, and correspondingly of $\delta$, 
depend strongly on the values of $\sin^2\theta_{23}$ and 
can differ significantly for the two neutrino mass orderings. 
The values of $\delta = 43.32^\circ$, $110.6^\circ$ and $120^\circ$ are 
strongly disfavored (if not ruled out) by the current data. 
It should be added that the difference between 
the predictions of $\cos\delta$ ($\delta$) 
for NO and IO neutrino mass spectra are due 
to the difference between the best fit values 
of $\sin^2\theta_{23}$ for the two spectra 
(see Table \ref{tab:parameters} and eq. (\ref{eq:s2th232018})).
For  $\sin^2\theta_{23} = 0.5$ we have for both spectra 
$\cos\delta=0$, or $\delta = \pi/2,3\pi/2$, 
with  $\delta = \pi/2$ strongly disfavored 
by the current data.

 It follows from the preceding results that the 
high precision measurement of 
$\sin^2\theta_{12}$ combined with the data on 
$\sin^2\theta_{13}$ will allow to critically test the 
predicted correlation between $\sin^2\theta_{12}$ and 
$\sin^2\theta_{13}$ of the considered $A_4$ model.
The high precision measurement of 
$\sin^2\theta_{23}$, the data on  $\sin^2\theta_{13}$
and a sufficiently precise determination of $\delta$ 
will make it possible to test the sum rule 
predictions for $\delta$ of the model. 
With the indicated tests the $A_4$ model of neutrino mixing 
discussed in the present subsection will
be either verified or ruled out.    

%
\subsection{Neutrino Mixing from $S_4$ Symmetry}
\label{subsec34}
%
%

 We will consider next a second rather simple example of 
generation of neutrino mixing based on the $S_4$ 
symmetry. 
We recall that the three $S_4$ generators  $S$, $T$ and $U$ satisfy 
the presentation rules given in eq. (\ref{S4STU}).
%
In the triplet representation of interest 
and in the basis employed by us $S$, $T$ 
and $U$ are given in eq. (\ref{S4STU3drep}).

 In this case let us assume that (see, e.g., \cite{Girardi:2015rwa})
\be
G_e = Z^T_3 = \{1,T,T^2\}\,,~~~
G_\nu = Z^{SU}_2 = \{1,SU\}\,,
\label{GeZ3GnuZ2}
\ee
%
where  $Z^{T}_3$, as we have discussed, is a 
$Z_3$  subgroup also of $S_4$ and 
$Z^{SU}_2$ is one of the 
$Z_2$ subgroups of $S_4$.
As in the case of $A_4$ symmetry 
considered in the preceding subsection,
$U_e$, which diagonalises 
$\rho(g_e)= T$, is effectively a unit $3\times3$ 
matrix and $M_e$ is a diagonal matrix containing 
the masses of the charged leptons.

 The matrix $U^\circ_\nu$, which diagonalises 
the element $\rho(g_{\nu}) = SU$ of $Z^{SU}_2$ 
(and $M^\dagger_{\nu}\,M_{\nu}$), with $S$ and $U$ 
given in eq. (\ref{S4STU3drep}), has the 
following general form:
\be
U^\circ_{\nu} = V_{\rm TBM}\,U_{23}(\theta^\nu_{23},\beta)\,,
\label{UonuA4}
\ee
%
where $V_{\rm TBM}$ is the TBM mixing matrix and 
\begin{equation}
U_{23}(\theta^\nu_{23},\beta) 
= 
 \begin{pmatrix}
1 & 0  & 0   \vspace{0.2cm} \\
0  & \cos \theta^{\nu}_{23} & \sin \theta^{\nu}_{23}\,e^{i\beta}   \\
0  & -\,\sin \theta^{\nu}_{23}\,e^{-i\beta}   &
\cos \theta^{\nu}_{23}
\end{pmatrix} \;,
\label{U23}
\end{equation}
%
the angle $\theta^{\nu}_{23}$ and the phase $\beta$ 
being arbitrary free parameters.
The form of $U^\circ_\nu$ follows from the fact that 
the element $\rho(g_\nu) = SU$, as is easy to verify, 
is diagonalised by  $V_{\rm TBM}$. However, in the resulting 
diagonal matrix the 2nd and the 3rd eigenvalues 
are degenerate and thus it is invariant with respect 
to a unitary transformation with   
$U_{23}(\theta^\nu_{23},\beta)$:
\be 
SU =\pm\, V_{\rm TBM}\,{\rm diag}(-1,1,1)\,V^T_{\rm TBM}
= \pm\, V_{\rm TBM}\,U_{23}(\theta^\nu_{23},\beta)\,{\rm diag}(-1,1,1)\,
(V_{\rm TBM}\,U_{23}(\theta^\nu_{23},\beta))^\dagger\,.
\label{SUdiag}
\ee
%

 We see that also in the model with $S_4$ symmetry under 
discussion, the underlying symmetry form 
of the PMNS matrix is again the TBM one, $V_{\rm TBM}$.
The matrix $U_{23}(\theta^\nu_{23},\beta)$ provides the 
necessary corrections to $V_{\rm TBM}$ leading, e.g., to 
$\theta_{13} \neq 0$. 

 Similarly to the model based on the 
$A_4$ symmetry discussed in the previous subsection, 
the $S_4$ model we are discussing contains two free 
parameters - the angle $\theta^\nu_{23}$ and the phase $\beta$.
However, as we show below, the testable phenomenological 
predictions of the model with $S_4$ symmetry differ significantly 
from the analogous predictions of the $A_4$ model.

 From eqs. (\ref{VTBM}), (\ref{U23}) and (\ref{UnuU0nu}) 
we get for the PMNS matrix:
\be
U_{\rm PMNS} 
= U^\circ_{\nu}\,P^\circ = V_{\rm TBM}\,U_{23}(\theta^\nu_{23},\beta)\,P^\circ  
=
\begin{pmatrix}
\sqrt{\dfrac{2}{3}} & \dfrac{c^\nu_{23}}{\sqrt{3}} & 
\dfrac{s^\nu_{23}}{\sqrt{3}}\,e^{i\beta} \vspace{0.2cm} \\
-\,\dfrac{1}{\sqrt{6}} & \dfrac{c^\nu_{23}}{\sqrt{3}} 
+ \dfrac{s^\nu_{23}}{\sqrt{2}}e^{-i\beta}  & 
-\,\dfrac{c^\nu_{23}}{\sqrt{2}} +
\dfrac{s^\nu_{23}}{\sqrt{3}}\,e^{i\beta} 
\vspace{0.2cm} \\
 -\,\dfrac{1}{\sqrt{6}}  & \dfrac{c^\nu_{23}}{\sqrt{3}} 
- \dfrac{s^\nu_{23}}{\sqrt{2}}e^{-i\beta} &
\dfrac{c^\nu_{23}}{\sqrt{2}} + \dfrac{s^\nu_{23}}{\sqrt{3}}\,e^{i\beta}
\end{pmatrix}\,P^\circ\;,
\label{S4PMNS}
\end{equation}
%
where $c^\nu_{23} \equiv \cos\theta^\nu_{23}$ and 
$s^\nu_{23} \equiv \sin\theta^\nu_{23}$.

 Proceeding as in subsection \ref{subsec33} we find:
\be 
\sin^2\theta_{13} = \dfrac{1}{3}\sin^2\theta^\nu_{23}\,,~~ 
\sin^2\theta_{12} = \dfrac{1}{3}\cos^2\theta^\nu_{23} = 
\dfrac{1 - 3\sin^2\theta_{13}}{3(1-\sin^2\theta_{13})}\,.
\label{S4s213s212}    
\ee
%
The neutrino mixing parameter $\sin^2\theta_{23}$ is 
determined by $\theta^\nu_{23}$ (or $\theta_{13}$) and $\beta$ and 
its value is not predicted:
\be 
\sin^2\theta_{23} = \dfrac{1}{c^2_{13}}\,
\large |-\,\dfrac{c^\nu_{23}}{\sqrt{2}} +
\dfrac{s^\nu_{23}}{\sqrt{3}}\,e^{i\beta}\, \large |^2 
=  \dfrac{1}{2} -  \sqrt{2}\,s_{13}\, 
\dfrac{(1 - 3\,s^2_{13})^{\frac{1}{2}}}{(1 - s^2_{13})}\,\cos\beta\,.
\label{S4s2231}
\ee
%
To leading order in $s_{13}$ we have:
\be 
 \dfrac{1}{2} -  \sqrt{2}\, s_{13}
\ltap  \sin^2\theta_{23} \ltap 
\dfrac{1}{2} +  \sqrt{2}\,s_{13}  
\,,~~{\rm or}~~
 0.293\ltap  \sin^2\theta_{23} \ltap 0.707\,,
\label{S4s2232}
\ee
%
where the numerical values are obtained for  
the maximal value of $\sin^2\theta_{13}$
allowed at $3\sigma$ C.L. The interval of values of  
$\sin^2\theta_{23}$ in eq. (\ref{S4s2232}) 
is larger than the $3\sigma$ experimentally allowed 
NO and IO intervals of values of $\sin^2\theta_{23}$ (see 
Table \ref{tab:parameters}).

The Dirac phase $\delta$ satisfies the following 
sum rule:
\be
\cos\delta = 
\dfrac{ \dfrac{1}{6} - c^2_{23} 
+ \dfrac{2}{3c^2_{13}}\,(c^2_{23} - s^2_{23}\,s^2_{13})}
{2c_{23}\,s_{23}\,s_{13}\,c_{12}s_{12}} = 
\dfrac{(-\,1 + 5s^2_{13})\,\cos2\theta_{23}}
{2\sqrt{2}\,\sin2\theta_{23}\,s_{13}\,(1 - 3s^2_{13})^{\frac{1}{2}}}\,, 
\label{cosdS4}
\ee
%
where we expressed  $c_{12}s_{12}$ in terms of $\theta_{13}$
using eq. (\ref{S4s213s212}), 
\be
c_{12}\,s_{12} = \dfrac{\sqrt{2}}{3c^2_{13}}\, (1-3s^2_{13})^{\frac{1}{2}}\,.
\label{c12s12th13}
\ee
%
We also have:
\be
\sin2\theta_{23}\, \sin\delta = \sin\beta\,.
\label{sindsinbetaS4}
\ee
%

 Similarly to the phases $\alpha$ of the $A_4$ model 
considered in the preceding subsection,
the phase $\beta$ of the discussed $S_4$ model 
contributes to the Majorana phase $\alpha_{31}$ 
in the standard parametrisation of the 
PMNS matrix (see eqs. (\ref{VP}) and (\ref{eq:Vpara})):
\be
\dfrac{\alpha_{31}}{2} = \dfrac{\xi_{31}}{2} + \beta_{2} + \beta_3\,,
\label{alpha31S4}
\ee
%
where 
\be
\label{beta2beta3}
\beta_2 = {\rm arg}
\big(-\,\dfrac{c^\nu_{23}}{\sqrt{2}} +
\dfrac{s^\nu_{23}}{\sqrt{3}}\,e^{i\beta}\big)\,,~~~
\beta_3 = {\rm arg}
\big(\dfrac{c^\nu_{23}}{\sqrt{2}} + \dfrac{s^\nu_{23}}{\sqrt{3}}\,e^{i\beta}\big)\,,
\ee
%
\begin{eqnarray}
\label{beta2}
\sin\beta_2 = \dfrac{s^\nu_{23}}{\sqrt{3}}\,\dfrac{\sin\beta}{s_{23}\,c_{13}} = 
 \tan\theta_{13}\,\cos\theta_{23}\,\sin\delta\,,\\[0.30cm]
\label{beta3}
\sin\beta_3 = \dfrac{s^\nu_{23}}{\sqrt{3}}\,\dfrac{\sin\beta}{c_{23}\,c_{13}} = 
 \tan\theta_{13}\,\sin\theta_{23}\,\sin\delta\,,
\end{eqnarray}
%
where we have used eqs. (\ref{S4PMNS})  and (\ref{sindsinbetaS4}) 
and  $\sin\delta$ in eqs. (\ref{beta2}) and (\ref{beta3})
can be considered as a function of $\theta_{23}$ and $\theta_{13}$ 
(via eq. (\ref{cosdS4})). 
We also have:
\be
\sin(\beta - \beta_2 - \beta_3) = -\,\sin\delta\,.
\label{sindsinalphab2b3S4}
\ee
%

 The model with $U_{\rm PMNS}= V_{\rm TBM}\,U_{23}(\theta^\nu_{23},\beta)$
was discussed on general phenomenological grounds in
\cite{Albright:2008rp},
where the predictions given in eqs. (\ref{S4s213s212}) and 
(\ref{cosdS4}) were obtained and the dependence of $\delta$ 
on $\sin^2\theta_{23}$ for a set of different values of 
$\theta_{13}$ was studied graphically.
The correlation between $\sin^2\theta_{21}$ and $\sin^2\theta_{13}$ 
and the sum rule for $\cos\delta$ can also can be obtained 
from the general results for the 
group $S_4$ derived in \cite{Girardi:2015rwa}~
\footnote{In \cite{Girardi:2015rwa} a different basis 
for the $S_4$ generators $S$, $T$ and $U$ has been employed.
The results of interest for, e.g., $\sin^2\theta_{12}$ 
in eqs. (\ref{S4s213s212})
and the sum rule for $\cos\delta$, eq. (\ref{cosdS4}), 
follow respectively from eq. (66)  in subsection 4.2 and 
Table 3 (Case B2)  in \cite{Girardi:2015rwa} by setting 
$\sin^2\theta^\circ_{12} = 1/6$ and $\sin^2\theta^\circ_{13} = 1/5$.
}.

 Thus, as in the $A_4$ model,
$\theta_{13}$ and $\theta_{23}$, or any pair 
of the four parameters $\theta_{12}$, $\theta_{23}$ 
$\theta_{13}$ and $\delta$, can be considered as 
the two independent parameters of the $S_4$ model.
The model predicts a correlation between 
the values of $\sin^2\theta_{12}$ and $\sin^2\theta_{13}$, 
which for the best fit value of $\sin^2\theta_{13}$ 
implies $\sin^2\theta_{12} = 0.319$. This prediction lies in the 
current $1\sigma$ allowed interval of values of  $\sin^2\theta_{12}$.
Using eqs. (\ref{cosdS4}),  (\ref{c12s12th13}) and the best fit values 
of $\sin^2\theta_{23}$ and  $\sin^2\theta_{13}$ 
from Table \ref{tab:parameters},
we also get the following predictions for 
$\cos\delta$ in the cases of NO and 
IO neutrino mass spectra:
\begin{eqnarray}
\label{cosdS4NO}
&\cos\delta = -\,0.338\,(0.402)\,,~
\delta = \pm 109.73^\circ\,(\pm 66.27^\circ)\,,~~{\rm NO~(IO)}\,.
\label{cosdS4IO}
\end{eqnarray}
%
Using instead the best fit values of $\sin^2\theta_{23}$ 
and $\sin^2\theta_{13}$ for NO (IO) spectrum from \cite{NuFITv32Jan2018}
we find rather different results due essentially to the difference 
in the best fit values of $\sin^2\theta_{23}$:
\begin{eqnarray}
\label{cosdS4NO2018}
& \cos\delta = 0.167\,(0.237)\,,~
\delta = \pm 80.38^\circ\,(\pm 76.30^\circ)\,,~~{\rm NO~(IO)}\,.
\label{cosdS4IO2018}
\end{eqnarray}
%
The values  $\delta = 109.73^\circ$, $66.27^\circ$, $80.38^\circ$ 
and $76.30^\circ$ are strongly disfavored by the current data. 
As in the $A_4$ model, the difference between 
the predictions of $\cos\delta$ ($\delta$) 
for NO and IO neutrino mass spectra 
are a consequence of the difference between the best fit values 
of $\sin^2\theta_{23}$ for the two spectra 
(see Table \ref{tab:parameters} and eq. (\ref{eq:s2th232018})).
For  $\sin^2\theta_{23} = 0.5$ we have for both spectra 
$\cos\delta=0$, or $\delta = \pi/2,3\pi/2$,  
also in the $S_4$ model,
with  $\delta = \pi/2$ strongly disfavored 
by the current data.

  As we have seen, the $A_4$ and $S_4$ models considered lead to 
largely different predictions for $\sin^2\theta_{12}$ and, 
if $\theta_{23} \neq \pi/4$, for $\cos\delta$ ($\delta$) as well. 
These predictions can be used to 
discriminate experimentally between the two  models.    
In both $A_4$ and $S_4$ models we have discussed the Majorana 
phases are not predicted.
%
\subsection{Comment on the Symmetry Breaking}
\label{subsec35}
%
%
  
 The discrete symmetry approach to neutrino mixing 
we have discussed so far allows to explain quantitatively 
the observed pattern of neutrino mixing. A complete 
self-consistent (renormalisable) theory based on this 
approach should include also a mechanism of neutrino 
mass generation as well as details of breaking of the 
flavour symmetry $G_f$ to the residual symmetries 
$G_e$ and $G_\nu$  in the charged lepton and 
neutrino sectors. As a rule, the non-Abelian 
flavour symmetry $G_f$ is broken spontaneously by 
a set of scalar fields, flavons, which are singlets with respect to 
the Standard Theory $SU(2)_L\times U(1)_{\rm Y_W}$ 
gauge symmetry but transform according 
certain irreducible representations of 
$G_f$, couple in a $G_f$-invariant manner to 
the LH lepton doublet fields and RH charged lepton 
$SU(2)_L$ singlet fields via 
Yukawa-type (typically non-renormalisable effective)  
interactions. These Yukawa-type effective interactions appear 
in the low-energy limit of a theory which is renormalisable at 
some high energy scale $\Lambda$ where the symmetry $G_f$ 
is exact (see, e.g., \cite{Altarelli:2010gt,King:2013eh,Meloni:2017cig,
Altarelli:2009gn,Altarelli:2012ss}).
The flavons develop non-zero vacuum 
expectation values in specific directions (``vacuum alignment'').
In the case of the $A_4$ model 
considered in subsection \ref{subsec33}, for example, 
the  $A_4$ symmetry breaking leading to 
$G_e = Z^T_3$ and $G_\nu = Z^S_2$ and generating 
Majorana mass term for the LH 
flavour neutrino fields  can be achieved 
i) by assigning $\tilde{e}_R(x)$, $\tilde{\mu}_R(x)$ and 
$\tilde{\tau}_R(x)$ to the three different singlet 
representations of $A_4$ 
${\bf 1}$, ${\bf 1''}$ and ${\bf 1'}$ (see Table 2), 
respectively,    
ii) by introducing two $A_4$ triplet and two $A_4$ singlet
flavon scalar fields, which develop vacuum expectation values    
in specific directions, and
iii) by using the rules of tensor products of irreducible 
representations for $A_4$ (for further details see, e.g.,  
\cite{Altarelli:2010gt,Ishimori:2010au,Tanimoto:2015nfa}).

   Discussing the flavon sectors of the models considered 
is beyond the scope of the present article. 
Examples of complete self-consistent (renormalisable) 
models, in which the breaking of the flavour symmetry $G_f$ 
to desired residual symmetries $G_e$ and $G_\nu$ 
with the help of sets of flavon fields 
developing non-zero vacuum expectation values 
in requisite directions and, thus,  generating 
$G_e-$ invariant charged lepton mass term
and $G_\nu-$ invariant neutrino Majorana mass term, 
include, e.g., the models in 
refs. \cite{Ding:2013hpa,Girardi:2013sza,Gehrlein:2014wda,
Altarelli:2012ss,King:2011zj}; 
for a review see, e.g., ref. \cite{King:2013eh}.
%
%
\subsection{Alternative Symmetry Forms of $U_{\nu}$: 
Bimaximal, Golden Ratio and Hexagonal Mixing 
}
\label{subsec36}
%
%

 Thus, TBM can only be an underlying approximate symmetry 
form of the PMNS neutrino mixing matrix. Other widely discussed 
underlying (approximate) symmetry forms of the PMNS matrix 
include: i) bimaximal (BM) mixing 
\footnote{Bimaximal mixing can also be  
a consequence of the conservation of the lepton charge
$L' = L_e - L_{\mu} - L_{\tau}$ (LC) \cite{SPPD82}, 
supplemented by $\mu - \tau$ symmetry.} \cite{BM},
ii) the golden ratio type A (GRA) mixing \cite{Datta:2003qg},
iii) the golden ratio type B (GRB)  mixing \cite{GRBM}, 
and 
iv) hexagonal (HG) mixing \cite{HGM}.
For all these forms, including the TBM one,
the matrix $U^\circ_{\nu}$ has the form:
$U^\circ_{\nu} = R_{23}(\theta^\nu_{23}) R_{13}(\theta^\nu_{13})
R_{12}(\theta^\nu_{12})$ with
$\theta^\nu_{23} = -\,\pi/4$ and  $\theta^\nu_{13} = 0$:
\begin{equation}
U^\circ_{\nu} =  R_{23} \left ( \theta^\nu_{23}=-\,\pi/4 \right)
R_{12}\left( \theta^\nu_{12}\right) = 
 \begin{pmatrix}
\cos \theta^{\nu}_{12} & \sin \theta^{\nu}_{12} & 0 \vspace{0.2cm} \\
- \dfrac{\sin \theta^{\nu}_{12}}{\sqrt{2}} &
\dfrac{\cos \theta^{\nu}_{12}}{\sqrt{2}} &
- \dfrac{1}{\sqrt{2}} \vspace{0.2cm} \\
- \dfrac{\sin \theta^{\nu}_{12}}{\sqrt{2}}  &
\dfrac{\cos \theta^{\nu}_{12}}{\sqrt{2}} &
\dfrac{1}{\sqrt{2}}
\end{pmatrix} \;.
\label{Unu12}
\end{equation}
%
The value of the angle  $\theta^{\nu}_{12}$, and thus of  
$\sin^2\theta^{\nu}_{12}$, depends on the symmetry 
form of $U^\circ_{\nu}$.
For the TBM, BM, GRA, GRB and HG forms we have: 
i)  $\sin^2\theta^{\nu}_{12} = 1/3$ (TBM),
ii)  $\sin^2\theta^{\nu}_{12} = 1/2$ (BM),
iii)  $\sin^2\theta^{\nu}_{12} =  (2 + \tilde{r})^{-1} \cong 0.276$ (GRA),
$\tilde{r}$ being the golden ratio, $\tilde{r} = (1 +\sqrt{5})/2$,
iv) $\sin^2\theta^{\nu}_{12} = (3 - \tilde{r})/4 \cong 0.345$ (GRB), and
v) $\sin^2\theta^{\nu}_{12} = 1/4$ (HG).

 As we have seen in subsections \ref{subsec32} and
\ref{subsec34}, the TBM form of $U^\circ_{\nu}$ 
can originate from $G_f=S_4$ symmetry \cite{S4} 
(with residual symmetry $G_\nu = Z^S_2\times Z^U_2$).
It can be obtained also from a $G_f = A_4$ symmetry \cite{A4}
(with $G_\nu= Z^S_2$ and the presence of accidental
 $\mu-\tau$ (i.e., $Z_2$) symmetry, see, e.g., \cite{Altarelli:2012ss}))
\footnote{
The TBM  form of $U^\circ_{\nu}$  can also be derived from 
$G_f = T'$ -  the double covering group of 
$A_4$ (see, e.g., \cite{Ishimori:2010au}) -  
with $G_\nu =   Z^{\rm S}_2$, provided  
the left-handed (LH) charged lepton and neutrino fields each 
transform as triplets  of $T^{\prime}$
(see, e.g.,  \cite{Girardi:2015rwa} for details).
Actually, as can be shown \cite{Feruglio:2007uu},
when working with 3-dimensional and 1-dimensional
representations of $T^{\prime}$, there is no way to distinguish 
$T^{\prime}$ from $A_4$.
}.

 The group $G_f = S_4$  can also be used 
to generate the BM from of $U^\circ _{\nu}$
(e.g., by choosing $G_{\nu} = Z_2$ combined with  
an accidental $\mu-\tau$ symmetry) 
\cite{S4,Altarelli:2009gn,Girardi:2015rwa}.

 The GRA form of $U^\circ_{\nu}$ can be obtained from the 
group $A_5$ \cite{A5}, which is the group 
of even permutations of five objects and is 
isomorphic to the group of rotational symmetries of the 
icosahedron. In this case  
$\sin^2 \theta^{\nu}_{12} = 1/(\tilde{r}\sqrt{5}) \cong 0.276$.

The GRB and HG forms of $U^\circ_{\nu}$ 
can be generated using the groups $G_f = D_{10}$ \cite{D10} and 
$G_f = D_{12}$ \cite{HGM}, respectively. The dihedral groups 
 $D_{10}$ and  $D_{12}$ are the groups of symmetries 
(rotations and reflections) of the regular 
decagon and dodecagon
\footnote{
The groups $D_{10}$ and $D_{12}$, as it is indicated in Table 
2, have 1-dimensional and 2-dimensional irreducible  
representations, but do not have  3-dimensional 
irreducible unitary representations. The problem of how  
the GRB and HG forms of $U^\circ_{\nu}$ 
can be generated using the groups $G_f = D_{10}$ and 
$G_f = D_{12}$, respectively, is discussed 
in refs. \cite{D10} and \cite{HGM} and we refer the interested 
reader to these articles.}. 
 $D_{10}$ and $D_{12}$
lead respectively to $\theta^{\nu}_{12} = \pi/5$ 
(or $\sin^2 \theta^{\nu}_{12} = (3 - \tilde{r})/4 \cong 0.345$)
and $\theta^{\nu}_{12} = \pi/6$ 
(or $\sin^2 \theta^{\nu}_{12} = 1/4$).
The angles $\pi/5$ and $\pi/6$ are the external angles 
of the decagon and dodecagon. 

   For all the five underlying symmetry forms of $U^\circ_{\nu}$  
listed above we have i) $\theta^{\nu}_{13} = 0$, 
which should be corrected 
to the measured value of $\theta_{13} \cong 0.15$, 
and ii) $\sin^2\theta^{\nu}_{23} = 0.5$, 
which might also need
to be corrected if it is firmly established that 
$\sin^2\theta_{23}$ deviates significantly from 0.5.
In the case of the BM form $\sin^2\theta_{12} = 0.5$, 
which is ruled out by the existing data 
and should be corrected. Finally, the value of  
$\sin^2\theta^{\nu}_{12}$ for the HG form lies outside the current 
$2\sigma$ allowed range of $\sin^2\theta_{12}$ and 
might need also to be corrected.

  The requisite corrections to the discussed 
underlying symmetry forms of the PMNS matrix 
can be generated in each specific case 
by ``decreasing'' the 
residual symmetry $G_\nu$ which leads to 
a given symmetry form.
In the case of $G_f = S_4$ ($G_f = A_4$), 
as we have seen, this can be achieved by ``decreasing'' 
$G_\nu$ from  $Z_2\times Z_2$ ($Z_2$ + the ``accidental''
 $\mu-\tau$ (i.e., $Z_2$)) symmetry to  $Z_2$,  
leading to additional ``correcting'' matrix factor 
in $U^\circ_{\nu}$. 

  As we have mentioned earlier, 
the corrections can also be provided by the 
matrix $U_e$. This approach was followed in 
\cite{Petcov:2014laa,Girardi:2015vha,Girardi:2015rwa,Girardi:2016zwz,Girardi:2014faa,Marzocca:2013cr} and corresponds to the case of $G_f$ 
i) either broken to $G_e = Z_2$, or 
ii) completely broken, by the charged lepton mass term. 
In this case the PMNS matrix has the following general form
\cite{Frampton:2004ud}:
\begin{equation}
U =  U_e^{\dagger}\, U_{\nu} = 
(\tilde{U}_{e})^\dagger\, \Psi U^\circ_{\nu} \, P^\circ\,.
\label{PMNS22} 
\end{equation}
%
Here $\tilde{U}_e$ is a $3\times 3$ unitary matrix 
and  $\Psi$ is a diagonal phase matrix.
The matrix $\tilde{U}_e$  was chosen in 
\cite{Petcov:2014laa,Girardi:2014faa,Marzocca:2013cr} 
to have the following two forms: 
\begin{eqnarray}
{\bf A0}:~ \tilde{U}_{e} = R^{-1}_{23}(\theta^e_{23})\,R^{-1}_{12}(\theta^e_{12})\,;~~
{\bf B0}:~\tilde{U}_{e}  = R^{-1}_{12}(\theta^e_{12})\,.
\label{Ue23122} 
\end{eqnarray}
%
where $\theta^e_{12}$ and $\theta^e_{23}$ are free real 
angle parameters. 
These two forms appear in a large class 
of theoretical models of flavour and studies, in which the generation 
of charged lepton masses is an integral part 
(see, e.g., \cite{Girardi:2013sza,FlavourG,Gehrlein:2014wda,Marzocca:2011dh}).
The phase matrix  $\Psi$ in cases {\bf A0} and {\bf B0} 
is given by \cite{Petcov:2014laa,Marzocca:2013cr}:
\begin{eqnarray}
{\bf A0}:~\Psi =
{\rm diag} \left(1,\text{e}^{-i \psi}, \text{e}^{-i \omega} \right)\,;~~
{\bf B0}:~\Psi =
{\rm diag} \left(1,\text{e}^{-i \psi}, 1 \right)\,.
\label{PsiAB2}
\end{eqnarray}
%
The phases $\omega$ and/or $\psi$ serve as a source 
for the Dirac CPV phase $\delta$ of the PMNS matrix
and contribute to the Majorana CPV phases of the PMNS matrix 
$\alpha_{21}$ and $\alpha_{31}$ \cite{Petcov:2014laa}.
We recall that the diagonal phase matrix $P^\circ$ in 
eq. (\ref{PMNS22}) is given in eq. (\ref{UnuU0nu}): 
it contains two phases, $\xi_{21}$ and $\xi_{31}$, 
which also contribute to the Majorana phases 
$\alpha_{21}$ and $\alpha_{31}$, respectively. 

\vspace{-0.2cm}
%
\subsection{Predictions for the Dirac CPV Phase}
\label{subsec37}
%
%
\subsubsection{The Cases of TBM, BM, GRA, GRB 
and HG Symmetry Forms Corrected by $U_e$}
\label{subsub371}
%
%

 Consider the case of the five 
underlying symmetry forms of 
$U^\circ_\nu$ - TBM, BM, GRA, GRB and HG -  corrected by 
the matrix $U_e$, with the PMNS matrix given 
in eq. (\ref{PMNS22}) and the matrices 
$\tilde{U}_{e}$ and $\Psi$ as given in 
eqs. (\ref{Ue23122}) and  (\ref{PsiAB2}).
In this  setting 
the Dirac phase $\delta$ of the PMNS matrix 
was shown to satisfy the following 
sum rule \cite{Petcov:2014laa}:
\begin{equation}
\cos\delta =  \frac{\tan\theta_{23}}{\sin2\theta_{12}\sin\theta_{13}}\,
\left [\cos2\theta^{\nu}_{12} + 
\left (\sin^2\theta_{12} - \cos^2\theta^{\nu}_{12} \right )\,
 \left (1 - \cot^2\theta_{23}\,\sin^2\theta_{13}\right )\right ]\,.
\label{cosdthnu2}
\end{equation}
%
Within the approach employed this sum rule is exact 
\footnote{The renormalisation group 
corrections to the sum rule for $\cos\delta$, 
eq. (\ref{cosdthnu2}), in the cases of
neutrino Majorana mass term generated by the Weinberg 
(dimension 5) operator added to i) the Standard Model, 
and ii) the minimal SUSY extension of the Standard Model,
have been investigated in \cite{Gehrlein:2016fms,Zhang:2016djh}. 
They were found in \cite{Gehrlein:2016fms}
to be negligible, e.g., when  the Weinberg 
operator was added to the Standard Model.
}
and is valid for any value of the angle $\theta^{\nu}_{23}$ 
\cite{Girardi:2015vha}
(and not only for  $\theta^{\nu}_{23} = -\,\pi/4$ 
of the five discussed symmetry forms of $U^\circ_\nu$).

 As we see, via the sum rule $\cos\delta$ 
is expressed in terms of the three 
neutrino mixing angles $\theta_{12}$, $\theta_{23}$, $\theta_{13}$ 
and one fixed (known) parameter $\theta^{\nu}_{12}$ 
which depends on the underlying symmetry form 
(TBM, BM, GRA, GRB, HG)
of the PMNS matrix.
The difference between the cases ${\bf A0}$ and ${\bf B0}$  
of forms of $\tilde{U}_{e}$ in eq.~(\ref{Ue23122}) is, in particular, 
in the correlation between 
the values of $\sin^2 \theta_{23}$ and  $\sin^2 \theta_{13}$ 
they lead to. In case ${\bf A0}$ of $\tilde{U}_{e}$,  
the values of $\sin^2 \theta_{23}$ and  $\sin^2 \theta_{13}$ 
are not correlated and $\sin^2 \theta_{23}$ can differ significantly 
from 0.5 \cite{Petcov:2014laa}.
For the form  ${\bf B0}$ of $\tilde{U}_{e}$ we have 
\cite{Petcov:2014laa}:
\begin{equation}
\sin^2 \theta_{23} = 
\frac{1}{2}\,  
\frac{1 - 2\,\sin^2 \theta_{13}}{1 - \sin^2 \theta_{13}} \cong 
 \frac{1}{2}\, (1 - \sin^2 \theta_{13})\,.
\label{eq:th1323B}
\end{equation}
%
Thus, in contrast to the case {\bf A0}, in case {\bf B0} 
the value of $\sin^2 \theta_{23}$ is correlated with the value of 
$\sin^2 \theta_{13}$ and as a consequence  $\sin^2 \theta_{23}$ 
can deviate from 0.5 insignificantly -  only by $0.5\sin^2 \theta_{13}$.

 Qualitatively, the result in eq. (\ref{cosdthnu2}) for $\delta$ 
can be understood as follows.
In the parametrisation defined in eq. (\ref{PMNS22}) 
with $U^\circ_{\nu}$, $\tilde{U}_{e}$ and $\Psi$ given 
in (\ref{Unu12}) and, e.g., by forms ${\bf B0}$ in  
eqs. (\ref{Ue23122}) and (\ref{PsiAB2}), we have:
\begin{align}
&U_{\rm PMNS} = R_{12}(\theta^e_{12}) \, \Psi \, 
R_{23}(\theta^{\nu}_{23}) \, R_{12}(\theta^{\nu}_{12}) \, P^\circ \,.
\label{eq:U12e23nu12nu2} 
\end{align}
%
The phase $\psi$ in the phase matrix $\Psi$ 
serves as a source for the Dirac phase $\delta$ 
(and gives a contribution to the Majorana phases 
$\alpha_{21,31}$ \cite{Petcov:2014laa}).
It follows from eq. (\ref{eq:U12e23nu12nu2}) 
that in the case under discussion, 
the three angles $\theta_{12}$, $\theta_{23}$, $\theta_{13}$  
and the Dirac phase $\delta$ of the standard parametrisation of 
$U_{\rm PMNS}$ are expressed in terms of the three parameters 
$\theta^e_{12}$, $\psi$ and $\theta^{\nu}_{12}$ 
($\theta^{\nu}_{23} = -\,\pi/4$). This suggests that it will be possible 
to express one of the four parameters $\theta_{12}$, $\theta_{23}$, 
$\theta_{13}$ and $\delta$, namely $\delta$, in terms of the other 
three, hence eq. (\ref{cosdthnu2}).
Although the case of $\tilde{U}_{e}$ having the form
${\bf A0}$ in eq. (\ref{Ue23122}) is somewhat more complicated, 
in what concerns $\cos\delta$ one arrives to the same 
conclusion and result \cite{Petcov:2014laa}.

  Given the values of $\sin\theta_{12}$, $\sin\theta_{23}$, $\sin\theta_{13}$
and $\theta^{\nu}_{12}$, $\cos\delta$ is determined uniquely 
by the sum rule (\ref{cosdthnu2}). This allows us to determine also 
$|\sin\delta|$ uniquely. However, in absence of additional 
information, ${\rm sgn}(\sin\delta)$ remains undetermined, which 
leads to a two-fold ambiguity in the determination of the value of 
$\delta$ from the given value of  $\cos\delta$.

    The fact that the value of the Dirac CPV phase
$\delta$ is determined (up to an ambiguity of
the sign of $\sin\delta$) by the values of the
three mixing angles  $\theta_{12}$, $\theta_{23}$
and $\theta_{13}$ of the PMNS matrix 
and the value of $\theta^{\nu}_{12}$ of 
the matrix $U^\circ_\nu$, eq. (\ref{Unu12}), 
is the most striking prediction of the
models considered. This result implies 
that in the schemes under discussion, 
the rephasing invariant $J_{\text{CP}}$
associated with the Dirac phase $\delta$, 
eq. (\ref{JCPstandparam}),
is also a function of the three angles
$\theta_{12}$, $\theta_{23}$ and $\theta_{13}$
of the PMNS matrix and of $\theta^{\nu}_{12}$:
\begin{equation}
J_{\text{CP}} = J_{\text{CP}}(\theta_{12},\theta_{23},\theta_{13},
\delta(\theta_{12},\theta_{23},\theta_{13},\theta^{\nu}_{12})) =
J_{\text{CP}}(\theta_{12},\theta_{23},\theta_{13},\theta^{\nu}_{12})\,.
\label{JCPNO2}
\end{equation}
%
This allows to obtain predictions for the 
values of $J_{\text{CP}}$ for the different symmetry forms of $\tilde{U}_\nu$
(specified by the value of $\theta^{\nu}_{12}$) 
using the current data on  $\theta_{12}$, $\theta_{23}$ and $\theta_{13}$.

 In \cite{Petcov:2014laa}, by using the sum rule in 
eq.~(\ref{cosdthnu2}), predictions for $\cos\delta$, $\delta$ and 
the $J_{\text{CP}}$ factor 
were obtained in the TBM, BM, GRA, GRB and HG cases for the 
b.f.v. of $\sin^2\theta_{12}$, 
$\sin^2\theta_{23}$ and  $\sin^2\theta_{13}$.
It was found that the predictions of $\cos\delta$
vary significantly with the symmetry form of 
$\tilde{U}_{\nu}$. For the b.f.v. 
of $\sin^2\theta_{12}=0.308$, $\sin^2\theta_{13}= 0.0234$ and
$\sin^2\theta_{23}= 0.437$~found for NO spectrum 
in \cite{Capozzi:2013csa}, for instance,
one gets\cite{Petcov:2014laa} $\cos\delta = (-0.0906)$, $(-1.16$),
$0.275$, $(-0.169)$ and $0.445$,
for the TBM, BM (LC), GRA, GRB and HG forms, respectively.
For the TBM, GRA, GRB and HG forms
these values correspond to  $\delta = \pm 95.2^{\circ}, \pm 74.0^{\circ}, 
\pm 99.7^{\circ}, \pm 63.6^{\circ}$.
For the b.f.v. given in Table \ref{tab:parameters}
and obtained in the global analysis 
\cite{Capozzi:2017ipn}
one finds 
in the cases of the  TBM, BM (LC), GRA, GRB and HG forms 
the values given in Table~\ref{tab:dcosd2017}.
\begin{table}[t]
\centering
\caption{Predicted values of $\cos\delta$ and $\delta$ 
for the five symmetry forms, 
 TBM, BM, GRA, GRB and HG, and $\tilde{U}_{e}$ given by 
 the form ${\bf A0}$ in eq. (\ref{Ue23122}),
 obtained using eq. (\ref{cosdthnu2}) 
 and the best fit values of 
 $\sin^2\theta_{12}$, $\sin^2\theta_{23}$ 
 and $\sin^2\theta_{13}$ for NO and IO neutrino 
 mass spectra from 
 ref. \cite{Capozzi:2017ipn}. 
(From refs. \cite{Girardi:2014faa,Agarwalla:2017wct}.)
}
\begin{tabular}{lcccc}
\hline
Scheme & $\cos\delta$ (NO) & $\delta$ (NO) & $\cos \delta$ (IO) & $\delta$ (IO)\\
\hline
TBM & $-0.16$ & $ \pm 99^\circ$ & $-0.27$ & $\pm 106^\circ$\\
BM (LC) & $-1.26$ & $\cos\delta-$unphysical & $-1.78$ &$\cos\delta-$unphysical\\
GRA & $\phantom{-}0.21$ & $\pm 78^\circ$ & $\phantom{-}0.24$ & $\pm 76^\circ $\\
GRB & $-0.24$ & $\pm 105^\circ$ & $-0.38$ &$\phantom{-}\pm 112^\circ$\\
HG & $\phantom{-}0.39$ & $\pm 67^\circ$ & $\phantom{-}0.48$ & $\pm 62^\circ$\\
\hline
\end{tabular}
\label{tab:dcosd2017}
\end{table} 
%
Due to the different NO and IO b.f.v. of $\sin^2\theta_{23}$,
the predicted values of $\cos\delta$ and $\delta$ 
for IO spectrum differ (in certain cases significantly) 
from those for the NO spectrum.   

Three comments are in order. First, according to the 
results found in \cite{Capozzi:2017ipn} and quoted in 
Table~\ref{tab:parameters},
the predicted values of $\delta$ lying in the first quadrant 
are strongly disfavored (if not ruled out) by the 
current data.
Second, the unphysical value of $\cos\delta$
in the  BM (LC) case  is a reflection of the fact that
the scheme under discussion with BM (LC) form of the matrix 
$U^\circ_{\nu}$
does not provide a good description of the current data on
$\theta_{12}$, $\theta_{23}$ and $\theta_{13}$~
\cite{Marzocca:2013cr}.
Physical values of $\cos \delta$ can be obtained 
in the case of the NO spectrum, 
e.g., for the b.f.v. of $\sin^2 \theta_{13}$
if the value of $\sin^2 \theta_{12}$ ($\sin^2 \theta_{23}$)
is larger (smaller) than the current  
best fit value 
\footnote{For, e.g.,  $\sin^2 \theta_{12} = 0.34$
allowed at $2\sigma$ by the current data,
we have $\cos \delta = -0.943$.
Similarly, for  $\sin^2 \theta_{12} = 0.32$,  
$\sin^2 \theta_{23}=0.41$ and 
$\sin\theta_{13}=0.158$ we have \cite{Petcov:2014laa}:
$\cos \delta = -0.978$.
} 
\cite{Petcov:2014laa,Girardi:2014faa}.
However, with the b.f.v. of 
$\sin^2\theta_{23}$ quoted in eq. (\ref{eq:s2th232018}), 
the BM (LC) form is strongly disfavored for both 
NO and IO spectra.

Third, the $A_4$ and $S_4$ models considered 
subsections \ref{subsec33} and \ref{subsec34} lead to 
 largely different predictions for $\sin^2\theta_{12}$ and, 
 if $\theta_{23} \neq \pi/4$, for $\cos\delta$ as well, 
which differ also from the predictions for $\cos\delta$  
we have obtained in the cases of the five different symmetry forms - 
TBM, BM, GRA, GRB, HG - and the matrix $\tilde{U}_{e}$ given by 
the forms ${\bf A0}$ and ${\bf B0}$ in eq. (\ref{Ue23122}).
These predictions can be used to discriminate experimentally 
between the different models. 

 The results quoted above imply~\cite{Petcov:2014laa}
that a measurement of $\cos\delta$ can allow  to
distinguish between at least some of the different 
symmetry forms of
$U^\circ_{\nu}$ provided 
$\theta_{12}$, $\theta_{13}$ and $\theta_{23}$ 
are known, and $\cos\delta$ is measured, 
with  sufficiently high precision 
\footnote{Detailed results on the dependence 
of the predictions for 
$\cos\delta$ on $\sin^2\theta_{12}$, 
 $\sin^2\theta_{23}$ and  $\sin^2\theta_{13}$ 
when the latter are varied in
their respective $3\sigma$ experimentally allowed 
ranges can be found in \cite{Girardi:2014faa}.}.
Even determining the sign of $\cos\delta$
will be sufficient to eliminate some of
the possible symmetry forms of $\tilde{U}_{\nu}$.
 
 These conclusions were confirmed by the statistical analyses 
performed in ref.~\cite{Girardi:2014faa} where
predictions of the sum rule (\ref{cosdthnu2}) for 
i) $\delta$, $\cos\delta$ and the rephasing
invariant $J_{\rm CP}$ using the ``data'' 
(best fit values and $\chi^2-$distributions) 
on $\sin^2\theta_{12}$, $\sin^2\theta_{13}$,
$\sin^2\theta_{23}$ and $\delta$ from 
\cite{Capozzi:2013csa},
and 
ii) for $\cos\delta$, using prospective uncertainties  
on $\sin^2\theta_{12}$, $\sin^2\theta_{13}$ and $\sin^2\theta_{23}$, 
were derived for the TBM, BM (LC), GRA, GRB and HG symmetry
forms of the matrix $\tilde{U}_{\nu}$. 
Both analyses were performed for the case of 
NO neutrino mass spectrum. 
The results for the IO spectrum are similar.
\begin{figure}[!t]
  \begin{center}
     \hspace{-0.9cm}
  \includegraphics[width=13cm]{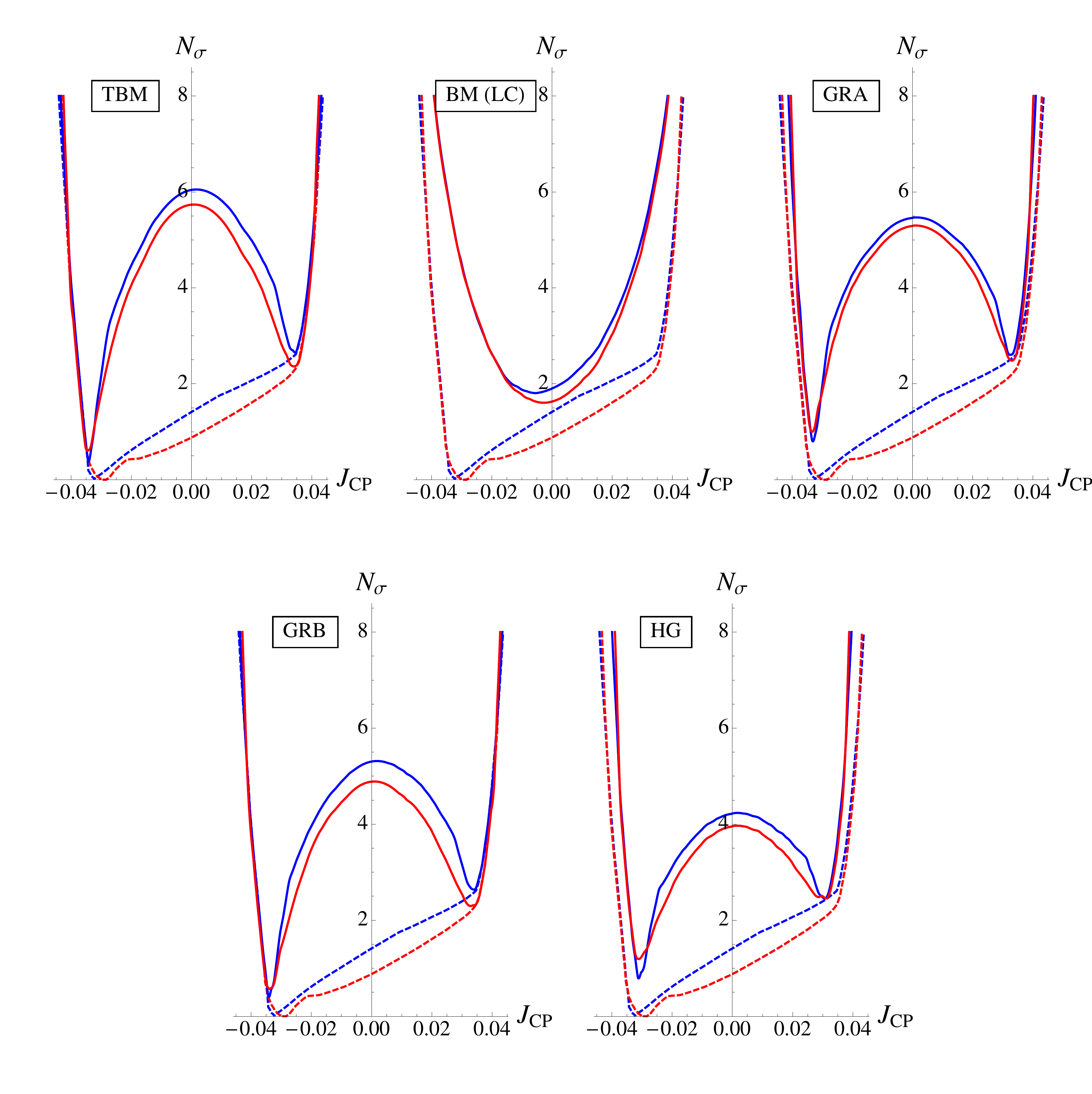}
     \end{center}
      
\vspace{-1.2cm} 
\caption{ 
\label{Fig:chi2JCP}
$N_{\sigma} \equiv \sqrt{\chi^2}$ as a function of $J_{\rm CP}$.
The dashed lines represent the results of the global fit
\cite{Capozzi:2013csa},
 while the solid lines represent the
results we obtain for the TBM, BM (LC), GRA
(upper left, central, right panels),
GRB and HG (lower left and right panels)
neutrino mixing symmetry forms.
The blue (red) lines are for NO (IO)
neutrino mass spectrum.
(From ref.~\protect\cite{Girardi:2014faa}.)
}
\end{figure}
%
\noindent 
  The aim of the first analysis, 
the results of which 
for $J_{\rm CP}$ are shown in Fig. \ref{Fig:chi2JCP} and 
are summarised in Table \ref{tab:Fig12}, 
was to derive the allowed ranges
for $\delta$ and $J_{\rm CP}$,
predicted on the basis of the current data on
the neutrino mixing parameters for each of the
symmetry forms of $U^\circ_{\nu}$ considered 
(see \cite{Girardi:2014faa} for details of the analysis).
It was found in \cite{Girardi:2014faa}, in particular, that
the CP-conserving value of $J_{\rm CP} = 0$ is
excluded  in the cases of the  TBM, GRA, GRB
and HG neutrino mixing symmetry forms, respectively,
at approximately $5\sigma$, $4\sigma$, $4\sigma$
and $3\sigma$ C.L. with respect to the C.L.
of the corresponding 
best fit values which all lie in the interval 
$J_{\rm CP} = (-0.034) - (-0.031)$ 
(see Table  \ref{tab:Fig12}). 
The best fit value for the BM (LC) form 
is much smaller and close to zero:
$J_{\rm CP} = (-5\times 10^{-3})$.
For the TBM, GRA, GRB and HG forms
at $3\sigma$ we have $0.020 \leq |J_{\rm CP}| \leq 0.039$.
Thus, for these four forms the CP violating 
effects in neutrino oscillations 
are predicted to be relatively large 
and observable in the T2HK and DUNE 
experiments \cite{DUNE2016,T2HK2015}.
\begin{table}[t]
\centering
\caption{ Best fit values  of $J_{\rm CP}$
and $\cos \delta$ and corresponding 3$\sigma$ ranges
(found fixing $\chi^2-\chi^2_{\rm min} = 9$)
for the five symmetry forms, 
TBM, BM, GRA, GRB and HG, and $\tilde{U}_{e}$ given by 
the form ${\bf A0}$ in eq. (\ref{Ue23122})
obtained using the data from \protect\cite{Capozzi:2013csa}
for NO neutrino mass spectrum.
(From ref. \protect\cite{Girardi:2014faa}, where results for 
IO spectrum are also given.)
}
\begin{tabular}{lcccc}
\toprule
Scheme & $J_{\rm CP}/10^{-2}$ (b.f.v.) & $J_{\rm CP}/10^{-2}$ ($3\sigma$ range) & $\cos \delta$ (b.f.v.) & $\cos \delta$ ($3\sigma$ range)\\
\midrule
TBM & $-3.4$ & $[-3.8,-2.8] \cup [3.1,3.6]$ & $-0.07$ & $[-0.47,\phantom{-}0.21]$\\
BM (LC) & $-0.5$ & $[-2.6,2.1]$ & $-0.99$ & $[-1.00,-0.72]$\\
GRA & $-3.3$ & $[-3.7,-2.7] \cup [3.0,3.5]$ & $\phantom{-}0.25$ & $[-0.08,\phantom{-}0.69]$\\
GRB & $-3.4$ & $[-3.9,-2.6] \cup [3.1,3.6]$ & $-0.15$ & $[-0.57,\phantom{-}0.13]$\\
HG & $-3.1$ & $[-3.5,-2.0] \cup [2.6,3.4]$ & $\phantom{-}0.47$ & $[\phantom{-}0.16,\phantom{-}0.80]$\\
\bottomrule
\end{tabular}
\label{tab:Fig12}
\end{table} 
%
\noindent 
These conclusions hold if one uses in the analysis 
the results on the neutrino mixing parameters and $\delta$, 
obtained in the more recent global analyses 
\cite{Capozzi:2017ipn,NuFITv32Jan2018,Capozzi:2018ubv}.

\begin{figure}[!t]
  \begin{center}
     \hspace{-0.9cm}
   \subfigure
  {\includegraphics[width=8cm]{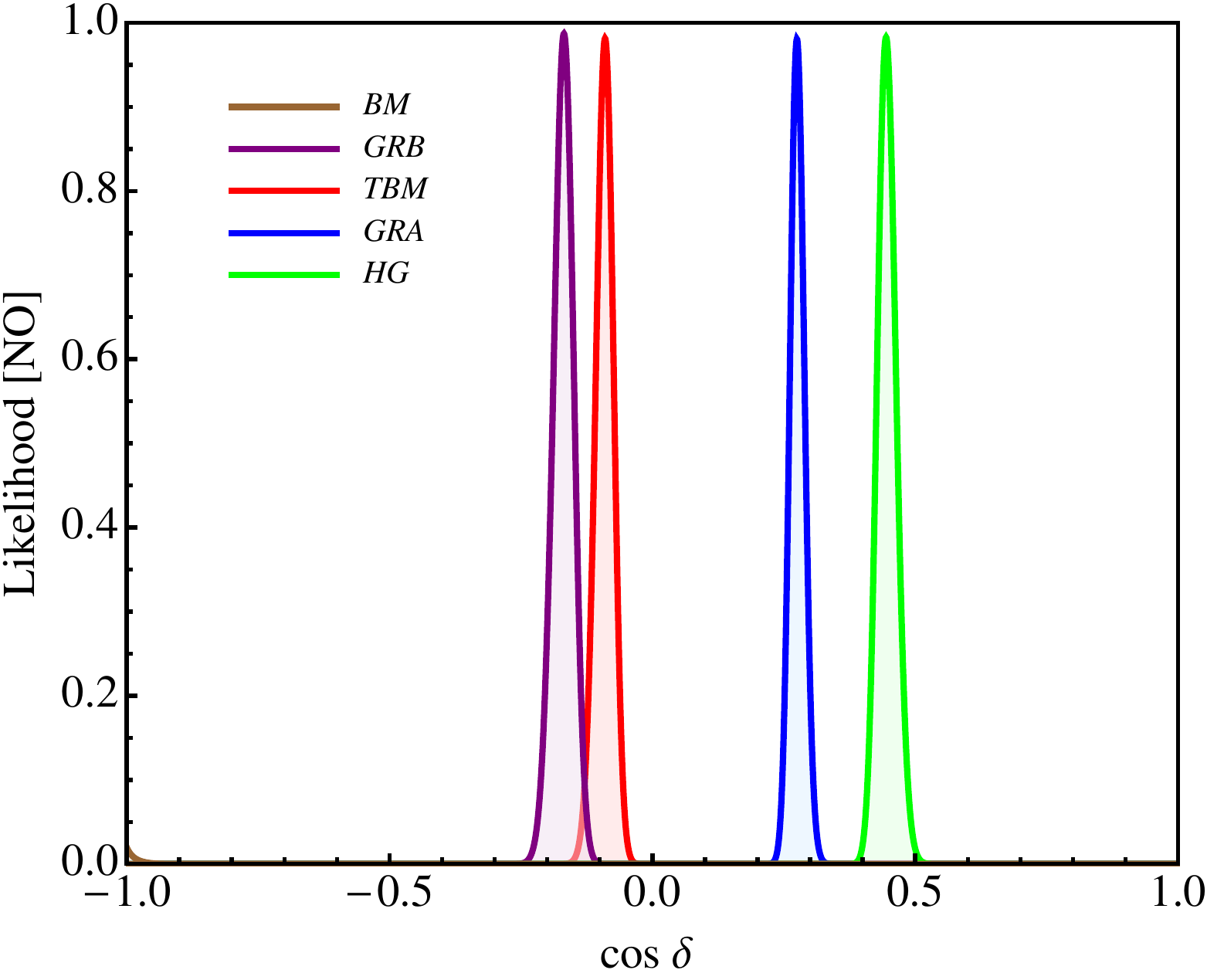}}
  {\includegraphics[width=8cm]{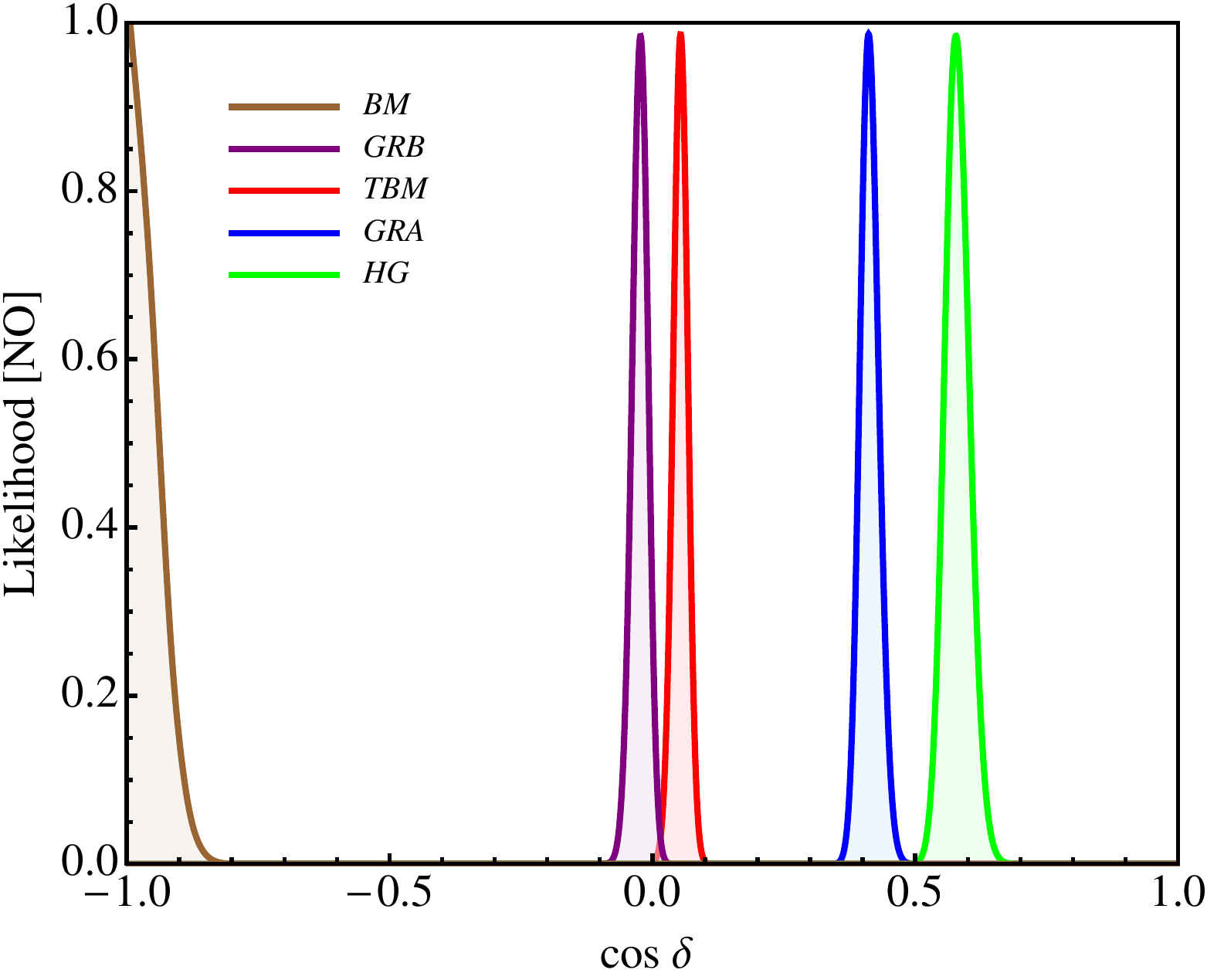}}
  \vspace{5mm}
     \end{center}
      
\vspace{-1.0cm} \caption{ 
\label{Fig:23}
The likelihood function versus $\cos\delta$ for NO
neutrino mass spectrum after marginalising over
$\sin^2\theta_{13}$ and $\sin^2\theta_{23}$,
for the TBM, BM (LC), GRA, GRB and HG symmetry forms
of the mixing matrix $U^\circ_{\nu}$. The figure is 
obtained by using the prospective $1\sigma$ uncertainties in the 
determination of $\sin^2\theta_{12}$, $\sin^2\theta_{13}$ and 
$\sin^2\theta_{23}$ within the Gaussian approximation.
In the left (right) panel $\sin^2 \theta_{12}$
is set to its b.f.v. of \protect\cite{Capozzi:2013csa}
$0.308$ (is set to $0.332$), 
the other mixing angles being fixed 
to their NO best fit values taken from 
\protect\cite{Capozzi:2013csa}.
See text for further details.
(From ref.~\protect\cite{Girardi:2014faa}.)
}
\end{figure}

 In Fig.~\ref{Fig:23} (left panel) we present the 
results of the statistical analysis of the 
predictions for $\cos\delta$,
namely the likelihood function
versus $\cos \delta$ within the Gaussian approximation
(see \cite{Girardi:2014faa} for details)
performed using the  b.f.v.
of the mixing angles for NO neutrino mass spectrum given in 
ref.~\cite{Capozzi:2013csa} and the prospective rather small 
$1\sigma$ uncertainties 
i) of  0.7\% on $\sin^2 \theta_{12}$, planned 
to be  reached in JUNO experiment \cite{JUNO},
ii) of 3\% on $\sin^2 \theta_{13}$, foreseen  
to be obtained in the  Daya Bay experiment 
\cite{Zhang:2015fya}, 
and iii) of 5\% on $\sin^2 \theta_{23}$, expected 
to be reached in the currently running 
and future planned long baseline neutrino 
oscillation experiments.
In the proposed upgrading of the currently taking 
data T2K experiment \cite{T2K20152016}, for example, 
$\theta_{23}$ is estimated to be determined with a 
$1\sigma$ error of $1.7^\circ$, $0.5^\circ$ and $0.7^\circ$ 
if the best fit value of $\sin^2\theta_{23} = 0.50$, $0.43$ and $0.60$, 
respectively. This implies that for these three values of 
 $\sin^2\theta_{23}$ the absolute (relative) $1\sigma$ error 
would be 0.0297 (5.94\%), 0.0086 (2\%) and 0.0120 (2\%).
This error on $\sin^2\theta_{23}$ is expected to  
be further reduced in the future planned T2HK \cite{T2HK2015} 
and DUNE \cite{DUNE2016} experiments.
%

 As we have already remarked,
 the BM (LC) case is very sensitive to the b.f.v. of $\sin^2 \theta_{12}$
and $\sin^2 \theta_{23}$ and is disfavored at more than $2\sigma$ for
the b.f.v. found in \cite{Capozzi:2013csa} for the NO spectrum. 
This case might turn out to be compatible with the data 
for larger (smaller) measured values of 
$\sin^2 \theta_{12}$ ($\sin^2 \theta_{23}$).
This is illustrated in Fig. \ref{Fig:23} (right panel).

 The measurement of $\sin^2 \theta_{12}$, $\sin^2 \theta_{13}$
and $\sin^2 \theta_{23}$ with the quoted precision will open up
the possibility to distinguish between the BM (LC), TBM/GRB, 
GRA and HG forms of $\tilde U_{\nu}$. Distinguishing between the 
TBM and GRB forms seems to require unrealistically high precision 
measurement of $\cos \delta$~
\footnote{Self-consistent models or theories of (lepton) flavour
which lead to the GRB form of $U^\circ_{\nu}$ might still be possible 
to distinguish from those leading to the TBM form using the specific 
predictions of the two types of models for the neutrino mixing angles.
The same observation applies to models 
which lead to the GRA and HG forms of $U^\circ_{\nu}$.}. 
Assuming that $|\cos \delta| < 0.93$, 
which means for 76\% of values of $\delta$, the error on 
$\delta$, $\Delta \delta$, for an error on $\cos \delta$, 
$\Delta(\cos \delta) = 0.10 \, (0.08)$, does not exceed 
$\Delta \delta \lesssim \Delta(\cos \delta) /\sqrt{1 - 0.93^2} 
= 16^{\circ} \, (12^{\circ})$.
This accuracy is planned to be reached in the future neutrino experiments
like T2HK, T2HKK (ESS$\nu$SB) \cite{T2HK2015,Abe:2016ero,Baussan:2013zcy}. 
Therefore a measurement of $\cos \delta$ in the quoted range 
with $\Delta(\cos \delta) = 0.08$
will allow one to distinguish between the TBM/GRB, BM (LC) and
GRA/HG forms at approximately $3\sigma$ C.L.,
if the precision achieved on $\sin^2 \theta_{12}$,
$\sin^2 \theta_{13}$ and $\sin^2 \theta_{23}$ is the same
as in Fig.~\ref{Fig:23}. 

  A more detailed study of the 
possibility to distinguish between  BM (LC), 
TBM, GRB, GRA and HG forms of $U^\circ_{\nu}$ using 
the prospective data from DUNE and T2HK experiments 
was performed in \cite{Agarwalla:2017wct}. 
\begin{figure}[!t]
  \begin{center}
     \hspace{-0.9cm}
  \includegraphics[width=15cm]{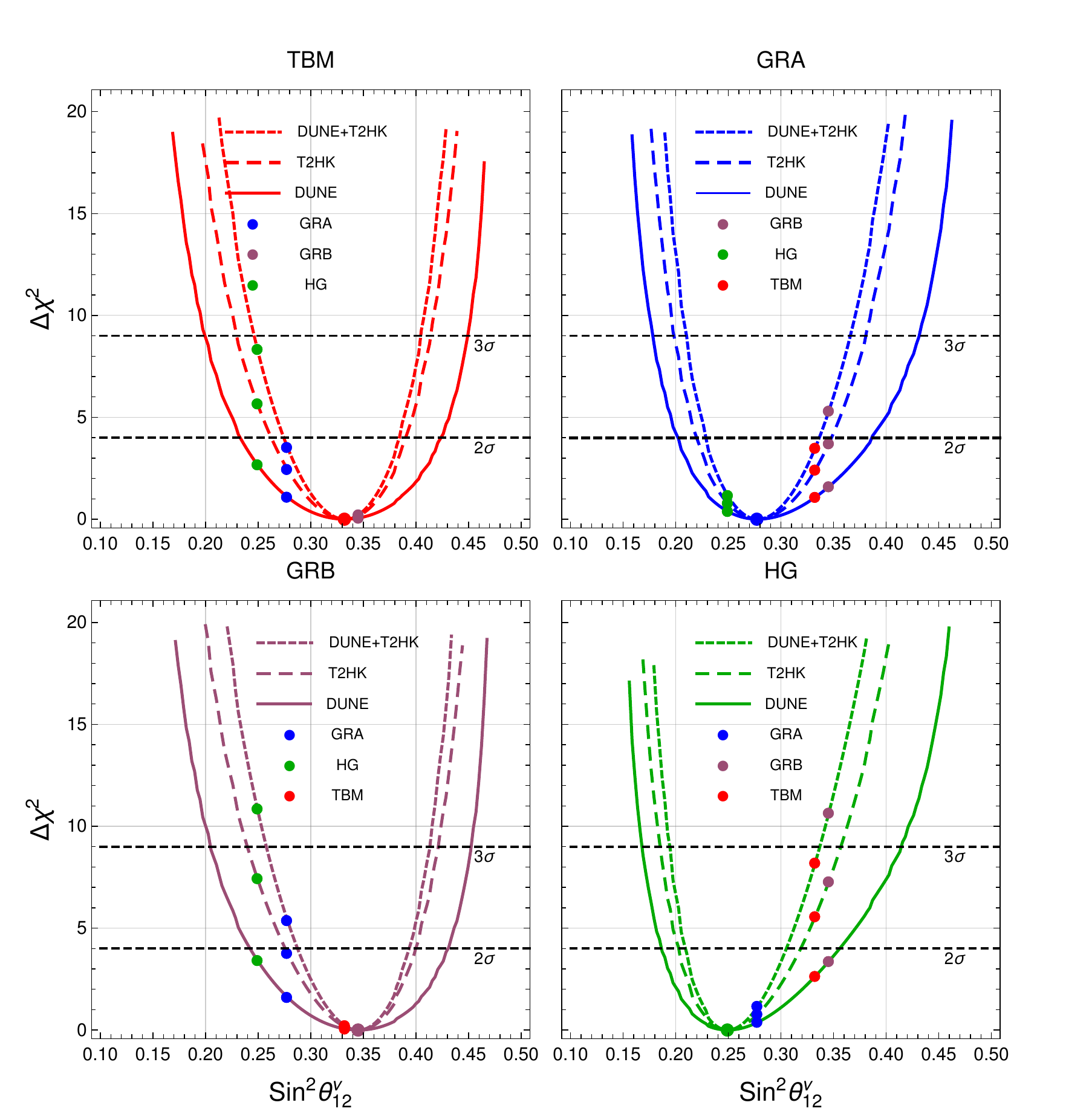}
     \end{center}
      
\vspace{-0.5cm} 
\caption{ 
\label{Fig:chis2thnu}
Sensitivities of the experiments DUNE, T2HK and 
their combined (prospective) data to the symmetry 
form parameter $\sin^2\theta^\nu_{12}$ allowing 
to distinguish between the TBM, GRA, GRB, and HG symmetry 
forms under the assumption that one of them is realised in Nature. 
In the top left and right panels the assumed 
true symmetry forms are respectively 
TBM ($\sin^2\theta^\nu_{12} = 1/3$)
and GRA ($\sin^2\theta^\nu_{12} = 0.276$), 
while in the bottom left and right
panel these forms are  GRB ($\sin^2\theta^\nu_{12} = 0.345$)  and 
HG ($\sin^2\theta^\nu_{12} = 0.25$). 
See text for further details.
(From ref.~\protect\cite{Agarwalla:2017wct}.)
}
\end{figure}
%
\noindent 
Some of the results of this study are illustrated in 
Fig. \ref{Fig:chis2thnu}. As is shown in \cite{Agarwalla:2017wct} 
and is indicated by Fig. \ref{Fig:chis2thnu},
the combined analysis of the data from the 
DUNE and T2HK experiments would allow 
to distinguish between TBM and HG (GRA)
symmetry forms of the PMNS matrix 
at approximately $3\sigma$ ($2\sigma$) confidence level; 
and the same data would allow to distinguish
between  GRB and HG (GRA) forms 
at more than $3\sigma$ (at approximately $2\sigma$) confidence level.
Using the data from the T2HK, T2HKK and DUNE experiments 
is expected to lead to a better discrimination between the different 
symmetry forms of $U_{\rm PMNS}$ owing to the better prospective sensitivity 
to $\delta$ of the combined data from the T2HK and T2HKK experiments.

In what concerns the BM (LC) form, 
as we have already discussed, it is not compatible with 
the best fit values of the neutrino mixing angles
(leading to $|\cos\delta| > 1$), but is viable 
if the $2\sigma$ ranges of the neutrino mixing angles 
are taken into account. If, e.g., one keeps $\sin^2\theta_{13}$ 
and $\sin^2\theta_{23}$ fixed at their best fit values for NO spectrum, 
one finds $\cos\delta = -\,1$, and thus a viable BM (LC) 
mixing form, for $\sin^2\theta_{12} =  0.334$,
which is the upper limit of the 
allowed $2\sigma$ range of $\sin^2\theta_{12}$  
(see Table \ref{tab:parameters}). 
For the indicated choice of values of 
$\sin^2\theta_{13}$, $\sin^2\theta_{23}$ and 
$\sin^2\theta_{12}$ the BM form, 
as was shown in \cite{Agarwalla:2017wct}, 
can be distinguished from the other 
four symmetry forms -- 
TBM, GRB, GRA and HG --
at more than $5\sigma$ using only 
the data from the DUNE experiment. 

\vspace{-0.3cm}
%
\subsubsection{Alternative Cases and the Power of Data}
\label{subsubsect372}
%

 In \cite{Girardi:2015vha} the 
analyses performed in 
\cite{Petcov:2014laa,Girardi:2014faa} was extended by obtaining  
sum rules for $\cos\delta$ for $U_{\rm PMNS}$ having 
the general form given in eq. (\ref{PMNS22}) and 
the following forms of 
$\tilde{U}_e$ and $U^\circ_\nu$~ 
\footnote{In \cite{Girardi:2015vha}
a systematic analysis of the forms of 
$\tilde{U}_e$ and  $U^\circ_\nu$, 
for which sum rules for $\cos\delta$ of the 
type of eq.~(\ref{cosdthnu2}) could be derived,
but did not exist in the literature, was performed.}: 
\begin{itemize}
\item[{\bf C0}.] $U^\circ_\nu = R_{23}(\theta^{\nu}_{23})R_{12}(\theta^{\nu}_{12})$ 
with  $\theta^{\nu}_{23} = -\pi/4$ and $\theta^{\nu}_{12}$
as dictated by TBM, BM, GRA, GRB or HG mixing, and 
i)  $\tilde{U}_e =  R^{-1}_{13}(\theta^{e}_{13})$  
($\Psi = {\rm diag}(1,1,e^{-i \omega})$),
ii) $\tilde{U}_e =  R^{-1}_{23}(\theta^{e}_{23})R^{-1}_{13}(\theta^{e}_{13})$
($\Psi =
{\rm diag}(1,e^{-i \psi},e^{-i \omega})$),
and iii) $\tilde{U}_e = R^{-1}_{13}(\theta^{e}_{13})R^{-1}_{12}(\theta^{e}_{12})$ 
($\Psi={\rm diag}(1,e^{-i \psi},e^{-i \omega})$); 
\vspace{-0.2cm}
\item[{\bf D0}.] $U^\circ_\nu = R_{23}(\theta^{\nu}_{23})R_{13}(\theta^{\nu}_{13}) 
 R_{12}(\theta^{\nu}_{12})$ with $\theta^{\nu}_{23}$, 
$\theta^{\nu}_{13}$ and $\theta^{\nu}_{12}$ 
fixed by arguments associated with symmetries, and 
iv) $\tilde{U}_e = R^{-1}_{12}(\theta^{e}_{12})$
($\Psi ={\rm diag}(1,e^{-i \psi}, 1)$),
and v) $\tilde{U}_e = R^{-1}_{13}(\theta^{e}_{13})$
($\Psi = {\rm diag}(1,1,e^{-i \omega})$).
\end{itemize}

\vspace{-0.3cm}
The sum rules for $\cos\delta$ were derived 
first for $\theta^{\nu}_{23} = -\,\pi/4$
for the cases listed in point {\bf C0}, 
and for the specific values of (some of) the angles 
in  $U^\circ_{\nu}$, characterising 
the cases listed in point {\bf D0}, as well as for 
arbitrary fixed values of all angles 
contained in $U^\circ_{\nu}$.
In certain models with $\sin^2\theta^\nu_{13}\neq 0$,
$\sin^2\theta_{23}$ is predicted to have specific values 
which differ significantly from those in case {\bf B0} 
\cite{Girardi:2015vha}:  
$\sin^2\theta_{23} = 0.455$; or 0.463; or 0.537; or 0.545, 
the uncertainties in these predictions being insignificant.

  Predictions for correlations between neutrino mixing angle 
values and/or sum rules for $\cos\delta$, which can be tested 
experimentally, were further derived in \cite{Girardi:2015rwa} 
for a large number of models based on $G_f = S_4$, $A_4$, $T^\prime$ and $A_5$ 
and all symmetry breaking patterns, i.e., all possible combinations 
of residual symmetries, which could lead to 
the correlations and sum rules of interest:\\
({\bf A}) $G_e = Z_2$ and $G_{\nu} = Z_n$, $n > 2$ 
or $Z_n \times Z_m$, $n,m \geq 2$;\\
({\bf B}) $G_e = Z_n$, $n > 2$ or $G_e = Z_n \times Z_m$, 
$n,m \geq 2$ and $G_{\nu} = Z_2$;\\
({\bf C}) $G_e = Z_2$ and $G_{\nu} = Z_2$;\\
({\bf D}) $G_e$ is fully broken and $G_{\nu} = Z_n$, $n > 2$ or 
$Z_n \times Z_m$, $n,m \geq 2$;\\
({\bf E}) $G_e = Z_n$, $n > 2$ or $Z_n \times Z_m$, $n,m \geq 2$ and 
$G_{\nu}$ is fully broken.

 The three LH neutrino fields and the three LH charged lepton fields 
were assumed in \cite{Girardi:2015rwa} 
to transform under the action of $G_f$  
by a 3-dimensional irreducible representation of $G_f$.  
In this case, as we have already remarked, 
the results obtained for  $A_4$ and $T^\prime$ coincide.
For each pattern, sum rules, i.e., 
relations between the neutrino mixing angles and/or between 
the neutrino mixing angles and the Dirac CPV phase $\delta$, 
when present, were derived.
We note that neutrino mixing sum rules can exist also 
in the case of pattern {\bf D} ({\bf E}) if due to additional
assumptions (e.g., additional symmetries)
the otherwise unconstrained unitary
matrix $U_e$ ($U_\nu$) is
constrained to have the specific form of
a matrix of $U(2)$ transformation in a plane or
of the product of two $U(2)$ transformations
in two different planes
\cite{Petcov:2014laa,Girardi:2015vha,Girardi:2015rwa,Marzocca:2013cr,Girardi:2014faa}.
Thus, the cases of patterns {\bf D} and {\bf E}
leading to interesting phenomenological predictions
are ``non-minimal'' from the point
of view of the symmetries employed
(see, e.g., 
\cite{FlavourG,Chen:2013wba, Girardi:2013sza,Gehrlein:2014wda}), 
compared to patterns {\bf A}, {\bf B} and {\bf C} characterised by 
non-trivial residual symmetries present in both 
charged lepton and neutrino sectors, which originate
from just one non-Abelian flavour symmetry.

 As was shown in \cite{Girardi:2015rwa},  
in the case of pattern {\bf A},  
$U^\circ_\nu$ is fixed by $G_\nu$. There are 
three different general 
sub-cases, A1,  A2 and  A3~, 
corresponding to $U_e$ determined up to 
a unitary rotation in the 
1-2, 1-3 and 2-3 plane, respectively.
In sub-cases A1 and A2 one obtains a correlation 
between $\sin^2\theta_{23}$ and 
$\sin^2\theta_{13}$ and a sum rule for  $\cos\delta$, 
while in sub-case A3, $\sin^2\th_{13}$ and $\sin^2\th_{12}$  are predicted 
and $\delta$ is not constrained: 
\be
\sin^2 \theta_{23} = 1 -
\frac{\cos^2 \theta^{\circ}_{13} \cos^2 \theta^{\circ}_{23}}
{1-\sin^2\th_{13}}\,,~~~{\rm A1}\,,
\label{eq:ss23A1}
\ee
\be
\cos \delta = \frac{\cos^2 \theta_{13} (\sin^2 \theta^{\circ}_{23} - \cos^2 \theta_{12}) + \cos^2 \theta^{\circ}_{13} \cos^2 \theta^{\circ}_{23} (\cos^2 \theta_{12} - \sin^2 \theta_{12} \sin^2 \theta_{13})}{\sin 2 \theta_{12} \sin \theta_{13} |\cos \theta^{\circ}_{13} \cos \theta^{\circ}_{23}| (\cos^2 \theta_{13} - \cos^2 \theta^{\circ}_{13} \cos^2 \theta^{\circ}_{23})^{\frac{1}{2}}}\,,
~{\rm A1}\,,
\label{eq:cosdeltaA1}
\ee
%
\be
\sin^2 \theta_{23} = 
\frac{\sin^2 \theta^{\circ}_{23}}{1 - \sin^2 \theta_{13}}\,,~~~{\rm A2}\,, 
\label{eq:ss23A2}
\ee
%
\be
\cos \delta = -\frac{\cos^2 \theta_{13} (\cos^2 \theta^{\circ}_{12} \cos^2 \theta^{\circ}_{23} - \cos^2 \theta_{12}) + \sin^2 \theta^{\circ}_{23} (\cos^2 \theta_{12} - \sin^2 \theta_{12} \sin^2 \theta_{13})}{\sin 2 \theta_{12} \sin \theta_{13} |\sin \theta^{\circ}_{23}| (\cos^2 \theta_{13} - \sin^2 \theta^{\circ}_{23})^{\frac{1}{2}}}\,,~{\rm A2}\,,
\label{eq:cosdeltaA2}
\ee
%
\be
\sin^2\th_{13} = \sin^2\th^\circ_{13}\,,
~\sin^2\th_{12} = \sin^2\th^\circ_{12}\,,~
\cos\delta - {\rm unconstrained}\,,~~{\rm A3}\,,
\label{eq:th13th12cosdA3}
\ee
%
where the angles $\th^\circ_{13}$, $\th^\circ_{23}$ and 
$\theta^{\circ}_{12}$ are fixed 
once the flavour symmetry group $G_f$ and 
the residual symmetry subgroups $G_e$ and $G_\nu$ are specified.

 In the case of pattern {\bf B}, 
of which there are also of three different sub-cases, 
B1, B2 and B3, corresponding to $U_e$ fixed by 
$G_e$ and  $U^\circ_\nu$ determined up to a unitary rotation 
in the 1-3, 2-3 and 1-2 plane, respectively,
there exist a correlation between $\sin^2\theta_{12}$ and 
$\sin^2\theta_{13}$ and a sum rule for  $\cos\delta$ (sub-cases B1, B2), 
or $\sin^2\th_{13}$ and $\sin^2\th_{23}$ are predicted 
while  $\delta$ remains unconstrained  (sub-case B3): 
\be
\sin^2 \theta_{12} = 
\frac{\sin^2 \theta^{\circ}_{12}}{1 - \sin^2 \theta_{13}}\,,~~~{\rm B1}\,,
\label{eq:ss12B1}
\ee
\be
\cos \delta = -\frac{\cos^2 \theta_{13} (\cos^2 \theta^{\circ}_{12} \cos^2 \theta^{\circ}_{23} - \cos^2 \theta_{23}) + \sin^2 \theta^{\circ}_{12} (\cos^2 \theta_{23} - \sin^2 \theta_{13} \sin^2 \theta_{23})}
{\sin 2 \theta_{23} \sin \theta_{13} |\sin \theta^{\circ}_{12}| (\cos^2 \theta_{13} - \sin^2 \theta^{\circ}_{12})^{\frac{1}{2}}}\,,~{\rm B1}\,,
\label{eq:cosdeltaB1}
\ee
%
\be
\sin^2 \theta_{12} = 1 - 
\frac{\cos^2\theta^{\circ}_{12} \cos^2\theta^{\circ}_{13}  }{1 - \sin^2\theta_{13}}\,,~~~{\rm B2}\,,
\label{eq:ss12B2}
\ee
\be
\cos \delta = 
\frac{\cos^2 \theta_{13} (\sin^2 \theta^{\circ}_{12} - \cos^2 \theta_{23}) 
+ \cos^2 \theta^{\circ}_{12} \cos^2 \theta^{\circ}_{13} 
( \cos^2 \theta_{23} - \sin^2 \theta_{13} \sin^2 \theta_{23} )}
{\sin 2 \theta_{23} \sin \theta_{13} 
| \cos \theta^{\circ}_{12} \cos \theta^{\circ}_{13}| 
(\cos^2 \theta_{13} - \cos^2 \theta^{\circ}_{12} 
\cos^2 \theta^{\circ}_{13} )^{\frac{1}{2}}}\,,~{\rm B2}\,,
\label{eq:cosdeltaB2}
\ee
%
\be
\sin^2\th_{13} = \sin^2\th^\circ_{13}\,,
~\sin^2\th_{23} = \sin^2\th^\circ_{23}\,,~
\cos\delta - {\rm unconstrained}\,,~~{\rm B3}\,,
\label{eq:th13th23cosdB3}
\ee
%
where, as in the case of pattern {\bf A},  
$\th^\circ_{12}$,  $\th^\circ_{23}$ and $\th^\circ_{13}$ 
are fixed once the symmetries are specified.

 Finally, in the case of pattern {\bf C}, 
of which there are altogether nine sub-cases, 
corresponding to $U_e$ and $U^\circ_\nu$, 
each determined by $G_e$ and $G_\nu$ up 
to a unitary rotations in the 
$i$-$j$ and $k$-$l$ planes, respectively, 
$i$-$j$=1-2,1-3,2-3,  $k$-$l$= 1-2,1-3,2-3,  
there is either a correlation between 
$\sin^2\th_{13}$ and $\sin^2\th_{12}$, or between 
$\sin^2\th_{13}$ and $\sin^2\th_{23}$, or 
else a sum rule for  $\cos\delta$.
We number them as in \cite{Girardi:2015rwa}, i.e., cases C1--C9.
Four of them lead to sum rules for 
$\cos\delta$, which have the form:
\begin{align}
&\text{C1, $(ij,kl) = (12,13)$:} \quad
\cos \delta = 
\dfrac{\sin^2 \theta^{\circ}_{23} - \cos^2 \theta_{12} \sin^2 \theta_{23} - \cos^2 \theta_{23} \sin^2 \theta_{12} \sin^2 \theta_{13}}
{\sin \theta_{13} \sin 2 \theta_{23} \sin \theta_{12} \cos \theta_{12}}\,, 
\label{eq:cosdeltaC1}\\[0.2cm]
&\text{C3, $(ij,kl) = (12,23)$:} \quad
\cos \delta = 
\dfrac{\sin^2 \theta_{12} \sin^2 \theta_{23} - \sin^2 \theta^{\circ}_{13} + \cos^2 \theta_{12} \cos^2 \theta_{23} \sin^2 \theta_{13}}
{\sin \theta_{13} \sin 2 \theta_{23} \sin \theta_{12} \cos \theta_{12}}\,, \\[0.2cm]
&\text{C4, $(ij,kl) = (13,23)$:} \quad 
\cos \delta = 
\dfrac{\sin^2 \theta^{\circ}_{12} - \cos^2 \theta_{23} \sin^2 \theta_{12} - \cos^2 \theta_{12} \sin^2 \theta_{13} \sin^2 \theta_{23}}
{\sin \theta_{13} \sin 2 \theta_{23} \sin \theta_{12} \cos \theta_{12}}\,, \\[0.2cm]
&\text{C8, $(ij,kl) = (13,13)$:} \quad 
\cos \delta = 
\dfrac{\cos^2 \theta_{12} \cos^2 \theta_{23} - \cos^2 \theta^{\circ}_{23} + 
\sin^2 \theta_{12} \sin^2 \theta_{23} \sin^2 \theta_{13}}
{\sin \theta_{13} \sin 2 \theta_{23} \sin \theta_{12} \cos \theta_{12}}\,.
\label{eq:cosdeltaC8}
\end{align}
%
The neutrino mixing angles in these cases should 
be treated as free parameters. 
Other two cases, C5 and C9, yield correlations between 
$\sin^2\th_{12}$ and $\sin^2\th_{13}$:
\begin{align}
&\text{C5, $(ij,kl) = (23,13)$:} \quad 
\sin^2 \theta_{12} = \frac{\sin^2 \theta^{\circ}_{12}}
{1 - \sin^2 \theta_{13}}\,,~\cos\delta - {\rm unconstrained}\,, \\[0.2cm]
&\text{C9, $(ij,kl) = (23,23)$:} \quad
\sin^2 \theta_{12} = \frac{\sin^2 \theta^\circ_{12} - \sin^2 \theta_{13}}
{1 - \sin^2 \theta_{13}}\,,~\cos\delta - {\rm unconstrained}\,.
\label{eq:ss12C9}
\end{align}
%
In cases C2 and C7, instead, there are correlations between
$\sin^2\th_{23}$ and $\sin^2\th_{13}$:
\begin{align}
&\text{C2, $(ij,kl) = (13,12)$:} \quad 
\sin^2 \theta_{23} = \frac{\sin^2 \theta^{\circ}_{23}}
{1 - \sin^2 \theta_{13}}\,,~\cos\delta - {\rm unconstrained}\,, 
\label{eq:ss23C2}\\[0.2cm]
&\text{C7, $(ij,kl) = (12,12)$:} \quad 
\sin^2 \theta_{23} = \frac{\sin^2 \theta^\circ_{23} - \sin^2 \theta_{13}}
{1 - \sin^2 \theta_{13}}\,,~\cos\delta - {\rm unconstrained}\,.
\label{eq:ss23C7}
\end{align}
%
Finally, in case C6, $(ij,kl) = (23,12)$, $\cos\delta$  
is unconstrained and $\sin^2\th_{13}$ is 
predicted to be equal to $\sin^2\th^\circ_{13}$. 
In cases C2, C5, C6, C7 and C9, as is indicated above,  
$\cos\delta$ remains unconstrained.

  Given the fact that the group  $A_4$ has 
eight Abelian subgroups (three $Z_2$, four $Z_3$ and one 
Klein group $K_4$ isomorphic to $Z_2\times Z_2$), 
the group  $S_4$ possesses  20 Abelian subgroups 
(nine $Z_2$, four $Z_3$, three $Z_4$ and four $Z_2\times Z_2$ groups, 
see, e.g., \cite{Tanimoto:2015nfa}),  
and $A_5$ has 36 Abelian subgroups (fifteen $Z_2$, ten $Z_3$, 
five $Z_2\times Z_2$ and six $Z_5$, see, e.g., \cite{Ding:2011cm})
the total number of the different 
residual symmetry patterns {\bf A}, {\bf B} and {\bf C} 
to be analysed is extremely large. 
For the group $A_4$ ($T^\prime$) alone there 
are altogether 64 cases (up to permutations of rows and columns of 
the predicted neutrino mixing matrix). 
It is quite remarkable that of these extremely 
large number of cases only a very limited number 
turned out to be phenomenologically 
viable, i.e., to be compatible with 
the existing data on the neutrino mixing angles 
\cite{Girardi:2015rwa,Petcov:2018snn}.
In the case of the group $G_f = A_4~(T^\prime)$, for example, 
only one case was found to be phenomenologically viable 
\cite{Girardi:2015rwa,Petcov:2018snn}, i.e., to be 
compatible with the experimentally determined 
values (including the $3\sigma$ uncertainties) 
of the three neutrino mixing parameters 
$\sin^2 \theta_{12}$, $\sin^2 \theta_{13}$ 
and $\sin^2 \theta_{23}$. 
Namely, this is case B1 with $(G_e,G_\nu) = (Z_3,Z_2)$, which yields 
$(\sin^2\th^\circ_{12},\sin^2\th^\circ_{23}) = (1/3,1/2)$ and corresponds to 
the TBM mixing matrix corrected from the right by the 
$U_{13}(\theta^\nu_{13},\alpha)$
transformation in the 1-3 plane (see sub-section 3.3).
The case B1 is common also to the two other groups $S_4$ and 
$A_5$. For $G_f = S_4$, there are 6 more viable cases. 
The $A_5$ flavour symmetry leads to 7 additional  
phenomenologically viable cases.

 One arrives at this results for the number 
of phenomenologically 
viable cases in the following way. 
For the groups $S_4$ and $A_5$ there are respectively 
altogether 8 and 13 cases,  
which are acceptable {\it a priori}, 
i.e., which lead to $U_{\rm PMNS}$ without zero entries.
They are summarised in Table~3 (for $S_4$) 
and Table~4 (for $A_5$) of Ref.~\cite{Petcov:2018snn}.
In Tables~3 and 4 in~\cite{Petcov:2018snn} 
the specific values of $\sin^2\theta^\circ_{ij}$ in 
each case are also given.
However, the case B1, as we have already noticed, 
is common to all the three flavour symmetry 
groups $A_4~(T^\prime)$, $S_4$ and $A_5$, 
while four cases, C1, C3, C4 and C8,  
are shared by $S_4$ and $A_5$.
Thus, there are 16 cases in total, which lead to different 
predictions for $\sin^2\th_{12}$ or 
$\sin^2\th_{23}$ and/or $\cos\delta$. 
A statistical analysis of these 
predictions showed \cite{Petcov:2018snn}
that two cases, namely, C4 
(for both $S_4$ and $A_5$) and B2A$_5$II 
(i.e., the second of the two B2 cases with 
$G_f = A_5$, characterised by different fixed  
values of $\theta^\circ_{12}$ and $\theta^\circ_{13}$) 
are globally disfavored at more than 
$3\sigma$ confidence level by the current data  
(including the uncertainties)  
on  $\sin^2\th_{12}$, $\sin^2\th_{23}$ and  
$\sin^2\th_{13}$ \cite{NuFITv32Jan2018}. 
As a consequence, only 14 cases altogether turned out 
to be phenomenologically viable at present.
Five of them, B1, B1$A_5$, B2$S_4$, B2$A_5$, C9$A_5$
\footnote{The notation X$G_f$ means case X, 
X=A1,A2,...,B1,...,C1,...,C9, corresponding to the group 
$G_f$, $G_f=A_4,S_4,A_5$. The group is not indicated 
in cases B1, C1, C3 and C8 (see below) because case B1 is common to 
the $A_4$, $S_4$ and $A_5$ groups, while each of the cases C1, C3 and C8
is shared by the $S_4$ and $A_5$ groups.},
lead to sharp predictions for $\sin^2\theta_{12}$, 
and four others, A1$A_5$, A2$A_5$, C2$S_4$, C7$S_4$, 
to similarly sharp predictions for $\sin^2\theta_{23}$.
The six phenomenologically viable cases ${\bf A}$ and  ${\bf B}$ 
lead also to predictions for $\cos\delta$, 
while five out of the eight viable cases ${\bf C}$,  
C1, C3, C3$A_5$ (which differs from C3 that is common to $S_4$ and $A_5$), 
C4$A_5$ (which differs from C4) and C8, also lead to 
predictions for $\cos\delta$. 

  Statistical analysis of these 14 cases was performed  
in \cite{Petcov:2018snn} using the best fit values 
of the three neutrino mixing
parameters $\sin^2\theta_{ij}$ from \cite{NuFITv32Jan2018}
and taking into account the prospective ($1\sigma$) 
uncertainties in the determination of the mixing angles, 
planned to be achieved in currently running 
(Daya Bay \cite{Zhang:2015fya}) and the next generation 
(JUNO \cite{JUNO}, T2HK \cite{T2HK2015}, DUNE \cite{DUNE2016}) of neutrino 
oscillation experiments: 3\% on $\sin^2\theta_{13}$ \cite{Zhang:2015fya},
0.7\% on  $\sin^2\theta_{12}$  \cite{JUNO} and  3\% on $\sin^2\theta_{23}$ 
\cite{DUNE2016,T2HK2015}. 
This analysis revealed that only six cases would be 
compatible with the indicated prospective data 
from the Daya Bay, JUNO, T2HK, DUNE neutrino 
oscillation experiments.

\begin{figure}[t]
\centering
\includegraphics[width=14.0cm]{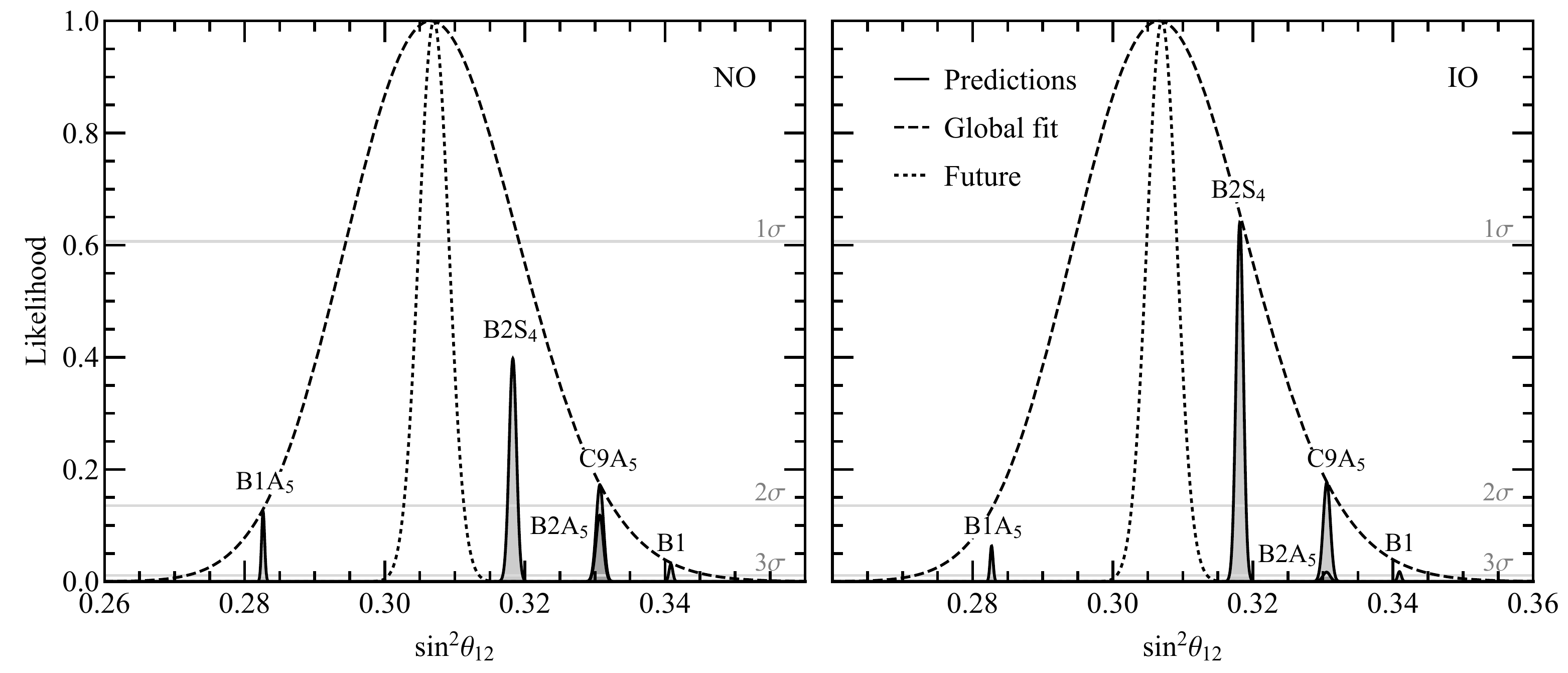}
\includegraphics[width=14.0cm]{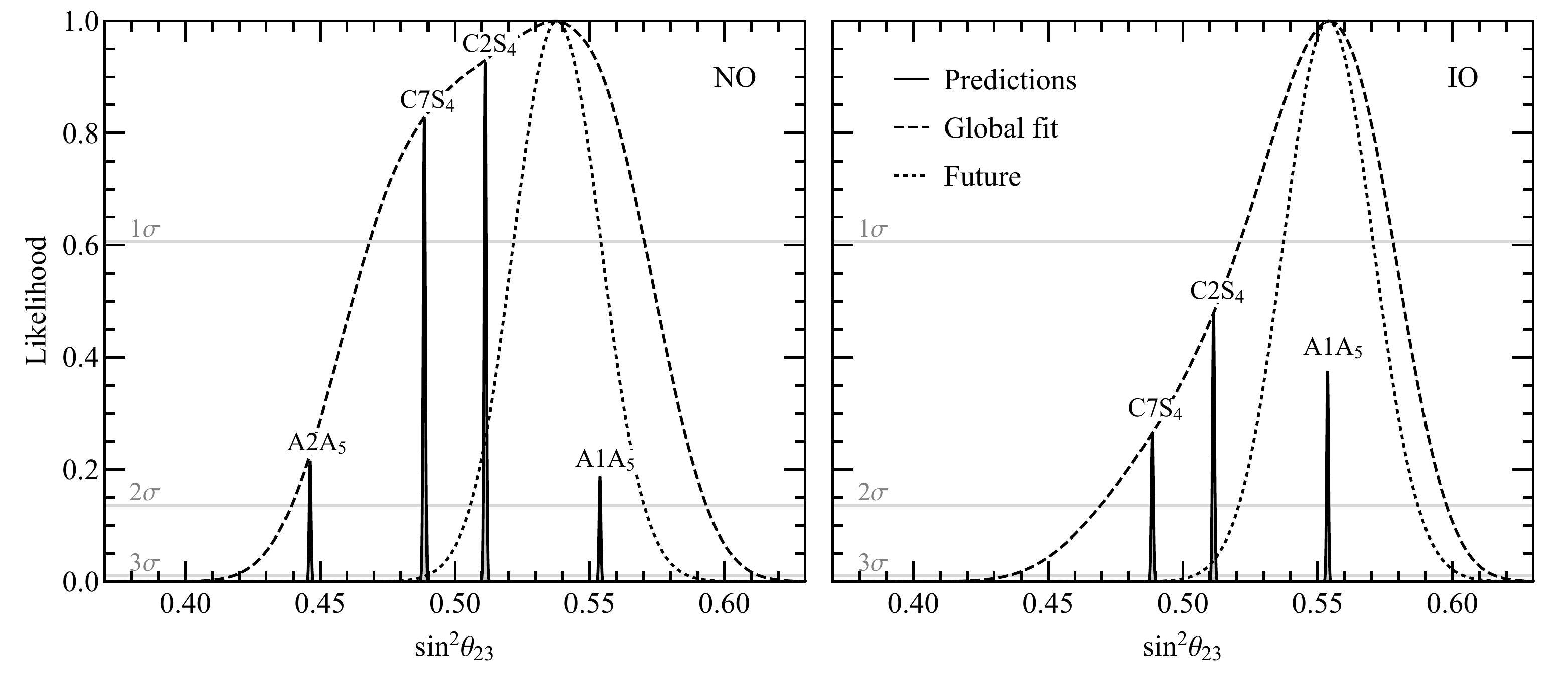}
\caption{Upper [lower] panel: 
predictions for $\sin^2\th_{12}$ [$\sin^2\th_{23}$] 
obtained using the current global data on the neutrino mixing parameters. 
``Future'' refers to 
the scenario with $\sin^2\th_{12}^\mathrm{bf} = 0.307$ 
[$\sin^2\th_{23}^\mathrm{bf} = 0.538~(0.554)$ for NO (IO)]
(current best fit values) and the relative $1\sigma$ uncertainty 
of $0.7\%$ [$3\%$] expected from JUNO [DUNE and T2HK]. 
See text for further details. 
(From Ref.~\cite{Petcov:2018snn}.)}
\label{fig:ss12ss23present}
\end{figure}

 In Fig.~\ref{fig:ss12ss23present}, we present 
the likelihood functions for $\sin^2\th_{12}$ and $\sin^2\th_{23}$, 
obtained for NO and IO spectra
in all the cases compatible at $3\sigma$ with 
the current global data~\cite{NuFITv32Jan2018}. 
The corresponding likelihood profiles are very narrow  
because their widths are determined by the small experimental 
uncertainty on $\sin^2\th_{13}$. 
In the upper (lower) panel, the dashed line corresponds to 
the likelihood for $\sin^2\th_{12}$ ($\sin^2\th_{23}$)  
extracted from the global analysis. 
%
\noindent
The dotted line represents 
the prospective precision on $\sin^2\th_{12}$ ($\sin^2\th_{23}$) 
corresponding to $1\sigma$ uncertainty of $0.7\%$ ($3\%$), 
which is planned to be achieved by JUNO \cite{JUNO} 
(DUNE \cite{DUNE2016} 
and 
T2HK \cite{T2HK2015}). 
It is obtained under the 
assumption that the best fit value(s) 
of $\sin^2\th_{12}$ ($\sin^2\th_{23}$) will not change in the future. 
If it is indeed the case, then, as is clear 
from Fig.~\ref{fig:ss12ss23present}, all five models 
leading to the predictions for $\sin^2\th_{12}$ 
will be ruled out by the JUNO measurement of this parameter.

\begin{figure}[t]
\centering
\includegraphics[width=14.0cm]{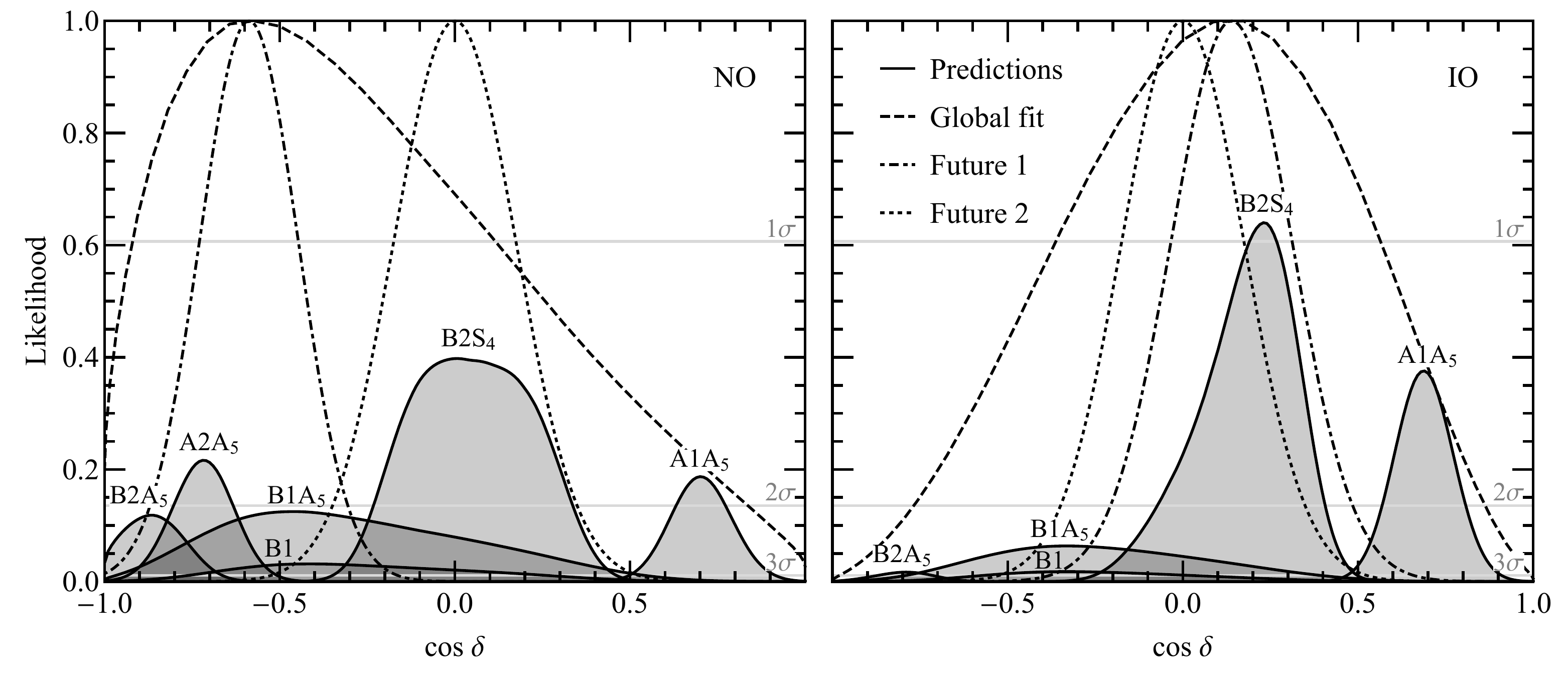}
\includegraphics[width=14.0cm]{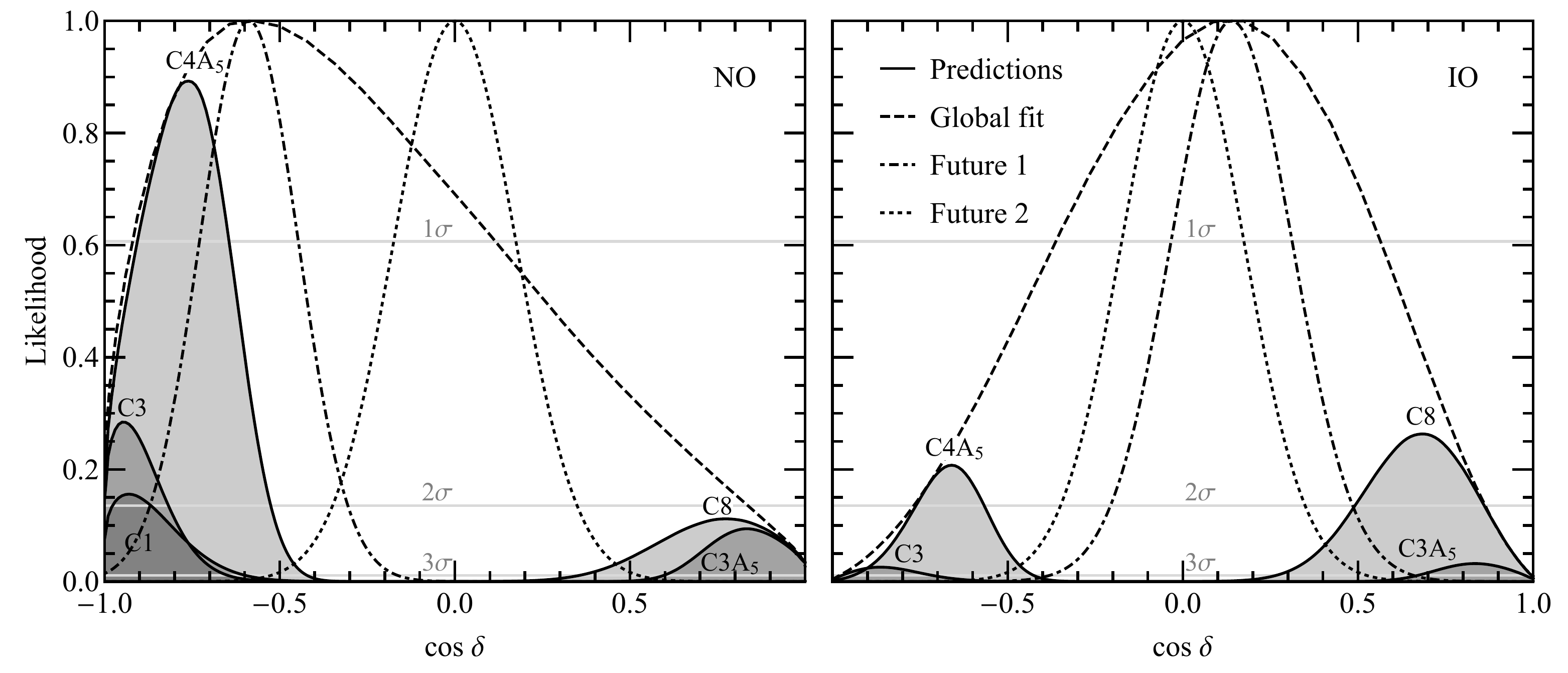}
\caption{Predictions for $\cos\delta$ obtained using 
the current global data on the neutrino mixing parameters. 
``Future 1'' refers to the scenario with
$\delta^{\rm bf} = 234^\circ~(278^\circ)$ 
for NO (IO) spectrum (current best fit values) 
and the $1\sigma$ uncertainty on $\delta$ of $10^\circ$. 
``Future 2'' corresponds to $\delta^{\rm bf} = 270^\circ$ and 
the $1\sigma$ uncertainty on $\delta$ of $10^\circ$. 
See text for further details. (From Ref.~\cite{Petcov:2018snn}.)
}
\label{fig:cosdeltaABCpresent}
\end{figure}

 The results of statistical analysis of the predictions 
for $\cos\delta$ are summarised 
in Fig.~\ref{fig:cosdeltaABCpresent}.
The dashed line stands for the likelihood 
extracted from the global analysis~\cite{NuFITv32Jan2018}. 
At present, all (almost all) values of $\cos\delta$ are 
allowed at $3\sigma$ for NO (IO) spectrum.
We also show the dash-dotted and dotted lines which 
represent two benchmark cases. 
The first case, marked as ``Future 1'', corresponds to  
the current best fit NO (IO) value~\cite{NuFITv32Jan2018}
$\delta^\mathrm{bf} = 234^\circ~(278^\circ)$ 
and the prospective $1\sigma$ uncertainty 
on $\delta$ of $10^\circ$.
The second case, ``Future 2'', corresponds to 
the potential best fit value 
$\delta^\mathrm{bf} = 270^\circ$ 
(for both NO and IO cases) 
and the same $10^\circ$ error on $\delta$. 
The likelihoods in cases C peak at values of 
$|\cos\delta|\sim 0.5 - 1$. 
Looking at the dotted line, we see that if in the 
future the best fit value of $\delta$ 
shifted to $270^\circ$ and the next generation of long-baseline experiments 
managed to achieve the $1\sigma$ uncertainty on $\delta$ 
of $10^\circ$, all cases C viable at the moment would be disfavored at 
approximately $3\sigma$ C.L. only by the measurement of $\delta$.
%

 The results of the studies \cite{Girardi:2015rwa,Petcov:2018snn}
summarised in the present subsection lead to the important 
conclusion that although the number of cases of 
non-Abelian discrete symmetry groups and their subgroups 
that can be used for description of lepton mixing is extremely 
large, only a very limited number 
survive when confronted with the existing data 
on the three neutrino mixing angles. 
This limited number of presently phenomenologically  
viable cases will be further considerably 
reduced by the precision measurements of the three 
neutrino mixing angles and the Dirac phase $\delta$ in the
currently running (Daya Bay) and future planned 
(JUNO, T2HK, T2HKK, DUNE) neutrino oscillation experiments.
As was shown in \cite{Petcov:2018snn} and 
we have briefly discussed, 
if the best fit values of $\sin^2\theta_{12}$,  
$\sin^2\theta_{23}$ and $\sin^2\theta_{13}$ 
as found in \cite{NuFITv32Jan2018} 
would not change significantly in the future,
only six cases would be compatible with 
the prospective data from the Daya Bay, JUNO, 
T2HK and DUNE neutrino oscillation experiments. 
This number would be further reduced by a precision 
measurement of the Dirac phase $\delta$.
%
%
\section{Flavour Symmetry Combined with Generalised CP Symmetry}
\label{sec4}
%
%

 In all models discussed by us the 
Majorana phases $\alpha_{21}$ and $\alpha_{31}$ 
remain undetermined. The values of the Majorana CPV 
phases are instead constrained to lie in 
certain narrow intervals, or are predicted,  
in theories which in addition to a flavour symmetry 
possess at a certain high-energy scale  
a ``generalised CP'' (GCP) symmetry \cite{Branco:1986gr}.
The GCP symmetry, as the term suggests, is a generalisation 
of the traditional (canonical) CP symmetry. 
The GCP symmetry should be implemented in a theory 
based on a discrete flavour symmetry in a way 
that  is consistent with the flavour symmetry 
\cite{Feruglio:2012cw,Holthausen:2012dk}.
At low energies the GCP symmetry is broken, in general, 
to residual CP symmetries of the charged lepton 
and neutrino sectors. 

  The GCP transformations are 
applied on  the LH and RH components of the 
charged lepton fields   $\tilde{l}_L(x)$ and  $\tilde{l}_R(x)$ and  
on the LH neutrino fields $\nu_{\tilde{l}L}(x)$ - the fields 
in terms of which the general charged lepton and neutrino Majorana mass 
terms are formed, eqs. (\ref{Le}) and (\ref{nuMajMass}).
The transformations of interest are defines as follows: 
\begin{align}
\tilde{l}_L(x) &\xrightarrow{CP} i (X_L)_{\tilde{l}\tilde{l}'} 
\gamma_0 C\, \overline{\tilde{l}'_L(x')}^T\,,~
\label{eq:CPlL} 
\\[0.2cm]
\tilde{l}_R(x) &\xrightarrow{CP} i (X_R)_{\tilde{l}\tilde{l}'} 
\gamma_0 C\, \overline{\tilde{l}'_R(x')}^T\,,~ 
\label{eq:CPlR} 
\\[0.2cm]
\nu_{\tilde{l}L}(x) &\xrightarrow{CP} i (X_L)_{\tilde{l}\tilde{l}'} \gamma_0 
 C\, \overline{\nu_{ \tilde{l}' L}(x')}^T\,, 
\label{eq:CPnuL}
\end{align}
%
where $\tilde{l} = \tilde{e},\tilde{\mu},\tilde{\tau}$,
$X_L$ and $X_R$ are $3\times 3$ unitary matrices and 
$x'= (t,-\mathbf{x})$. 
The transformations of $\tilde{l}_L(x)$ and 
$\nu_{\tilde{l}L}(x)$ should involve the same matrix $X_L$ 
in order to ensure the CP invariance of the CC weak 
interaction Lagrangian, expressed in terms of the 
SM  $SU(2)_L$ lepton doublet fields 
$\tilde{l}_L(x)$ and $\nu_{\tilde{l}L}(x)$: 
\be
\mathcal{L_{\rm CC}} = 
-\,\frac{g}{\sqrt{2}}\, \sum_{\tilde{l} = \tilde{e},\tilde{\mu},\tilde{\tau}} 
\overline{\tilde{l}_L(x)}\, \gamma_\alpha\, 
\nu_{\tilde{l}L}(x)\, W^{\alpha\dagger}(x) + \mathrm{h.c.}\,.
\label{eq:LCC0}
\ee
%
The GCP symmetry will hold then in the lepton sector if it is 
a symmetry of the charged lepton and neutrino mass terms, 
 eqs. (\ref{Le}) and (\ref{nuMajMass}),
i.e.,  if the charged lepton and neutrino Majorana mass matrices 
satisfy the following constraints: 
\begin{align}
X_L^\dagger\, M_e\, X_R = M_e^*\,,~or\\[0.2cm]
X_L^\dagger\, M_e\,M^\dagger_e X_L = (M_e\,M^\dagger_e)^*\,,\\[0.2cm]
X_L^T\, M_\nu\, X_L = M_\nu^*\,.
\label{eq:GCPMeMnu}
\end{align}
%

 In the presence of flavour symmetry the form of the 
GCP transformations is significantly constrained. 
Indeed,  consider a
GCP transformation on a generic field  $\varphi(x)$ 
which is assigned to an r-dimensional irreducible unitary 
representation  $\rho_r(g)$  of $G_f$:
\be
\varphi(x) \xrightarrow{CP} X_r\, \varphi^*(x')\,, 
\label{eq:GCPTransformation}
\ee
%
where $X_r$ is a unitary matrix.  The action of 
the CP transformation on the spinor indices in the case 
of $\varphi$ being a spinor  (shown explicitly in 
eqs.~\eqref{eq:CPlL}--\eqref{eq:CPnuL}) 
has been omitted here for simplicity. 
If both the flavour symmetry and the GCP symmetry hold, 
the theory under study should be invariant also under the 
following sequence of transformations:
a GCP transformation, followed by a flavour symmetry 
transformation, which in turn is followed by an (inverse) 
GCP transformation, i.e., under
\be
\varphi(x) \xrightarrow{CP} X_r\, \varphi^*(x') \xrightarrow{G_f} 
X_r\, \rho_r(g_f)^*\, \varphi^*(x') \xrightarrow{CP^{-1}} 
X_r\, \rho_r(g_f)^*\, X_r^{-1}\, \varphi(x)\,.
\ee
%
In order for the theory to be invariant under this sequence 
of transformations the resulting transformation should be
a flavour symmetry transformation of $\varphi(x)$
corresponding to an element $g'_f$ of  $G_f$, which can differ 
from $g_f$:
\be
X_r\, \rho_r(g_f)^*\, X_r^{-1} 
= \rho_r(g'_f)\,, \qquad g_f\,,\, g'_f \in G_f\,.
\label{eq:consistcond}
\ee
%
This equation represents a {\it consistency condition} which has to 
be satisfied in order for the implementation of the GCP symmetry 
in the theory to be compatible with the 
presence of a flavour symmetry \cite{Feruglio:2012cw,Holthausen:2012dk}. 
For a discrete flavour symmetry group $G_f$, the consistency 
condition (\ref{eq:consistcond}) will hold if it is satisfied 
by the group's generators.

 Let us denote by $H_\textrm{CP} = \{X_L\}$ the full 
set of 
GCP transformations $X_L$ acting on 
$\nu_{\tilde{l}L}(x)$ and $\tilde{l}_L(x)$, 
which are compatible with a given flavour symmetry group 
$G_f$, i.e., which satisfy the consistency condition 
(\ref{eq:consistcond}) in which $X_r$ is replaced by $X_L$ 
and  $\rho_r(g_f)$ is the irreducible unitary representation of 
$G_f$ to which $\nu_{\tilde{l}L}(x)$ and $\tilde{l}_L(x)$ 
are assigned.
We denote further by $H^\nu_\textrm{CP} = \{X_\nu\}$ and 
$H^\ell_\textrm{CP} = \{X_\ell\}$ the sets of GCP transformations 
$X_\nu$ and $X_\ell$ which are compatible respectively 
with the residual flavour symmetries $G_{\nu}$ 
and $G_{\ell}$ of the neutrino and charged lepton sectors, i.e., 
of the neutrino Majorana and charged lepton mass terms. 
$X_\nu$ and $X_\ell$ satisfy consistency conditions 
in which $\rho_r(g_f)$ ($\rho_r(g'_f)$) is replaced respectively 
by  $\rho_r(g_\nu)$ ( $\rho_r(g'_\nu)$)
and $\rho_r(g_\ell)$ ($\rho_r(g'_\ell)$), 
where $g_\nu,g'_\nu \in G_{\nu}$ and $g_\ell,g'_\ell \in G_{\ell}$.

 The sets $H_\textrm{CP}$, $H^\nu_\textrm{CP}$ and $H^\ell_\textrm{CP}$
are not groups by themselves. They are sets of GCP transformations,
which always involve conjugation of the fields they act upon;
$H^\nu_\textrm{CP}$ and $H^\ell_\textrm{CP}$ are subsets of  $H_\textrm{CP}$.
These sets become groups only if they are extended by (at least) 
an identity element which does not conjugate fields 
(see, for instance, Appendix B of ref. \cite{Feruglio:2012cw}).
When one writes a semi-direct product of $G_f$ and $H_{\rm CP}$, 
$G_f \rtimes H_{\rm CP}$, and the semi-direct 
\footnote{In the case of $G_\nu$ or $G_e$ being a $Z_2$ symmetry, 
the corresponding product becomes direct.}
 products of $G_\nu$ and $H_{\rm CP}^\nu$, $G_\nu \rtimes H_{\rm CP}^\nu$, 
and of $G_\ell$ and $H_{\rm CP}^\ell$, 
$G_\ell \rtimes H_{\rm CP}^\ell$, it is always implicitly assumed 
that $H_{\rm CP}$, $H_{\rm CP}^\nu$ and $H_{\rm CP}^\ell$
are appropriate groups, which are obtained from a single 
generating GCP transformation, as explained in Appendix B of
ref. \cite{Feruglio:2012cw}.
In this case $H^\nu_\textrm{CP}$ and $H^\ell_\textrm{CP}$ 
are subgroups of $H_\textrm{CP}$.

%
\subsection{Implications for the Majorana Phases}
\label{use:subsec41}
%

 As we have discussed in section \ref{subsec31}, 
the unitary matrix $U_{\nu}$ which diagonalises the 
neutrino Majorana mass matrix $M_\nu$ and enters into 
the expression for the PMNS matrix, eq. (\ref{UPMNSUeUnu}),
is related to the unitary matrix $U^\circ_{\nu}$ 
diagonalising $M^\dagger_{\nu}\,M_{\nu}$ and  $\rho_r(g_\nu)$  
in the following way: $U_\nu \equiv  U^\circ_\nu P^\circ$, where 
$P^\circ={\rm diag}(1,e^{i\frac{\xi_{21}}{2}},e^{i\frac{\xi_{31}}{2}})$ 
(see eq. (\ref{UnuU0nu})). The phases $\xi_{21}$ and $\xi_{31}$ 
contribute respectively to the Majorana phases $\alpha_{21}$ and 
$\alpha_{31}$ of the PMNS matrix, eq. (\ref{VP}). 
These phases remain undetermined by the flavour symmetries 
under discussion. We will  consider next the implications of 
a residual GCP symmetry $H_{\rm CP}^\nu \subset H_{\rm CP}$, 
which is preserved in the neutrino sector, 
for the determination of  $\xi_{21}$ and $\xi_{31}$, and thus 
of the Majorana phases  $\alpha_{21}$ and $\alpha_{31}$.

 In the case of a residual GCP symmetry $H_{\rm CP}^\nu$,
the neutrino Majorana mass matrix satisfies the 
condition given in eq. (\ref{eq:GCPMeMnu}),  
$X_L \in H_{\rm CP}^\nu$ being the GCP transformation 
defined in eq. (\ref{eq:CPnuL}).
Using eq.~(\ref{UnuMnu}) we find: 
\begin{equation}
(X_L^{\rm d})^T\, M_\nu^{\rm d}\, X_L^{\rm d} = M_\nu^{\rm d}\,,~~\text{with}~~
M_\nu^{\rm d} = {\rm diag}(m_1,m_2,m_3)\,,~~
X_L^{\rm d} = U_\nu^\dagger \, X_L \, U_\nu^*\,.
\label{eq:XLd}
\end{equation}
%
For $m_1 \neq m_2 \neq m_3$ and 
\footnote{It follows from the neutrino oscillation data 
that $m_1 \neq m_2 \neq m_3$, and that at least 
two of the three neutrino masses,  $m_{2,3}$ ($m_{1,2}$) 
in the case of the NO (IO) spectrum, are non-zero.
However, even if $m_1 = 0$ ($m_3 = 0$) at tree level 
and the zero value is not protected by a symmetry, 
$m_1$ ($m_3$) will get a non-zero contribution at 
least at two loop level \cite{CNLSPST83} 
and in the framework of a self-consistent (renormalisable) 
theory of neutrino mass generation this higher order contribution 
will be finite.
}
${\rm min}(m_j)\neq 0$, $j=1,2,3$,
as it is not difficult to show 
(see, e.g., \cite{Feruglio:2012cw,Girardi:2016zwz,Everett:2015oka}),
the unitary matrix $X_L^{\rm d}$ can have only 
the following form:
\begin{equation}
X_L^{\rm d} =  \diag(\pm1, \pm1, \pm1)\,,
\end{equation}
%
where the signs of the three non-zero entries in 
$X_L^{\rm d}$ are not correlated. 
Further, using that $U_\nu =  U^\circ_\nu P^\circ$, 
we obtain from eq. (\ref{eq:XLd}) \cite{Everett:2015oka}: 
\begin{equation}
(U^\circ_\nu)^\dagger \, X_L \, (U^\circ_\nu)^* = 
 P^\circ \, X_L^{\rm d} \, P^\circ =
{\rm diag} \left (\pm 1,\pm e^{i \xi_{21}}, \pm e^{i \xi_{31}}\right )\,.
\label{eq:xiphases}
\end{equation}
%
Thus, we come to the conclusion that the phases $\xi_{21}$ and 
$\xi_{31}$ will be determined provided 
i) the matrix $U^\circ_\nu$, which diagonalises 
$M^\dagger_{\nu}\,M_{\nu}$ and  $\rho_r(g_\nu)$ 
(see eq. ({\ref{Uonurhognu}) and the related discussion)
is fixed by the residual flavour symmetry $G_\nu$, and 
ii) the GCP transformations 
$X_L \in H_{\rm CP}^\nu$, which are consistent with $G_\nu$, 
are identified.

%
\subsection{Concrete Examples of Symmetries}
\label{use:subsec42}
%

 Now we turn to concrete examples \cite{Girardi:2016zwz}. 
For $G_f = A_4$ we choose to work in 
the Altarelli-Feruglio basis \cite{Altarelli:2005yx}, 
in which the $S$ and $T$ generators have the form given in 
eq. (\ref{S4STU3drep}). Preserving the $S$ generator, i.e., 
choosing $G_{\nu} = Z^S_2 = \{1,S\}$, leads to 
$U^\circ_\nu = V_{\rm TBM}$, provided there is 
an additional accidental  $\mu$\,--\,$\tau$ 
symmetry \cite{Altarelli:2010gt}.
The twelve GCP transformations consistent with 
the $A_4$ flavour symmetry for the triplet representation 
in the chosen basis have been found in \cite{Ding:2013bpa}, 
solving the consistency condition 
\begin{equation}
X_L \, \rho^*(g) \, X^{-1}_L = \rho (g^\prime)\,, 
\quad g, g^\prime \in A_4\,.
\label{eq:consistcond2}
\end{equation}
%
These transformations can be summarised in a compact way as follows:
\begin{equation}
X_L = \rho (g)\,, \quad g \in A_4\,,
\end{equation}
%
i.e., the GCP transformations consistent with the 
$A_4$ flavour symmetry are of the same form as the flavour 
symmetry group transformations \cite{Ding:2013bpa}. 
They are given in Table~1 in \cite{Ding:2013bpa} 
together with the elements $\hat S$ and $\hat T$ 
to which the generators $S$ and $T$ 
of $A_4$ are mapped by the consistency condition 
in eq.~(\ref{eq:consistcond2}).
Further, since in our case the residual flavour symmetry 
$G_\nu = Z^S_2 \times Z_2$, 
where the $Z^S_2$ factor corresponds to the preserved 
$S$ generator, only those $X$ are acceptable, 
for which $\hat S = S$. From Table~1 in \cite{Ding:2013bpa} 
it follows that there are four such GCP transformations, 
namely, $\rho(E)$, $\rho(S)$, $\rho(T^2ST)$ and $\rho(TST^2)$, 
where $E$ is the identity element of the group.
The last two transformations are not symmetric in the chosen basis, 
and, as shown in \cite{Ding:2013bpa}, 
lead to partially degenerate neutrino mass spectrum 
with two equal masses  (see also \cite{Feruglio:2012cw}), 
which is ruled out by the existing neutrino oscillation data.
Thus, we are left with two allowed generalised CP transformations, 
$\rho(E)$ and $\rho(S)$, for which we have:
\begin{align}
& V_{\rm TBM}^{\dagger} \, \rho(E) \, V^*_{\rm TBM} = \rho(E) = 
\diag(1, 1, 1)\,, \\
& V_{\rm TBM}^{\dagger} \, \rho(S) \, V^*_{\rm TBM} = \diag(-1, 1, -1)\,.
\end{align}
%
Finally, according to eq.~(\ref{eq:xiphases}), this implies 
that the phases $\xi_{21}$ and  $\xi_{31}$ can be 
either $0$ or $\pi$. The same conclusion holds for a $T^{\prime}$ 
flavour symmetry, because restricting ourselves 
to the triplet representation
for the LH charged lepton and neutrino
fields, there is no way to distinguish $T^{\prime}$ from $A_4$
\cite{Feruglio:2007uu}.

 In the case of $G_f = S_4$ considered in \cite{Girardi:2016zwz}, the 
authors chose to work with the two generators $S$ and $T$ of $S_4$ 
in the basis given in \cite{Altarelli:2009gn}. In this case the generators 
$S$ and $T$ satisfy the following presentation rules: 
\be 
S^2 = T^4 = (ST)^3 = (TS)^3 = {\bf 1}\,.
\label{S4ST}
\ee
%
Although the presentation rules for $S$ and $T$ 
given above differ from the presentation rules 
when the third generator $U$ for $S_4$ is employed,
eq. (\ref{S4STU}), we will keep the notation $S$ and $T$
for the two generators satisfying the presentation 
rules (\ref{S4ST}) in the discussion which follows.
In the basis chosen in \cite{Altarelli:2009gn} and used 
in \cite{Girardi:2016zwz}, $S$ and $T$  have the following form 
in the two triplet representations of interest:
\be
S = \pm\,
\begin{pmatrix}
0 &-\,\frac{1}{\sqrt{2}} &-\,\frac{1}{\sqrt{2}} \\
-\,\frac{1}{\sqrt{2}} & \frac{1}{2} & -\,\frac{1}{2} \\
-\,\frac{1}{\sqrt{2}} & -\,\frac{1}{2}  & \frac{1}{2}
\end{pmatrix}\,,
\quad
T = 
\pm\,\begin{pmatrix}
-\,1 & 0 & 0 \\
0 & -\,i & 0 \\
0 & 0 & i
\end{pmatrix}\,,
\label{S4ST3drep}
\ee
%

The residual symmetry $G_{\nu} =  Z^S_2 \times Z_2$, 
where the $Z^S_2$ factor corresponds to preserved 
$S$ generator in the chosen basis 
and the second one arises accidentally 
(corresponding to a $\mu$\,--\,$\tau$ symmetry), 
leads to the bimaximal mixing,   
$U^\circ_\nu = V_{\rm BM}$ \cite{Altarelli:2009gn}.
As in the previous example, the GCP transformations 
consistent with the $S_4$ flavour symmetry are 
of the same form as the flavour symmetry group transformations 
\cite{Holthausen:2012dk}.
Solving the consistency condition in eq.~(\ref{eq:consistcond2}),
in \cite{Girardi:2016zwz} ten symmetric GCP transformations 
consistent with the $S_4$ flavour symmetry for the 
triplet representation in the chosen basis were found. 
They are summarised in Table~\ref{tab:GCPS4} together with the 
elements $\hat T$ and $\hat S$ to which 
the consistency condition maps the group generators $T$ and $S$.
\begin{table}[h]
\centering
\begin{tabular}{lcc}
\toprule
 $g$, $X = \rho(g)$  & $T \rightarrow \hat T$ & $S \rightarrow \hat S$ \\
 \midrule
$(ST^2)^2$ 
& $T$ & $S$ \\
$T^3$
& $T^3$ & $T^3ST$ \\
$E$ 
& $T^3$ & $S$ \\ 
$T$ 
& $T^3$ & $TST^3$ \\
$T^2ST^2$ 
& $STS$ & $S$ \\ 
$ST^2S$ 
& $T$ & $T^2ST^2$ \\
$S$ 
& $TST$ & $S$ \\ 
$T^2$ 
& $T^3$ & $T^2ST^2$ \\
$STS$ 
& $ST^2$ & $ST^2ST$ \\ 
$TST$ 
& $T^2S$ & $TST^2S$ \\
\bottomrule
\end{tabular}
\caption{The ten symmetric generalised CP transformations $X = \rho(g)$ 
consistent with the $S_4$ flavour symmetry for the triplet representation 
$\rho$ in the chosen basis \cite{Altarelli:2009gn}, determined by 
the consistency condition in eq.~(\ref{eq:consistcond}). 
The mapping $(T,S) \rightarrow (\hat T, \hat S)$ is realised 
via the consistency condition applied to the group generators 
$T$ and $S$, i.e., $X \rho^*(T) X^{-1} = \rho(\hat T)$ and 
$X \rho^*(S) X^{-1} = \rho(\hat S)$. $E$ denotes the identity 
element of $S_4$.
(From \protect\cite{Girardi:2016zwz}.) 
\label{tab:GCPS4}}
\end{table}
%
 
 From  Table \ref{tab:GCPS4} we see that there are four 
symmetric GCP transformations consistent with 
the preserved $S$ generator, namely,
$\rho(E)$, $\rho(S)$, $\rho(T^2ST^2)$ and $\rho(ST^2ST^2)$. 
Substituting them and $U^\circ_\nu = V_{\rm BM}$ 
in eq.~(\ref{eq:xiphases}), one finds \cite{Girardi:2016zwz}: 
\begin{align}
& V_{\rm BM}^{\dagger} \, \rho(E) \, V^*_{\rm BM}  = \rho(E) = \diag(1,1,1)\,, \\
& V_{\rm BM}^{\dagger} \, \rho(S) \, V^*_{\rm BM}  = \diag(1,-1,1)\,, \\
& V_{\rm BM}^{\dagger} \, \rho(T^2ST^2) \, V^*_{\rm BM}  = \diag(-1,1,1)\,, \\
& V_{\rm BM}^{\dagger} \, \rho(ST^2ST^2) \, V^*_{\rm BM}  = \diag(-1,-1,1)\,.
\end{align}
%
Therefore also in this case the phases $\xi_{21}$ and  $\xi_{31}$
are fixed by the residual GCP symmetry to be either
$0$ or $\pi$. As was shown in \cite{Girardi:2016zwz},
these results for the phases $\xi_{21}$ and  $\xi_{31}$ hold also
for $G_f = A_5$ and $G_\nu = Z_2\times Z_2$, generated 
by $\tilde{S}$ and $\tilde{T}^3\tilde{S}\tilde{T}^2\tilde{S}\tilde{T}^3$ 
and leading to the GRA mixing, $U^\circ_\nu = U_{\rm GRA}$ 
(see section \ref{subsec36}), 
when the flavour symmetry is combined with the GCP symmetry.

  If the matrix $U_e$ originating 
from the charged lepton sector is non-trivial, as like 
in the cases {\bf A} and {\bf B} defined by 
equations (\ref{PMNS22}), (\ref{Ue23122}) and (\ref{PsiAB2}),
the Majorana phases $\alpha_{21}$ and $\alpha_{31}$ 
of the PMNS matrix receive also contributions from the phases 
associated with $U_e$ in the PMNS matrix \cite{Petcov:2014laa}.
In the specific examples of the forms {\bf A} and {\bf B} 
of $U_e$,  the phases $\omega$ and/or $\psi$ 
of the matrix $\Psi$ in eqs. (\ref{PMNS22}) and (\ref{PsiAB2}), 
as we have already remarked, serve as a source for the Dirac CPV 
phase $\delta$ of the PMNS matrix and contribute to the Majorana phases 
$\alpha_{21}$ and $\alpha_{31}$ \cite{Petcov:2014laa}.
In these cases the Majorana phases $\alpha_{21}$ and $\alpha_{31}$ 
are determined by the sums respectively of the phases 
$\xi_{21}$ and  $\xi_{31}$ and of the indicated 
additional contributions due to the phases 
in the matrix $\Psi$. As a consequence, $\alpha_{21}$ and $\alpha_{31}$ 
have non-trivial values which differ from $0$ or $\pi$ 
even when  $\xi_{21}$ and  $\xi_{31}$
are fixed by the employed residual GCP symmetry 
to be either $0$ or $\pi$ \cite{Petcov:2014laa,Girardi:2016zwz}.

As we have indicated earlier, the Majorana phases play
important role, e.g, in $|\Delta L| = 2$
processes like neutrinoless double beta 
($(\beta\beta)_{0\nu}$-)
decay $(A,Z) \rightarrow (A,Z+2) + e^- + e^-$,
$L$ being the total lepton charge,
in which the Majorana nature of
massive neutrinos $\nu_i$ manifests itself 
(see, e.g, ref. \cite{Olive:2016xmw}). 
Determining the values of the Majorana phases 
allows to make predictions for the basic 
$(\beta\beta)_{0\nu}-$ decay parameter --
the effective neutrino Majorana mass 
(see, e.g, refs. 
\cite{STPNuNature2013,Petcov:2014laa,Girardi:2016zwz,Penedo:2017vtf}). 
%
%
\subsection{Examples of Models}
\label{use:subsec43}
%
%

 In the scenarios involving a GCP symmetry, 
which were most widely explored so far 
(see, e.g., \cite{Feruglio:2012cw,Ding:2013hpa,Ding:2013bpa,Li:2015jxa,DiIura:2015kfa,Ballett:2015wia}), a non-Abelian flavour symmetry $G_f$ 
consistently combined with a 
GCP symmetry $H_{\rm CP}$ is broken to residual Abelian symmetries 
$G_e = Z_n$, $n > 2$, or $Z_m \times Z_k$, $m,\,k \geq 2$,  
and $G_\nu = Z_2 \times H^\nu_{\rm CP}$ of the charged lepton and 
neutrino mass terms, respectively~%
\footnote{We note that in refs.~\cite{Ding:2013bpa,Li:2015jxa} 
the residual symmetry $G_e$ of the charged lepton mass term 
is augmented with a remnant CP symmetry $H^\ell_{\rm CP}$ as well.
}. 
The factor $H^\nu_{\rm CP}$ in $G_\nu$ stands 
for a remnant GCP symmetry of the neutrino mass term. 
In such a set-up, $G_e$ fixes completely the 
form of the unitary matrix $U_e$ 
which diagonalises the product $M_e M_e^\dagger$ 
and enters into the expression of the PMNS matrix. 
At the same time, $G_\nu$ fixes the unitary 
matrix $U_\nu$, diagonalising the neutrino Majorana mass 
matrix $M_\nu$ up to a single free real parameter~---~%
a rotation angle $\th^\nu$. Given the fact that the 
PMNS neutrino mixing matrix $U_{\rm PMNS}$ is given by the product 
$U_{\rm PMNS} = U_e^\dagger\,U_\nu$,
all three neutrino mixing angles
are expressed in terms of this rotation angle.
In this class of models one obtains 
specific correlations between the values of the 
three neutrino mixing angles, while 
the leptonic CPV phases are typically 
predicted to be exactly 
$0$ or $\pi$, or else $\pi/2$ or $3\pi/2$. 
For example, in the set-up considered in \cite{Feruglio:2012cw} 
and based on $G_f \rtimes H_{\rm CP} = S_4 \rtimes H_{\rm CP}$ 
broken to $G_e = Z_3^T$ and $G_{\nu} = Z_2^S \times H^\nu_{\rm CP}$ with 
$H^\nu_{\rm CP} = \{U,SU\}$~
\footnote{We recall that $S$, $T$ and $U$ are the generators of $S_4$ 
in the basis for its 3-dimensional representation 
specified in eq. (\ref{S4STU3drep}).}, 
the authors find:
\begin{align}
\sin^2\th_{13} = \frac{2}{3}\sin^2\th^\nu\,,\quad 
&\sin^2\th_{12} = \frac{1}{2 + \cos2\th^\nu} = \frac{1}{3\left(1 - \sin^2\th_{13}\right)}\,,\quad
\sin^2\th_{23} = \frac{1}{2}\,,
\label{Fer2012}\\[2mm]
&|\sin\delta| = 1\,,\quad
\sin\alpha_{21} = \sin\alpha_{31} = 0\,.
\end{align}
%
It follows, in particular, from the results on the neutrino 
oscillation parameters~---~best fit values, $2\sigma$ and $3\sigma$ 
allowed ranges~---~%
obtained in the global fit of neutrino oscillation 
data \cite{Capozzi:2017ipn} and summarised in Table~\ref{tab:parameters},  
as well as in the more recent analyses
\cite{NuFITv32Jan2018,Capozzi:2018ubv},
that the prediction quoted in eq.~(\ref{Fer2012}) 
for $\sin^2\theta_{12}$ lies outside of its currently 
allowed $2\sigma$ range
\footnote{We have used the best fit value of 
$\sin^2\theta_{13}$ to obtain the prediction of 
$\sin^2\theta_{12} = 0.341$ leading to the quoted conclusion.
Using the  $2\sigma$ allowed range for $\sin^2\theta_{13}$ 
leads to a minimal value of $\sin^2\theta_{12} = 0.340$,  
which is still larger than the maximal allowed value of 
$\sin^2\theta_{12}$ at $2\sigma$ C.L., 
but inside its $3\sigma$ allowed range.}.
In what concerns the prediction $\sin^2\theta_{23} = 1/2$,
according to \cite{NuFITv32Jan2018,Capozzi:2018ubv}, 
it lies within the $1\sigma$ ($2\sigma$) allowed range of 
$\sin^2\theta_{23}$ for NO (IO) spectrum.

 Other examples of one (real angle) parameter models 
based on the flavour symmetry groups $A_4$, $S_4$ and $A_5$
combined with GCP symmetry can be found, e.g., in 
refs. \cite{Ding:2013hpa,Ding:2013bpa,Li:2015jxa,DiIura:2015kfa,Ballett:2015wia,Yao:2016zev,Feruglio:2013hia}. Most of them share some of 
the properties of the model discussed above: 
the correlation between  $\sin^2\theta_{12}$ and $\sin^2\theta_{13}$, 
the predictions that $|\sin\delta| = 1$, 
$\sin\alpha_{21} = \sin\alpha_{31} = 0$.
Some of the models predict $\sin\delta = 0$, which is 
disfavored by the results of the global neutrino data analyses 
(see Table \ref{tab:parameters}).
In certain set-ups the Majorana phases $\alpha_{21}$ and $\alpha_{31}$
take non-trivial values while $|\sin\delta| = 1$.

 In a class of models based on the groups $G_f = \Delta(3n^2)$ and  
$G_f = \Delta(6n^2)$ of flavour symmetry combined with GCP symmetry  
the neutrino mixing angles and the CPV phases are functions 
of one real angle and one or two discrete phase parameters, 
which depend on the parameter $n$ which characterises 
the size of the group $G_f$ (see, e.g., 
\cite{Hagedorn:2014wha,Hagedorn:2016lva,Ding:2014ssa,Ding:2014ora,Ding:2015rwa}). 
Due to the presence of the additional discrete valued phases,
the CPV phase $\delta$ can have non-trivial and non-maximal values,
i.e., one can have $|\sin\delta| \neq 1,0$. 
In this class of models, as a rule, there exist 
correlations between the values of $\sin^2\theta_{12}$, 
$\sin^2\theta_{23}$ and of $\sin^2\theta_{13}$ 
in the form of, e.g., the following  
relations \cite{Ding:2014ora,Ding:2015rwa}: 
$3\cos^2\theta_{13}\sin^2\theta_{12} = 1$ and 
$\sin^2\theta_{23} = 0.5 \pm 0.5\tan\theta_{13}\sqrt{2-\tan^2\theta_{13}}$.
The first correlation $3\cos^2\theta_{13}\sin^2\theta_{12} = 1$ 
is typical for the models under discussion, 
while the second one or similar occur in most of them. 
For the best fit value of $\sin^2\theta_{13}$ given 
in Table ~\ref{tab:parameters},
the quoted relations lead to the predictions:
$\sin^2\theta_{12} = 0.340$ and $\sin^2\theta_{23} = 0.396$ or 0.604.
As we have already noticed, the value $\sin^2\theta_{12} = 0.340$ 
is outside the $2\sigma$, but within the $3\sigma$ allowed ranges of 
$\sin^2\theta_{12}$ found in 
\cite{Capozzi:2017ipn,NuFITv32Jan2018,Capozzi:2018ubv}. 
As it follows from the results reported, e.g., 
in \cite{Capozzi:2018ubv}, 
both predicted values of  $\sin^2\theta_{23}$ 
lie outside the $3\sigma$ allowed range of  
$\sin^2\theta_{23}$.
 
 Theoretical models based on the approach to 
neutrino mixing that combines discrete 
symmetries and GCP invariance, 
in which the neutrino mixing angles and the leptonic 
CPV phases are functions of two or three parameters have also 
been considered in the literature 
(see, e.g., 
\cite{Girardi:2013sza,Turner:2015uta,Girardi:2016zwz,Lu:2016jit,Penedo:2017vtf}).
In these models the residual symmetry $G_e$ of the charged lepton mass term 
is typically assumed to be a $Z_2$ symmetry or to be fully broken.
In spite of the larger number of parameters in terms of which 
the neutrino mixing angles and the leptonic CPV phases are 
expressed, the values of the CPV phases are still predicted 
to be correlated with the values of the three neutrino mixing angles.
A set-up with $G_e = Z_2 \times H^e_{\rm CP}$ and 
$G_\nu = Z_2 \times H^\nu_{\rm CP}$ has been 
considered in \cite{Lu:2016jit}. 
The resulting PMNS matrix in such a scheme depends on two 
free real parameters~---~two angles $\th^\nu$ and $\th^e$.
The authors have obtained  several 
phenomenologically viable neutrino mixing patterns 
from $G_f = S_4$ combined with $H_{\rm CP}$, 
broken to all possible residual symmetries of the type 
indicated above. Models allowing for three free parameters 
(two real angles and one phase) have been investigated  in 
\cite{Girardi:2013sza,Turner:2015uta,Girardi:2016zwz,Penedo:2017vtf}.
In, e.g., \cite{Turner:2015uta},
the author has considered $G_f = A_5$ combined with $H_{\rm CP}$, 
which are broken to $G_e = Z_2$ and $G_\nu = Z_2 \times H^\nu_{\rm CP}$.
In this case, the matrix $U_e$ depends on an angle $\th^e$ and 
a phase $\delta^e$, while the matrix $U_\nu$ depends on an angle $\th^\nu$.
In these two scenarios the leptonic CPV phases
possess  non-trivial values.

\vspace{-0.3cm}
\section{ Outlook}
\label{sect:outlook}
%

\vspace{-0.3cm}
The results obtained in 
refs.~\cite{Petcov:2014laa,Girardi:2015vha,Girardi:2015rwa,Feruglio:2012cw,Girardi:2013sza,Ballett:2015wia,Girardi:2016zwz,Penedo:2017vtf,Hanlon:2013ska,Girardi:2014faa,Agarwalla:2017wct,Petcov:2018snn,Altarelli:2009gn}
and in many other studies (quoted in the present and the cited articles) 
show that a sufficiently precise measurement of 
the Dirac phase $\delta$ of the PMNS neutrino mixing matrix
in the current and future neutrino oscillation experiments,
combined with planned improvements of the precision
on the neutrino mixing angles, can provide unique information 
about the possible discrete symmetry origin of 
the observed pattern of neutrino mixing and, correspondingly,
about the existence of new fundamental symmetry in the lepton sector. 
Thus, these experiments will not simply provide 
a high precision data on the neutrino mixing and 
Dirac CPV parameters, but will probe at fundamental level 
the origin of the observed form of neutrino mixing.  
These future data will show, in particular, 
whether Nature followed the discrete symmetry approach for fixing 
the values of the three neutrino mixing angles 
and of the Dirac and Majorana CP violation phases 
of the PMNS neutrino mixing matrix. 
We are looking forward to these data and
to the future exciting developments in neutrino physics.

\vspace{0.3cm}
{\bf Acknowledgements.}
I would like to thank my former Ph.D. students I. Girardi and 
A.V. Titov for the fruitful and enjoyable collaboration 
and the numerous illuminating discussions of the problems 
considered in the present article. This work was supported in part 
by the INFN program on Theoretical Astroparticle Physics (TASP),
by the European Union Horizon 2020 research and innovation programme
under the  Marie Sklodowska-Curie grants 674896 and 690575, 
and by the  World Premier International Research Center 
Initiative (WPI Initiative, MEXT), Japan.


\end{document}